\documentclass[a4paper,fleqn,usenatbib]{mnras}

\usepackage[T1]{fontenc}
\usepackage{ae,aecompl}

\usepackage{graphicx}	
\usepackage{amsmath}	
\usepackage{amssymb}	

\title[Turbulence in relativistic plasma]{Numerical investigation of kinetic turbulence in relativistic pair plasmas I: Turbulence statistics}

\author[V. Zhdankin et al.]{
Vladimir Zhdankin,$^{1}$\thanks{E-mail: zhdankin@jila.colorado.edu}
Dmitri A. Uzdensky,$^{2,3}$
Gregory R. Werner,$^{2}$
\newauthor
Mitchell C. Begelman$^{1,4}$
\\
$^{1}$JILA, NIST and University of Colorado, 440 UCB, Boulder, Colorado 80309, USA\\
$^{2}$Center for Integrated Plasma Studies, Department of Physics, 390 UCB, University of Colorado, Boulder, CO 80309, USA\\
$^{3}$Institute for Advanced Study, 1 Einstein Dr., Princeton, NJ 08540, USA\\
$^{4}$Department of Astrophysical and Planetary Sciences, 391 UCB, Boulder, CO 80309, USA\\
}

\date{Accepted XXX. Received YYY; in original form ZZZ}

\pubyear{2017}

\begin{document}
\label{firstpage}
\pagerange{\pageref{firstpage}--\pageref{lastpage}}
\maketitle

\begin{abstract}
We describe results from particle-in-cell simulations of driven turbulence in collisionless, magnetized, relativistic pair plasma. This physical regime provides a simple setting for investigating the basic properties of kinetic turbulence and is relevant for high-energy astrophysical systems such as pulsar wind nebulae and astrophysical jets. In this paper, we investigate the statistics of turbulent fluctuations in simulations on lattices of up to $1024^3$ cells and containing up to $2 \times 10^{11}$ particles. Due to the absence of a cooling mechanism in our simulations, turbulent energy dissipation reduces the magnetization parameter to order unity within a few dynamical times, causing turbulent motions to become sub-relativistic. In the developed stage, our results agree with predictions from magnetohydrodynamic turbulence phenomenology at inertial-range scales, including a power-law magnetic energy spectrum with index near $-5/3$, scale-dependent anisotropy of fluctuations described by critical balance, log-normal distributions for particle density and internal energy density (related by a $4/3$ adiabatic index, as predicted for an ultra-relativistic ideal gas), and the presence of intermittency. We also present possible signatures of a kinetic cascade by measuring power-law spectra for the magnetic, electric, and density fluctuations at sub-Larmor scales.
\end{abstract}

\begin{keywords}
turbulence -- plasmas -- MHD
\end{keywords}




\section{Introduction}

High-energy astrophysical systems are often composed of low-density, high-temperature plasmas with large-scale motions driven by a variety of mechanisms (e.g., gravitational interactions, shocks, shear flows, thermal instabilities). Under most circumstances, turbulence is inevitable in such situations. High-resolution, multi-wavelength images reveal that systems such as the Crab nebula \citep{hester_2008} and the Messier 87 jet \citep{hines_etal_1989, sparks_etal_1996} are manifestly turbulent, while there are strong theoretical reasons for expecting turbulence in other systems such as black-hole accretion flows \citep{balbus_hawley_1998} and intracluster gas \citep{begelman_fabian_1990}. Turbulent dissipation and particle acceleration is also an attractive mechanism for explaining radiative signatures of distant gamma-ray bursts \citep{kumar_narayan_2009, narayan_kumar_2009, lazar_etal_2009, zhang_etal_2009}. Understanding the fundamental properties of astrophysical turbulence is important for modeling these systems and for interpreting observations.

The basic properties of turbulence in high-energy astrophysical systems remain poorly constrained by both theory and observation, in contrast to turbulence associated with terrestrial fluids, laboratory experiments, and heliospheric plasmas. The plasmas in high-energy astrophysical systems are often characterized by low collisionality and relativistic temperatures, and can develop strong magnetic fields, relativistic bulk motions, shocks, localized dissipative events, and a hard nonthermal particle distribution (responsible for radiative emissions across a wide range of observable frequencies). This leads to complexity that, in principle, can be manifest both at small microphysical scales and at large inertial-range scales.

A proper, first-principles investigation of turbulence in collisionless plasmas requires a kinetic treatment, as opposed to the commonly-used magnetohydrodynamic (MHD) approximation. This is especially crucial for describing small-scale effects such as energy dissipation, transport phenomena, kinetic instabilities, and radiation emission. For high-energy astrophysical systems, collisionless plasma physics has been recognized to play a central role in processes such as electron-ion heating partition \citep[e.g.,][]{begelman_chiueh_1988, quataert_1998, quataert_gruzinov_1999, howes_2010}, particle diffusion, angular momentum transport \citep{sharma_etal_2006, sharma_etal_2007, riquelme_etal_2012, hoshino_2013, hoshino_2015, kunz_stone_quataert_2016}, heat conduction \citep{narayan_medvedev_2001}, and particle acceleration.

Significant recent progress toward understanding kinetic turbulence in plasmas has been made with massively parallel numerical simulations \citep[e.g.,][]{tatsuno_etal_2009, howes_etal_2011, wan_etal_2012, tenbarge_howes_2013, karimabadi_etal_2013, wu_etal_2013, makwana_etal_2015, wan_etal_2015, parashar_etal_2015, roytershteyn_etal_2015, told_etal_2015, kunz_stone_quataert_2016, franci_etal_2016, cerri_etal_2016}. These studies focused mainly on the non-relativistic regime and were often performed under various approximations (e.g., reduced dimensionality or hybrid-kinetic framework) due to computational constraints. Somewhat surprisingly, there appear to have been no systematic studies of kinetic turbulence in the relativistic regime, despite it being a more tractable numerical problem.

In this paper, we describe results from particle-in-cell (PIC) simulations of driven turbulence in magnetized, collisionless, relativistically-hot electron-positron (pair) plasmas. These simulations are sufficiently large to properly capture the transition from an inertial-range MHD cascade at large scales to a kinetic cascade at sub-Larmor scales. The simulations were first introduced in our recent letter, \cite{zhdankin_etal_2017}, where we demonstrated that turbulence can efficiently accelerate particles to a nonthermal, power-law energy distribution. In this follow-up paper, we discuss the turbulence statistics of these simulations in more detail. We plan to follow this paper by a second paper which will address the particle statistics, acceleration mechanisms, and astrophysical implications.

To the best of our knowledge, this work represents the first systematic numerical investigation of turbulence in this physical regime, which is relevant for understanding relativistic astrophysical systems such as pulsar wind nebulae and astrophysical jets made of pair plasmas \citep{rees_gunn_1974, reynolds_etal_1996}. We characterize turbulence statistics in the inertial range and in the kinetic range using standard methodologies, including probability density functions, Fourier power spectra, and structure functions. We find that statistics in the inertial range agree with predictions from phenomenological theories of MHD turbulence \citep{goldreich_sridhar_1995, thompson_blaes_1998}, including equipartition of magnetic and bulk kinetic energies, an inertial-range energy spectrum that approaches a power-law with $-5/3$ index, anisotropy of magnetic fluctuations in agreement with critical balance, log-normal distributions for particle density and internal energy per particle (related by a $4/3$ adiabatic index), and the presence of intermittency. We also show that the turbulence statistics become qualitatively different in the kinetic range, where energy is dissipated by collisionless plasma processes. In particular, we perform new measurements of steep power-law spectra at sub-Larmor scales.

This paper is organized as follows. We provide an overview of theoretical background and the literature in Section~\ref{sec2}. We then describe the numerical background and list of simulations in Section~\ref{sec3}. We present our findings in Section~\ref{sec4}: visuals are described in Section~\ref{sec:visuals}, evolution of parameters and energies in \ref{sec:evolution} and \ref{sec:energetics}, probability density functions in \ref{sec:pdfs}, power spectra in \ref{sec:spectra}, and structure functions in \ref{sec:sf} and \ref{sec:anisotropy}. Finally, we close with a discussion in Section~\ref{sec5} and a conclusion in Section~\ref{sec6}.

\section{Background} \label{sec2}

\subsection{Turbulence preliminaries} \label{sec:turb_back}

Turbulence in a collisionless plasma can be divided into two physical regimes based on the spatial scale of fluctuations. The inertial range involves fluctuations that are much larger than the characteristic plasma kinetic scales (i.e., particle Larmor radius and plasma skin depth) and is rigorously described by MHD models for a wide class of underlying particle distributions \citep{schekochihin_etal_2009, kunz_etal_2015}. The kinetic range encompasses fluctuations that are smaller than the kinetic scales and demands a more complete physical model (e.g., the Vlasov-Maxwell system of equations). These two regimes can be further partitioned into multiple subregimes, such as the weak and strong turbulence regimes for MHD or the sub-ion and sub-electron regimes for the kinetic cascade. In this section, we provide a brief qualitative summary of turbulence in these two regimes.

The theory for relativistic turbulence is substantially less developed than for the non-relativistic case, with only a handful of rigorous analytic works on the topic for hydrodynamic fluids \citep{fouxon_oz_2010, liu_oz_2011, eling_etal_2011, eyink_drivas_2017} and MHD \citep{thompson_blaes_1998}. There is uncertainty in basic notions such as the relevance of an energy cascade \citep{fouxon_oz_2010}, the proper quantities with which to characterize the inertial range, and the role of compressive fluctuations\footnote{Relativistic turbulence necessarily becomes compressible since the speed of sound is bounded by the speed of light (specifically, $c_s = c/\sqrt{3}$ in an ultra-relativistic fluid). Highly compressible (i.e., supersonic) turbulence is nontrivial to describe even in the non-relativistic regime \citep[e.g.,][]{cho_lazarian_2003, kritsuk_etal_2007, galtier_banerjee_2011, federrath_2013} and is richer than the incompressible case due to the freedom to use a variety of equations of states (e.g., isothermal, adiabatic) to describe different classes of fluids.}. Numerical simulations of relativistic hydrodynamic and MHD turbulence have demonstrated that energy spectra and structure functions are broadly similar to the non-relativistic case \citep[e.g.,][]{cho_2005, zrake_macfadyen_2011, zrake_macfadyen_2012, radice_rezzolla_2013, cho_lazarian_2013, zrake_2014, zrake_east_2016, takamoto_lazarian_2016, takamoto_lazarian_2017}, although the inertial range in these simulations is often limited.

Therefore, for the present work, we will focus on comparing inertial-range statistics to existing non-relativistic MHD turbulence phenomenology. This is reasonable for our simulations because the magnetization is of order unity, so that bulk motions are only marginally relativistic at large scales and become increasingly sub-relativistic at smaller scales. The plasmas in our simulations, however, do have an ultra-relativistic temperature, which alters the equation of state from the non-relativistic case. In principle, this could affect the bulk properties of the turbulence, although these changes are small unless there is a significant compressive component.

The classical phenomenology of strong incompressible MHD turbulence in the non-relativistic regime is based on the theory of \cite{goldreich_sridhar_1995}, which has accumulated support from numerical simulations \citep[e.g.,][]{cho_vishniac_2000, maron_goldreich_2001, cho_etal_2002, muller_etal_2003, beresnyak_2014} and solar wind measurements \citep[e.g.,][]{horbury_etal_2008, wicks_etal_2010}. According to the Goldreich-Sridhar model, turbulent fluctuations in the magnetic field and fluid velocity, denoted $\delta \boldsymbol{B}_k$ and $\delta \boldsymbol{v}_k$ respectively, cascade from small wavevectors $\boldsymbol{k}$ (large scales) to large $\boldsymbol{k}$ (small scales), with nonlinear energy transfer caused by interactions between counter-propagating Alfv\'{e}n wave packets that travel along the background magnetic field $\boldsymbol{B}_0$. This process transfers energy through the inertial range to progressively smaller scales. For scales sufficiently deep in the inertial range ($kL/2\pi \gg 1$, where $L$ is the energy injection scale), turbulence is strong, i.e., the interactions are predominantly nonlinear. The turbulence also becomes anisotropic due to the presence of the large-scale magnetic field, necessitating the decomposition of the fluctuation wavevector into components parallel and perpendicular to $\boldsymbol{B}_0$, denoted $k_\parallel$ and $k_\perp$, respectively. Each stage of the strong turbulence energy cascade is conjectured to satisfy critical balance, which is the condition that linear timescales $\tau_A \sim 1/ k_\parallel v_A$ (associated with shear Alfv\'{e}n waves that propagate at the Alfv\'{e}n velocity $v_A$) match nonlinear timescales $\tau_{nl} \sim 1 / k_\perp \delta v_k$. This leads to a scale-dependent anisotropy of the turbulent fluctuations such that $k_\parallel \sim k_\perp^{2/3} L^{-1/3}$. The inertial-range perpendicular energy spectrum is predicted to be $E(k_\perp) \sim k_\perp^{-5/3}$, while the parallel energy spectrum is given by $E(k_\parallel) \sim k_\parallel^{-2}$. The energy spectrum has comparable contributions from magnetic energy and kinetic energy.

More recent phenomenological models of MHD turbulence account for possible correlations between the turbulent flow and magnetic field fluctuations \citep{boldyrev_2005, boldyrev_2006, chandran_etal_2015, mallet_schekochihin_2016}. In particular, scale-dependent dynamic alignment may cause the energy spectrum to become shallower (with index approaching $-3/2$) for sufficiently strong guide field or large system size, as previously demonstrated in MHD simulations \citep{mason_etal_2006, mason_etal_2008, perez_etal_2012}. Due to relatively small size and modest guide field, the simulations described in this paper are unlikely to have reached this asymptotic regime, if at all present under these physical conditions.

The Goldreich-Sridhar phenomenology was extended to the limit of ultra-relativistic strong MHD turbulence by \cite{thompson_blaes_1998}. In this case, the plasma magnetization is assumed to be very large ($\sigma \gg 1$), or equivalently, the Alfv\'{e}n velocity approaches the speed of light ($v_A \to c$). In this limit, the plasma has negligible inertia (and can thus be treated as a massless fermion fluid), allowing the system to be described by force-free MHD. \cite{thompson_blaes_1998} described the nonlinear mode interactions for long-wavelength, low-frequency pertubations (corresponding to Alfv\'{e}n, slow, and fast modes). Assuming critical balance, they showed that the inertial-range nonlinear interactions are predominantly Alfv\'{e}nic and produce a magnetic energy spectrum identical to the non-relativistic case.

We now proceed to discuss the kinetic range. \cite{schekochihin_etal_2009} analytically investigated kinetic cascades in non-relativistic electron-proton plasmas using the gyrokinetic framework, which is appropriate for describing strongly anisotropic turbulent fluctuations with frequencies below the proton cyclotron frequency (generally satisfied deep within an Alfv\'{e}nic cascade). It was shown that the turbulent cascade can continue through the kinetic range, down to scales at which weak collisionality irreversibly converts energy to heat (and therefore produces entropy). It was pointed out that there are several possible classes of kinetic cascades, with the realized outcome depending on physical parameters. The main cases were that of (1) a kinetic Alfv\'{e}n wave (KAW) cascade (naturally continuing the MHD Alfv\'{e}nic cascade below kinetic scales), in which the power spectra of electric, density, and magnetic fluctuations were calculated to be $E_E \sim k_\perp^{-1/3}$, $E_n \sim k_\perp^{-7/3}$, and $E_B \sim k_\perp^{-7/3}$, and (2) an entropy cascade, which develops fine structure in particle velocity space until collisions damp the fluctuations, characterized by $E_E \sim k_\perp^{-4/3}$, $E_n \sim k_\perp^{-10/3}$, and $E_B \sim k_\perp^{-16/3}$. Measurements in the solar wind are consistent with the KAW cascade at scales below the ion gyroradius \citep{alexandrova_etal_2009, sahraoui_etal_2009}, which may continue as an entropy cascade at scales below the electron gyroradius. It is reasonable to expect similar kinetic cascades to occur in a collisionless relativistic pair plasma, although the phenomenology must be extended to this regime.

Finally, we end this subsection with a comment regarding the connection between the inertial range and kinetic range. The division of turbulence into these two regimes is convenient, but the coupling between the two can be highly nontrivial. For example, in high-beta plasmas, kinetic effects can build up a pressure anisotropy that destabilizes large-scale, high-amplitude shear Alfv\'{e}n waves to the firehose instability, conceivably causing a direct transfer of energy from large scales to kinetic scales, bypassing the turbulent cascade \citep{squire_quataert_shekochihin_2016, squire_etal_2017}. In a similar vein, in low-beta plasmas with dynamic alignment, the inertial-range energy cascade may be mediated by the collisionless tearing instability at small scales, rather than classical nonlinear eddy interactions. This leads to a steeper spectrum (the so-called disruption range) that extends beyond kinetic scales into the inertial range \citep{loureiro_boldyrev_2017, mallet_etal_2017}. As a third example, the extent of the MHD inertial range governs the degree of anisotropy and spatial inhomogeneity (via intermittency) of small-scale fluctuations, which can influence the nature of the kinetic fluctuations and the dominant dissipation channels. These aspects of turbulence are highly nonlinear and are therefore well-suited for study by kinetic simulation.

\subsection{Vlasov-Maxwell equations} \label{sec:vlasov}

The kinetic dynamics of a collisionless pair plasma can be described by the Vlasov-Maxwell system of equations\footnote{For simplicity, we present the vector formulation rather than the covariant tensor formulation \citep[e.g.,][]{brizard_chan_1999}.}. Let $f^\pm(\boldsymbol{x},\boldsymbol{p},t)$ be the  positron (electron) distribution function at position $\boldsymbol{x}$, momentum $\boldsymbol{p}$, and time $t$; for later use, we also denote the total particle distribution function $f = f^+ + f^-$. The Vlasov-Maxwell equations describe the evolution of $f^\pm$, along with the magnetic field $\boldsymbol{B}(\boldsymbol{x},t)$ and electric field $\boldsymbol{E}(\boldsymbol{x},t)$, and are given by
\begin{align}
\partial_t f^\pm &= -\boldsymbol{v} \cdot \nabla f^\pm \mp e \left( \boldsymbol{E} + \frac{\boldsymbol{v} \times \boldsymbol{B}}{c} \right) \cdot \frac{\partial f^\pm}{\partial \boldsymbol{p}} \nonumber \\
\partial_t \boldsymbol{E} &= c \nabla \times \boldsymbol{B} - 4 \pi \boldsymbol{J} \nonumber \\
 \partial_t \boldsymbol{B} &= - c \nabla \times \boldsymbol{E} \, , \label{eq:vlasov}
\end{align}
along with the constraints $\nabla \cdot \boldsymbol{E} = 4 \pi \rho$ and $\nabla \cdot \boldsymbol{B} = 0$. Here, the particle velocity is given by $\boldsymbol{v} = \boldsymbol{p} c / \sqrt{m^2 c^2 + p^2}$ (where $m$ is the electron rest mass and $c$ is the speed of light), while the charge density $\rho(\boldsymbol{x},t)$ and current density $\boldsymbol{J}(\boldsymbol{x},t)$ are given by
\begin{align}
\rho &= e \int d^3 p (f^+ - f^-) \nonumber \\
\boldsymbol{J} &= e \int d^3 p \boldsymbol{v} (f^+ - f^-)
\end{align}
where $e$ is the elementary charge.

For an ultra-relativistic pair plasma, such that the mean particle Lorentz factor is~$\bar{\gamma} \gg 1$ (where $\gamma = \sqrt{1 + p^2/m^2c^2}$ is the individual particle Lorentz factor), the characteristic kinetic scales are given by the Larmor radius $\rho_e = \bar{\gamma} m c^2 / e B_{\rm rms}$ and plasma skin depth $d_e = \sqrt{\bar{\gamma} m c^2 / 4 \pi n_0 e^2}$, given mean total particle density~$n_0$ and characteristic (rms) magnetic field~$B_{\rm rms}$. The Debye length is given by $\lambda_D = \sqrt{T_e / 4\pi n_0 e^2}$, where $T_e$ is the temperature of electrons and positrons (assumed to be equal). For an ultra-relativistic Maxwell-J\"{u}ttner distribution, $T_e = \bar{\gamma} m c^2/3$, so that the Debye length is always comparable to the skin depth, $\lambda_D/d_e = 1/\sqrt{3}$. Given the three characteristic scales $L$, $\rho_e$, and $d_e$ in the system, one can form two free dimensionless parameters, which we take to be the system size relative to the Larmor radius~$L / \rho_e$ and the nominal magnetization\footnote{It is also common to define the magnetization as $B_{\rm rms}^2/4\pi \bar{w}_e$ where $\bar{w}_e$ is the mean relativistic enthalpy density, given by $\bar{w}_e = (4/3) n_0 \bar{\gamma} m c^2$ for ultra-relativistic particles, which yields a different pre-factor from the definition used in this paper.} $\sigma = B_{\rm rms}^2 / 4 \pi n_0 \bar{\gamma} m c^2 = (d_e/\rho_e)^2$. These are the two parameters which we vary in this study. We note that the magnetization is related to plasma beta, $\beta = 8\pi n T_e/B_{\rm rms}^2$, by $\sigma = 3/2\beta$ for an isotropic Maxwell-J\"{u}ttner particle distribution.

For completeness, we note that there are several covariant quantities derived from the Maxwell-Vlasov system, which take a fundamental role in fluid theories. One of these is the stress-energy tensor ${\mathcal T}^{\mu\nu} = {\mathcal T}^{\mu\nu}_M + {\mathcal T}^{\mu\nu}_V$, where the Maxwell and Vlasov contributions are given by \citep[e.g.,][]{weinberg_1972}
\begin{align}
{\mathcal T}^{\mu\nu}_M &= \frac{1}{16 \pi} F_{\alpha\beta} F^{\alpha\beta} g^{\mu\nu} - \frac{1}{4\pi} F^{\mu\alpha} g_{\alpha\beta} F^{\nu\beta} \nonumber \\
{\mathcal T}^{\mu\nu}_V &= \int d^3p \frac{p^{\mu} p^{\nu} c}{\sqrt{m^2c^2 + p^2}} f \, ,
\end{align}
where $F^{\mu\nu} = \partial^\mu A^\nu - \partial^\nu A^\mu$ is the electromagnetic tensor, $A^\mu = (\phi, \boldsymbol{A})$ is the electromagnetic four-potential (satisfying $\boldsymbol{E} = - \nabla \phi - \partial_t \boldsymbol{A}/c$ and $\boldsymbol{B} = \nabla \times \boldsymbol{A}$), $g^{\mu\nu}$ is the Minkowski metric, and $p^{\mu} = (\gamma m c,\boldsymbol{p})$ is the particle four-momentum. Conservation of energy and momentum is given by $\partial {\mathcal T}^{\mu\nu}/\partial x^\mu = 0$, which serves as the fluid equation of motion. Another covariant quantity is the number density four-current, given by
\begin{align}
{\mathcal N}^\mu = \int d^3p \frac{p^\mu}{\sqrt{m^2 c^2 + p^2}} f \, .
\end{align}
Conservation of particle number is given by~$\partial {\mathcal N}^\mu / \partial x^\mu = 0$. Although these covariant quantities can be used to characterize the turbulence, we instead focus on the non-covariant quantities defined on the simulation lattice, as described in the following subsection.

\subsection{Turbulence characterization} \label{sec:defs}

To characterize turbulence, we measure statistics of $\boldsymbol{B}$, $\boldsymbol{E}$, and various fluid quantities obtained from the particle distribution function $f$ by integrating over momentum space. The basic fluid quantities include the lab-frame particle density $n(\boldsymbol{x},t)$, fluid velocity $\boldsymbol{v}_f(\boldsymbol{x},t)$, fluid energy density ${\mathcal E}_f(\boldsymbol{x},t)$, and fluid momentum density ${\boldsymbol{\mathcal  P}}_f(\boldsymbol{x},t)$. These are given by
\begin{align}
n(\boldsymbol{x},t) &= \int d^3p f(\boldsymbol{p}, \boldsymbol{x},t) \nonumber \\
\boldsymbol{v}_f(\boldsymbol{x},t) &= \frac{1}{n(\boldsymbol{x},t)} \int d^3p \frac{\boldsymbol{p} c}{\sqrt{m^2c^2 + p^2}} f (\boldsymbol{p},\boldsymbol{x},t) \nonumber \\
{\mathcal E}_f(\boldsymbol{x},t) &= \int d^3p \sqrt{m^2 c^4 + p^2 c^2} f (\boldsymbol{p}, \boldsymbol{x},t) \nonumber \\
{\boldsymbol{\mathcal  P}}_f(\boldsymbol{x},t) &= \int d^3p \boldsymbol{p} f (\boldsymbol{p},\boldsymbol{x},t) \, .
\end{align}
In terms of the covariant quantities, ${\mathcal N}^\mu = (n, n \boldsymbol{v}_f/c)$ and ${\mathcal T}_V^{0\mu} = ({\mathcal E}_f, \boldsymbol{\mathcal P}_f c)$. We will not consider higher-order moments of the distribution function, which include the pressure tensor and heat flux, in the present work.

The total energy in the system is given by the sum of the magnetic, electric, and kinetic energies:
\begin{align}
E_{\rm tot}(t) &= \int d^3x {\mathcal T}^{00} \nonumber \\
&= \int d^3x \left[ \frac{|\boldsymbol{B}(\boldsymbol{x},t)|^2}{8\pi} +  \frac{|\boldsymbol{E}(\boldsymbol{x},t)|^2}{8\pi} + {\mathcal E}_f(\boldsymbol{x},t) \right] \, . \label{eq:tot_energy}
\end{align}
We denote the mean-field magnetic energy density ${\mathcal E}_{\rm mean} = B_0^2/8\pi$ and the turbulent magnetic energy density ${\mathcal E}_{\rm mag}(\boldsymbol{x},t) = |\delta\boldsymbol{B}(\boldsymbol{x},t)|^2/8\pi$, where $\boldsymbol{B}_0$ is the mean magnetic field and $\delta \boldsymbol{B} = \boldsymbol{B} - \boldsymbol{B}_0$ is the fluctuating part. We also denote the electric energy density ${\mathcal E}_{\rm elec}(\boldsymbol{x},t) = |\boldsymbol{E}(\boldsymbol{x},t)|^2/8\pi$. The total turbulent magnetic energy is given by $E_{\rm mag}(t) = \int d^3x {\mathcal E}_{\rm mag}(\boldsymbol{x},t)$, electric energy by $E_{\rm elec}(t) = \int d^3x {\mathcal E}_{\rm elec}(\boldsymbol{x},t)$, and kinetic energy by $E_{\rm kin}(t) = \int d^3x {\mathcal E}_f(\boldsymbol{x},t)$ (which is simply a sum of all particle kinetic energies). In the absence of external energy sources or sinks (such as for decaying turbulence in a closed domain), the total energy $E_{\rm tot}$ is conserved. However, since the turbulence in our simulations has an energy source (external driving) but no energy sink, we instead have
\begin{align}
\frac{d E_{\rm tot}}{dt} = \dot{E}_{\rm inj} \, ,
\end{align}
where $\dot{E}_{\rm inj}(t)$ is the energy injection rate due to the external driving.

It is also beneficial to further decompose the total particle kinetic energy~$E_{\rm kin}$ into internal fluid energy and bulk fluid kinetic energy. This is complicated by the absence of a standard model for dissipative relativistic fluids \citep[][]{eckart_1940, landau_lifshitz_1959, israel_stewart_1979}; see \cite{andersson_comer_2007} for a review. In this work, we apply the following ad-hoc prescription. We subdivide the domain into cells (associated with, e.g., simulation lattice cells); in each cell, one can form a four-momentum corresponding to the system of particles in the cell, $p^\mu_{\rm cell} = \Sigma_{i} p^\mu_i$, where the summation is taken over all particles (electrons and positrons) in the cell. The scalar $(p^\mu_{\rm cell} p_{{\rm cell},\mu}/c^2)^{1/2}$ can then be interpreted as the center-of-momentum frame energy of the system of particles in the cell. One can then associate the fluid internal energy density ${\mathcal E}_{\rm int}$ with this scalar (divided by the fixed cell volume), while the fluid bulk kinetic energy density ${\mathcal E}_{\rm bulk}$ is given by the remainder; in terms of fluid quantities, this translates to
\begin{align}
{\mathcal E}_{\rm int} &= \sqrt{{\mathcal E}^2_f - |{\boldsymbol{\mathcal  P}}_f|^2 c^2} \nonumber \\
{\mathcal E}_{\rm bulk} &= {\mathcal E}_f - {\mathcal E}_{\rm int} = |\boldsymbol{{\mathcal W}}|^2 \, ,
\end{align} 
where we defined $\boldsymbol{{\mathcal W}} \equiv {\boldsymbol{\mathcal  P}}_f c / [{\mathcal E}_f + ({\mathcal E}_f^2 - |{\boldsymbol{\mathcal  P}}_f|^2 c^2)^{1/2}]^{1/2}$ (which has dimensions compatible with $\boldsymbol{B}$). The corresponding total internal energy and total bulk fluid kinetic energy are denoted by $E_{\rm int}(t) = \int d^3x {\mathcal E}_{\rm int}(\boldsymbol{x},t)$ and $E_{\rm bulk}(t) = \int d^3x {\mathcal E}_{\rm bulk}(\boldsymbol{x},t)$. In the non-relativistic limit, ${\mathcal E}_{\rm int} \to \int d^3p [m c^2 + |\boldsymbol{p} - \boldsymbol{p}_f(\boldsymbol{x},t)|^2/2 m] f(\boldsymbol{p}, \boldsymbol{x},t)$, where $\boldsymbol{p}_f(\boldsymbol{x},t) = \boldsymbol{\mathcal P}_f(\boldsymbol{x},t) / n(\boldsymbol{x},t)$, while $\boldsymbol{{\mathcal W}} \to (n / 2 m)^{1/2} \boldsymbol{p}_f$.

The prescription proposed above for decomposing particle kinetic energy into bulk fluid energy and internal energy is not Lorentz covariant; one may instead use a covariant definition such as ${\mathcal N}^\mu {\mathcal N}^\nu {\mathcal T}_{\mu\nu}/n^2$ \cite[e.g.,][]{eckart_1940}, which requires measuring the pressure tensor. Our method is convenient for estimating energies on the lattice of a numerical simulation, and has the following simple interpretation. In each cell, the fluid can be characterized by an effective velocity,
\begin{align}
\frac{\boldsymbol{V}_{\rm eff}}{c} = \frac{{\boldsymbol{\mathcal  P}}_f c}{{\mathcal E}_f} = \frac{{\boldsymbol{\mathcal  P}}_f c}{\sqrt{{\mathcal E}_{\rm int}^2 + |{\boldsymbol{\mathcal  P}}_f c|^2}} \, , 
\end{align}
and corresponding effective bulk Lorentz factor $\Gamma_{\rm eff} = 1/\sqrt{1 - V_{\rm eff}^2/c^2}$. The internal energy is then an effective rest mass density in the frame moving with $V_{\rm eff}$, so that ${\mathcal E}_{\rm int}  = {\mathcal M}_{\rm eff} c^2 = {\mathcal E}_f/\Gamma_{\rm eff}$, and the bulk energy is an effective kinetic energy, ${\mathcal E}_{\rm bulk} = (\Gamma_{\rm eff} - 1) {\mathcal M}_{\rm eff}c^2$.
 
\section{Simulations} \label{sec3}

\subsection{Simulation details} \label{sec:sim}

In this work, we apply PIC simulations to model turbulence in collisionless, relativistic pair plasmas from first principles. PIC simulations provide numerical approximations to the solutions of Vlasov-Maxwell equations (Eqs.~\ref{eq:vlasov}) by evolving a large population of particles rather than evolving the distribution functions $f^\pm$ directly. Hence, electrons and positrons are evolved by the Lorentz force,
\begin{align}
\frac{d\boldsymbol{p}_i}{dt} &= q_i \left[\boldsymbol{E}(\boldsymbol{x}_i,t) + \frac{\boldsymbol{v}_i}{c} \times \boldsymbol{B}(\boldsymbol{x}_i,t) \right] \nonumber \\
\frac{d\boldsymbol{x}_i}{dt} &= \boldsymbol{v}_i 
\end{align}
where $\boldsymbol{x}_i$ and $\boldsymbol{p}_i$ are the position and momentum of the $i$th particle, $q_i$ is its charge ($-e$ for electrons, $e$ for positrons), and $\boldsymbol{v}_i = \boldsymbol{p}_i c / \sqrt{m^2 c^2 + p_i^2}$ is the particle velocity. The magnetic field and electric field are obtained by solving the Maxwell equations on the lattice, using charge and current densities obtained from the particles in each cell.

We performed our simulations using the electromagnetic PIC code \textsc{Zeltron} \citep{cerutti_etal_2013}. \textsc{Zeltron} evolves the Vlasov-Maxwell system in time, discretizing the electromagnetic fields, charge density and current density on a regular Cartesian grid and approximating the distribution function in Monte Carlo fashion by a collection of representative macroparticles. The subsequent evolution is essentially a time-integration via the method of characteristics of the Vlasov-Maxwell equations, where the macroparticle trajectories follow physical particle trajectories. Using robust, well-tested PIC methods \citep{BirdsallPIC}, \textsc{Zeltron} evolves the fields on the standard Yee mesh \citep{Yee-1966}, leapfrogging the electric and magnetic fields as well as macroparticle positions and velocities. Macroparticles are moved via the standard relativistic Boris push \citep{Boris-1970,Vay-2008}. To avoid small numerical errors in the electric field building up over long times, \textsc{Zeltron} uses a divergence-cleaning algorithm, adjusting the electric field slightly at each timestep to maintain Gauss's law throughout the simulation. \textsc{Zeltron} is parallelized using MPI (message passing interface) and spatial domain decomposition.

We choose a periodic cubic domain of size $L^3$ with uniform background magnetic field $\boldsymbol{B}_0 = B_0 \hat{\boldsymbol{z}}$. Since MHD turbulence is generally considered to be sensitive to dimensionality \citep[e.g.,][]{howes_2015}, we focus on 3D simulations. We initialize the simulations with zero electromagnetic fluctuations ($\delta \boldsymbol{B} = \boldsymbol{E} = 0$) and particles sampled from a uniform non-drifting Maxwell-J\"{u}ttner distribution,
\begin{align}
f_0(\boldsymbol{p}) = \frac{1}{4\pi m^3 c^3 \theta K_2(1/\theta)} \exp{\left( - \frac{\sqrt{1 + p^2/m^2c^2}}{\theta}\right)} \, ,
\end{align}
where $K_2$ is the modified Bessel function of the second kind. We choose an initial temperature $\theta \equiv T_e/mc^2 = 100$, yielding particles with a mean Lorentz factor $\bar{\gamma}_0 \approx 300$. This initial stable thermal equilibrium is then disrupted by external driving. To drive strong, critically-balanced turbulence that naturally occurs within the MHD cascade, we apply a random fluctuating external current density $\boldsymbol{J}_{\rm ext}$ in the form of an oscillating Langevin antenna \citep{tenbarge_etal_2014}. We drive $J_{{\rm ext},z}$ at eight large-scale modes, $\boldsymbol{k}_0 L / 2 \pi \in \{ (1,0,\pm1), (0,1,\pm1), (-1,0,\pm1), (0,-1,\pm1) \}$, and each of $J_{{\rm ext},x}$ and $J_{{\rm ext},y}$ in four modes to enforce $\nabla \cdot \boldsymbol{J}_{\rm ext} = 0$ (necessary to satisfy charge conservation). We choose a driving frequency of $\omega_0 = 0.6 \cdot 2 \pi v_{A0} / \sqrt{3} L$ and decorrelation rate $\Gamma_0 = 0.5 \cdot 2 \pi v_{A0} / \sqrt{3} L$, where $v_{A0} = c \sqrt{\sigma_0/(\sigma_0 + 4/3)}$ is the initial relativistic Alfv\'{e}n velocity\footnote{Alternatively, it may also be reasonable to have a time-dependent driving frequency based on the instantaneous Alfv\'{e}n velocity, to account for the large-scale eddy turnover time increasing as the plasma heats up.} \citep{sakai_kawata_1980, gedalin_1993}. We tune the driving amplitude such that rms magnetic fluctuations are comparable to background field, $\delta B_{\rm rms} \sim B_0$. The energy injection rate in our simulations is given by $\dot{E}_{\rm inj} = - \int d^3x \boldsymbol{E} \cdot \boldsymbol{J}_{\rm ext}$, which is statistically constant in time during developed turbulence.

As noted in Sec.~\ref{sec:vlasov}, the dimensionless physical parameters in the system are the magnetization $\sigma$ and the ratio of system size to Larmor radius $L/\rho_{e}$. We emphasize that our simulations do not achieve a strict statistical steady state, because our numerical set-up includes an energy source (external driving) but no energy sink. For this reason, we parameterize our simulations by the initial magnetization $\sigma_0 \equiv \sigma(t=0)$ and initial ratio of system size to Larmor radius $L/\rho_{e0} \equiv L/\rho_{e}(t=0)$. During fully-developed turbulence, which begins after a few dynamical times and ends before heating causes the growth of~$\rho_e$ to suppress the inertial range (i.e., $\rho_e \sim L/2\pi$), the dimensionless parameters differ from these initial values. As we will show, in practice, some of the turbulence statistics (e.g., magnetic field fluctuations) are insensitive to the time dependence of these physical parameters, while others (e.g., electric field fluctuations) exhibit a secular evolution.

\subsection{List of simulations}

\begin{table*}[h!b!p!]
\caption{List of largest simulations \newline} \label{sims}
\begin{tabular}{|c|c|c|c|c|c|c|c|} 
	\hline
\hspace{0.5 mm} Case \hspace{0.5 mm}  & \hspace{1 mm} $N^3$ \hspace{1 mm}   & \hspace{1 mm} $L/2\pi\rho_{e0}$ \hspace{1 mm}  &   \hspace{1 mm} $\sigma_0$ \hspace{1 mm} &   \hspace{1 mm} $Tc/L$ \hspace{1 mm}  & \hspace{1 mm} $N_{\rm ppc}$ \hspace{1 mm} & \hspace{1 mm} $R_{{\rm err},T} (\%)$ \hspace{1 mm} & \hspace{1 mm} $R_{{\rm err},L/c} (\%)$ \hspace{1 mm} \\
	\hline
A2 & $1024^3$ & 108.6 & $0.5$ & $22.3$ & 128 & $3.7\%$ & $0.16\%$ \\
A4 & $1024^3$ & 108.6 & $2$ & $13.4$ & 192 & $3.0\%$ & $0.23\%$ \\
B1 & $768^3$ & 61.1 & $0.25$ & $22.3$ & 256 & $3.7\%$ & $0.16\%$ \\
B2 & $768^3$ & 81.5 & $0.5$ & $10.1$ & 256 & $0.6\%$ & $0.06\%$ \\
B3 & $768^3$ & 81.5 & $1$ & $11.2$ & 128 & $2.3\%$ & $0.21\%$  \\
B4 & $768^3$ & 81.5 & $2$ & $9.2$ & 128 & $3.3\%$ & $0.36\%$ \\
B5 & $768^3$ & 81.5 & $4$ & $13.4$ & 96 & $2.4\%$ & $0.18\%$ \\
C1 & $512^3$ & 40.7 & $0.25$ & $22.3$ & 256 & $2.3\%$ & $0.10\%$ \\
C2 & $512^3$ & 54.3 & $0.5$ & $17.9$ & 128 & $0.8\%$ & $0.05\%$ \\
C3 & $512^3$ & 54.3 & $1$ & $14.1$ & 128 & $2.0\%$ & $0.14\%$ \\
C4 & $512^3$ & 54.3 & $2$ & $15.1$ & 128 & $2.6\%$ & $0.17\%$ \\
C5 & $512^3$ & 54.3 & $4$ & $15.6$ & 128 & $2.5\%$ & $0.16\%$ \\
	\hline
\end{tabular}
\centering
\label{table-sims}
\end{table*}

We performed a series of simulations on lattices of~$N^3$ cells, where $N \in \{256,384,512,768,1024\}$. The simulations have varying physical parameters~$\sigma_0$ and~$L/\rho_{e0}$, as well as varying durations $T$ and number of particles per cell $N_{\rm ppc}$. The largest of our simulations are listed in Table~\ref{table-sims}; in addition, we did a more thorough parameter scan with $256^3$ and $384^3$ simulations, which are not listed. Our scan in magnetization covers $\sigma_0 \in \{ 0.25, 0.5, 1, 2, 4\}$. For simulations with $\sigma_0 \ge 0.5$, we chose $\rho_{e0} = 1.5 \Delta x$ (where $\Delta x$ is the lattice cell size), corresponding to a ratio of driving scale to initial Larmor radius of $L/2\pi\rho_{e0} \in \{108.6,81.5,54.3,40.7,27.2\}$ for respective $N \in \{1024,768,512,384,256\}$. For simulations with $\sigma_0 = 0.25$, we choose $\rho_{e0} = 2 \Delta x$ so that the plasma skin depth and hence the Debye length is marginally resolved ($d_{e0} = \Delta x$). We run all cases for a duration of at least $9 L/c$. The Alfv\'{e}n crossing time $\tau_A = L/v_A$ is longer than the light crossing time $L/c$ and slowly increases in time; for example, at $\sigma = 0.25$, $\tau_A = 2.5 L/c$. Motivated by convergence studies and demands on energy conservation, we choose $N_{\rm ppc} \ge 128$ for all simulations except for case B5, in which $N_{\rm ppc} = 96$ was chosen to compensate for load imbalance issues; we discuss convergence with respect to $N_{\rm ppc}$ in Appendix~\ref{sec:ppc}. Since energy conservation is not enforced by the numerical scheme, the deviation from exact energy conservation is a measure of numerical error. We list the maximum relative error in energy conservation in Table~\ref{table-sims}, defined by~$R_{{\rm err},T} = \max_{0<t<T}{[|E_{\rm tot}(t)-E_{\rm inj}(t)-E_{\rm tot}(0)|/E_{\rm tot}(0)]}$, where $E_{\rm inj}$ is the amount of injected energy from the external driving. For reference, we also note the typical error per light crossing time,~$R_{{\rm err},L/c} = \max_{0<t<T}{[|E_{\rm tot}(t)-E_{\rm inj}(t)-E_{\rm tot}(0)|/E_{\rm tot}(0)]}(L/cT)$, which better represents the relevant error. The error per simulation tends to be on order of a few percent.

There are three notable simulations which constitute our most robust data sets at large system size. The first, Case A2, is our fiducial case (with $1024^3$ cells, $\sigma_0 = 0.5$, and $L/2\pi\rho_{e0} = 108.6$). Unless otherwise mentioned, we will describe results from this simulation. This case has approximately $1.4 \times 10^{11}$ total particles and duration $T = 22.3 L/c = 11.6 L/v_{A0}$. The second, Case A4, is a representative high-$\sigma_0$ simulation (with $1024^3$ cells, $\sigma_0 = 2$, $L/2\pi\rho_{e0} = 108.6$, and $\sim 2 \times 10^{11}$ particles). The third, Case B1, is well-suited for studying turbulence statistics at low $\sigma$ and for comparing to non-relativistic phenomenology (with $768^3$ cells, $\sigma_0 = 0.25$, $L/2\pi\rho_{e0} = 61.1$, and duration $T = 22.3 L/c = 8.9 L/v_{A0}$).

\section{Results} \label{sec4}

\subsection{Visuals} \label{sec:visuals}

 \begin{figure*}
  \includegraphics[width=\columnwidth]{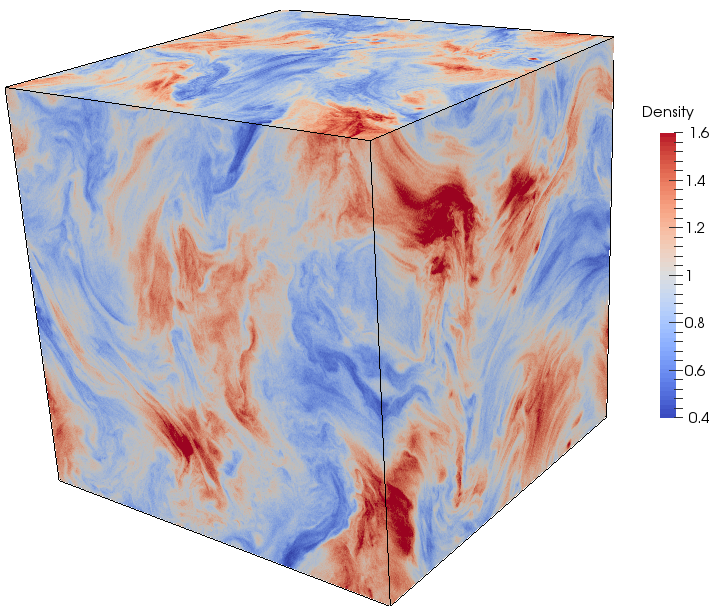}
   \includegraphics[width=\columnwidth]{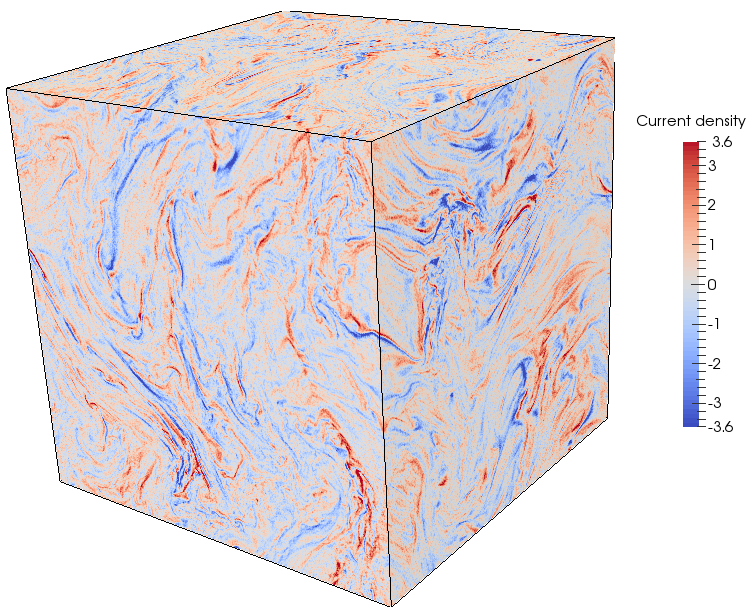}
    \includegraphics[width=\columnwidth]{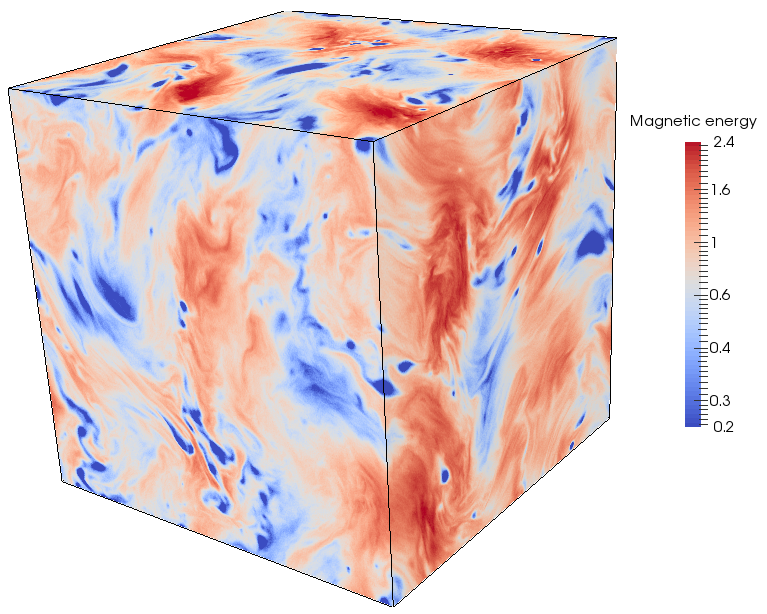}
     \includegraphics[width=\columnwidth]{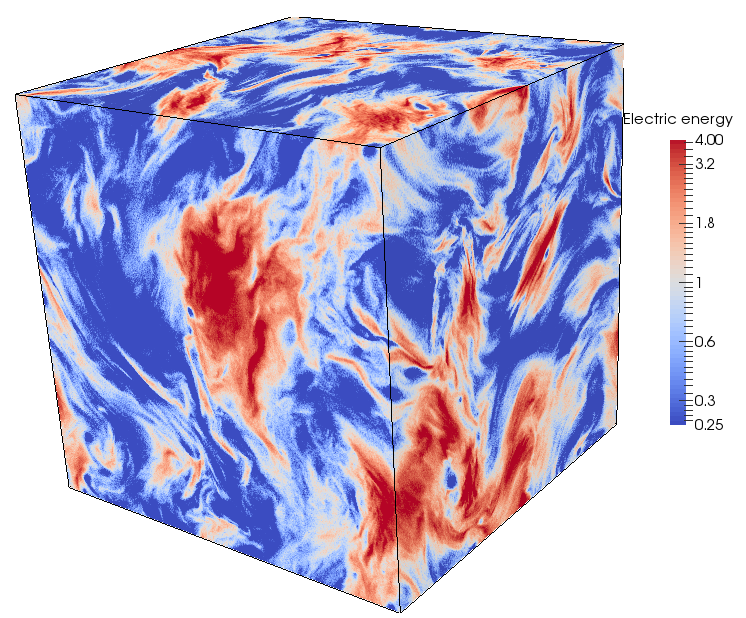}
      \centering
   \caption{\label{fig:images0} Surface visualizations of various quantities. From left to right, top to bottom: particle density $n$, electric current density $J_z$ (component parallel to $\boldsymbol{B}_0$), magnetic energy density ${\mathcal E}_{\rm mag,tot}$, and electric energy density ${\mathcal E}_{\rm elec}$. All quantities are normalized to the mean value, except for current density, which is normalized to the rms value.}
 \end{figure*}

Before beginning the quantitative analysis, we first present some visuals of fully-developed turbulence in our fiducial simulation. In Fig.~\ref{fig:images0}, we show surface images of several quantities on the boundary of the domain: particle density $n$, electric current density $\boldsymbol{J}$, total magnetic energy density ${\mathcal E}_{\rm mag,tot} = B^2/8\pi$, and electric energy density ${\mathcal E}_{\rm elec}$. These quantities all show qualitative differences. For example, the particle density is dominated by large-scale structure, while the current density is dominated by small-scale structure. The magnetic energy exhibits both large-scale structure (in the form of cloud-like structures) and small-scale structure (in the form of magnetic holes).

\begin{figure*}
  \includegraphics[width=\columnwidth]{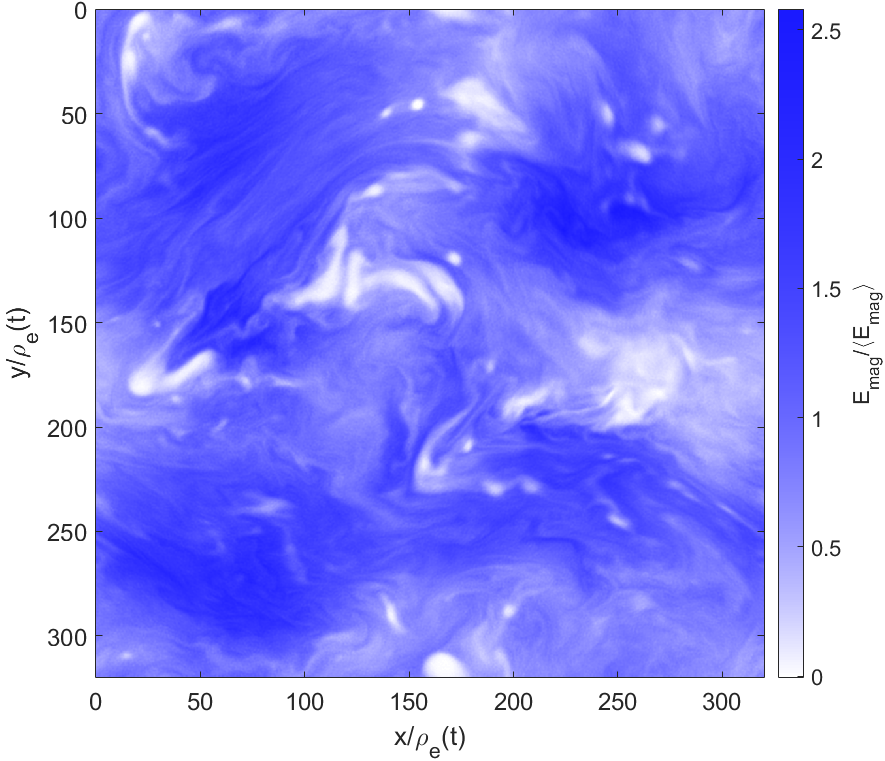}
   \includegraphics[width=\columnwidth]{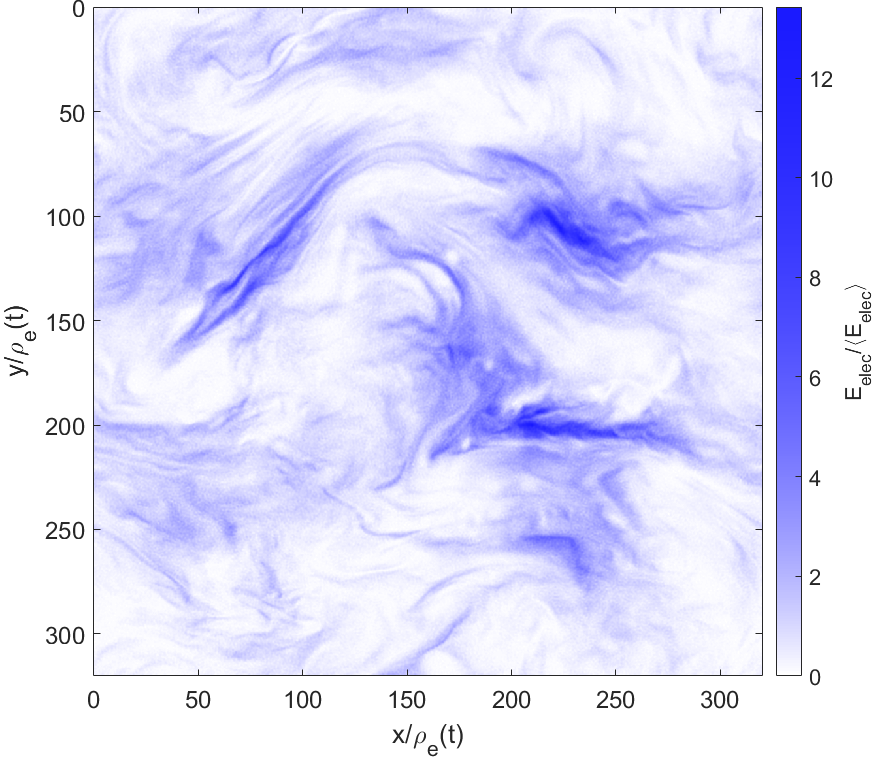}
    \includegraphics[width=\columnwidth]{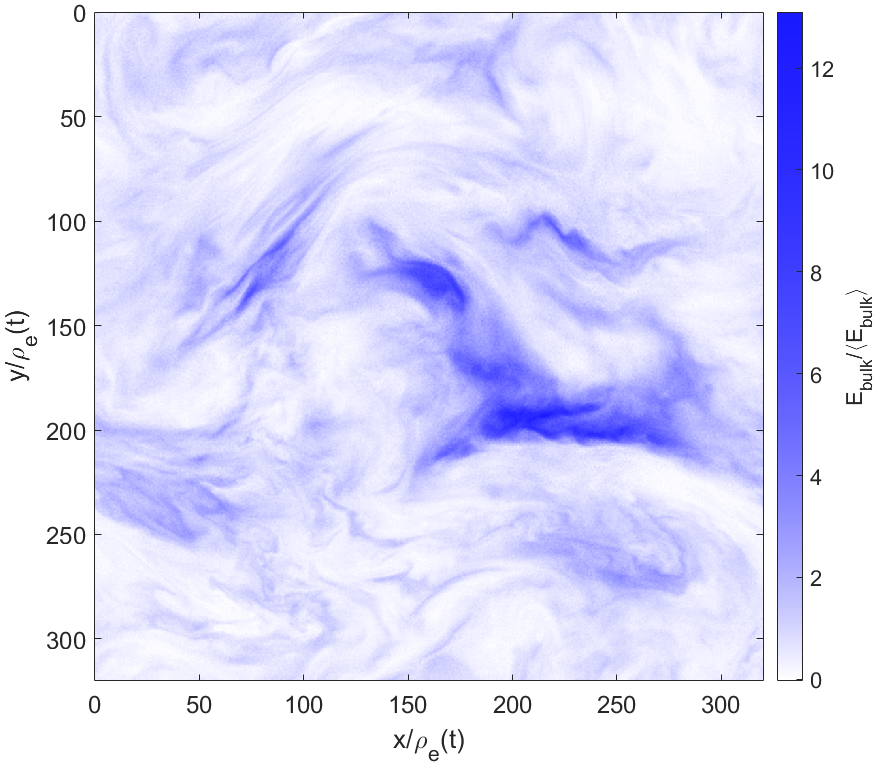}
     \includegraphics[width=\columnwidth]{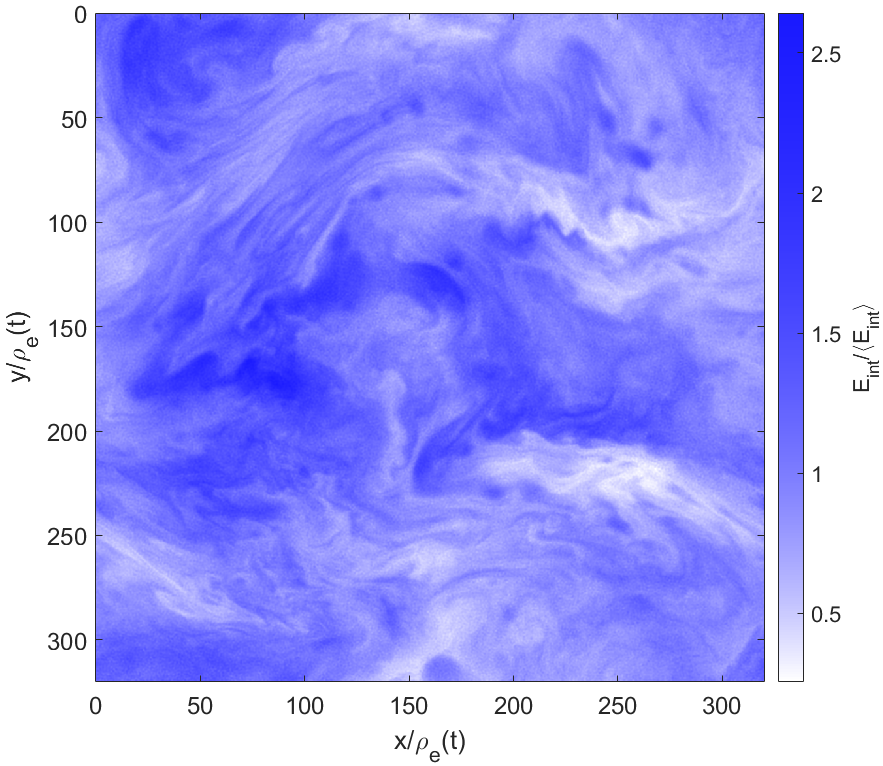}
      \centering
   \caption{\label{fig:images1} Visualization of the energy contributions in an arbitrary $xy$ plane (perpendicular to $\boldsymbol{B}_0$). From left to right, top to bottom: magnetic energy density ${\mathcal E}_{\rm mag}$, electric energy density ${\mathcal E}_{\rm elec}$, bulk fluid energy density ${\mathcal E}_{\rm bulk}$, and internal energy density ${\mathcal E}_{\rm int}$.}
 \end{figure*}

Next, we examine 2D images for various quantities in an arbitrary $xy$ slice of our fiducial simulation. In Fig.~\ref{fig:images1}, we show the energy densities. The magnetic energy density ${\mathcal E}_{\rm mag,tot}$ is marked by round coherent structures inside of which the magnetic field essentially vanishes. These structures tend to coincide with regions of high internal energy density ${\mathcal E}_{\rm int}$, implying that they are pressure-balanced magnetic holes, as also seen in nonrelativistic kinetic turbulence simulations \citep{roytershteyn_etal_2015} and in solar wind observations \citep{turner_etal_1977}. The electric energy density ${\mathcal E}_{\rm elec}$ is strongly correlated with the bulk fluid kinetic energy ${\mathcal E}_{\rm bulk}$, with both marked by sheet-like structures.

  \begin{figure*}
     \includegraphics[width=0.65\columnwidth]{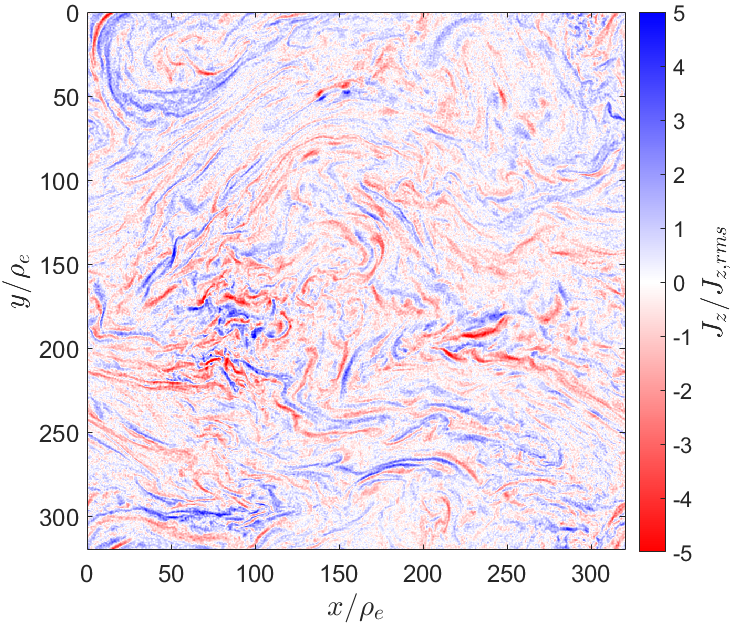}
     \includegraphics[width=0.65\columnwidth]{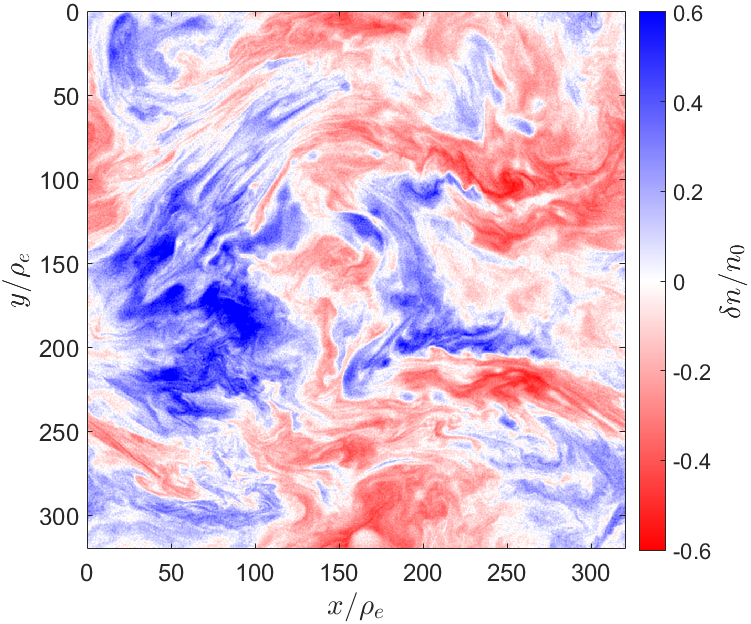}
      \includegraphics[width=0.65\columnwidth]{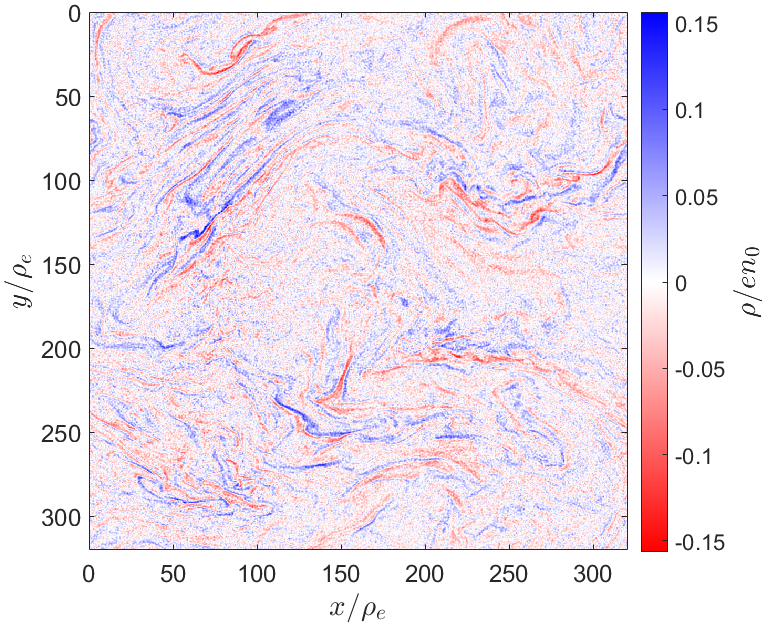}
     \includegraphics[width=0.65\columnwidth]{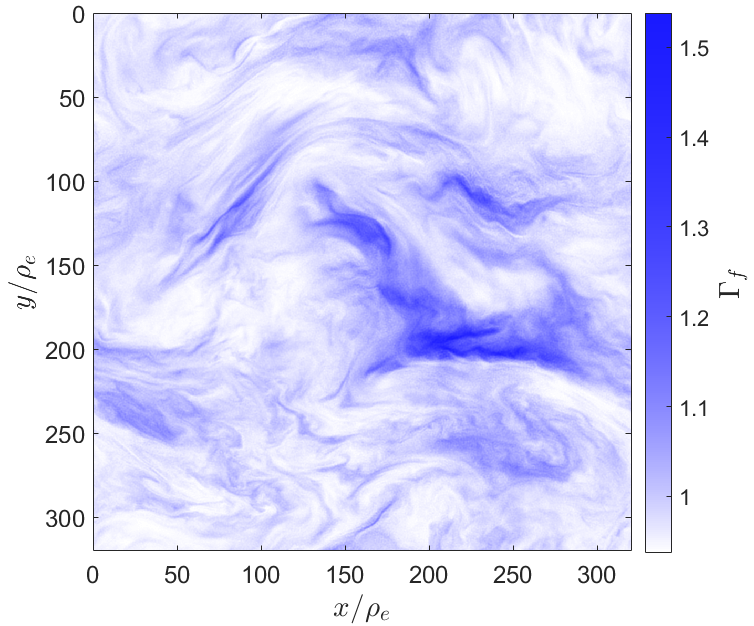}
     \includegraphics[width=0.65\columnwidth]{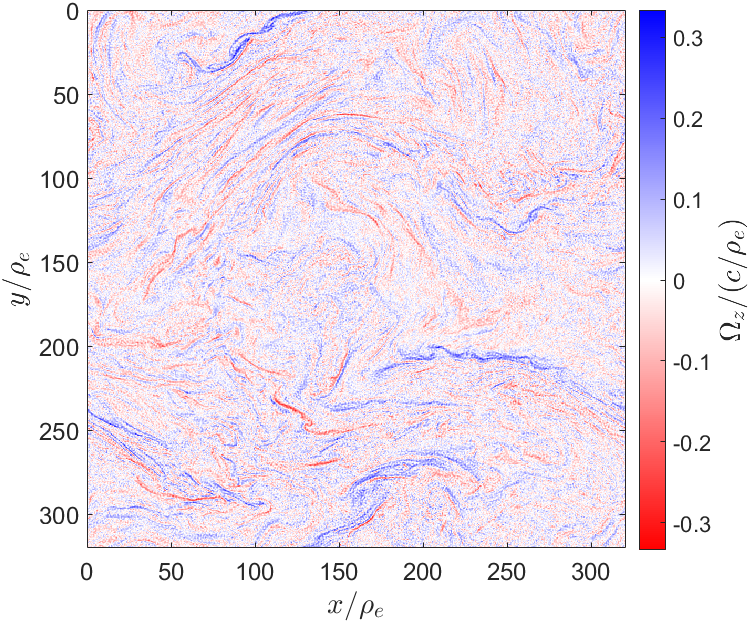}
     \includegraphics[width=0.65\columnwidth]{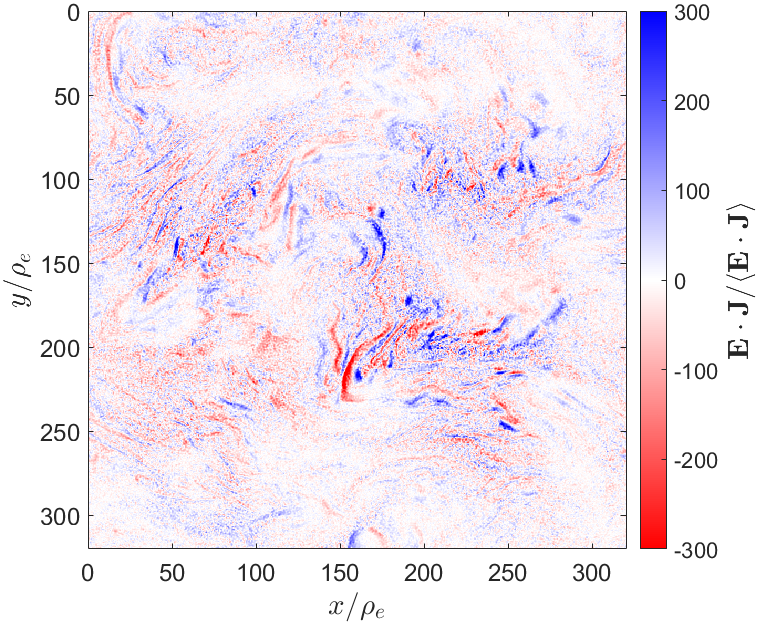}
      \centering
   \caption{\label{fig:images2} Visualization of various quantities in the same planes as in Fig.~\ref{fig:images1}. From left to right, top to bottom: current density $J_z$ (component parallel to $\boldsymbol{B}_0$, normalized to rms value), particle density fluctuations $\delta n/n_0$, charge density $\rho/e n_0$, bulk flow Lorentz factor $\Gamma_f$, vorticity $\Omega_z \rho_e/c$, and electromagnetic dissipation proxy $\boldsymbol{E}\cdot\boldsymbol{J}$ (normalized to mean).}
 \end{figure*}

In Fig.~\ref{fig:images2}, we show similar 2D images for the current density, particle density, charge density, bulk flow Lorentz factor, vorticity, and an electromagnetic dissipation proxy. The current density $\boldsymbol{J}$ is characterized by kinetic-scale structure in the form of current sheets, which have thicknesses near the kinetic scales and lengths spanning a range of scales up to the driving scale. These structures are known to be a consequence of the intermittency of MHD turbulence \citep[see, e.g.,][and references therein]{zhdankin_etal_2016b}. The particle density, on the other hand, is characterized by large, irregular cloud-like structures. Density fluctuations $\delta n = n - n_0$ reach up to $\sim 50\%$ of the mean density. The charge density $\rho$ is characterized by thin sheet-like structures, much like in the current density; we find that the charge density can approach $\rho \sim 0.15 e n_0$. The bulk flow Lorentz factor $\Gamma_f = 1/\sqrt{1 - v_f^2/c^2}$ is typically close to $1$, implying non-relativistic bulk fluid motions, but can reach as high as $\sim 1.7$ in localized jets. Unsurprisingly, the Lorentz factors are closely correlated with the bulk fluid energy density. The vorticity $\boldsymbol{\Omega} = \nabla \times \boldsymbol{v}_f$ is characterized by intermittent vorticity sheets, with similar morphology to current sheets, as also observed in MHD turbulence \citep[][]{zhdankin_etal_2016b}. In contrast, the divergence of the flow, $\nabla \cdot \boldsymbol{v}_f$, is essentially uniform and dominated by particle noise (not shown), implying that shocks are not a significant component of the turbulence. The quantity $\boldsymbol{E}\cdot\boldsymbol{J}$, which can be used as an electromagnetic dissipation proxy (representing energy exchange between electromagnetic fields and particles), shows very intense, localized structures. These structures have amplitudes that exceed the mean dissipation rate, $\langle\boldsymbol{E}\cdot\boldsymbol{J}\rangle$, by a factor of $\sim 100$. Finally, the non-ideal term $\boldsymbol{E}\cdot\boldsymbol{B}$ (not shown) is predominantly uniform and dominated by particle noise.

\subsection{Evolution} \label{sec:evolution}
 
 \begin{figure}
\includegraphics[width=0.9\columnwidth]{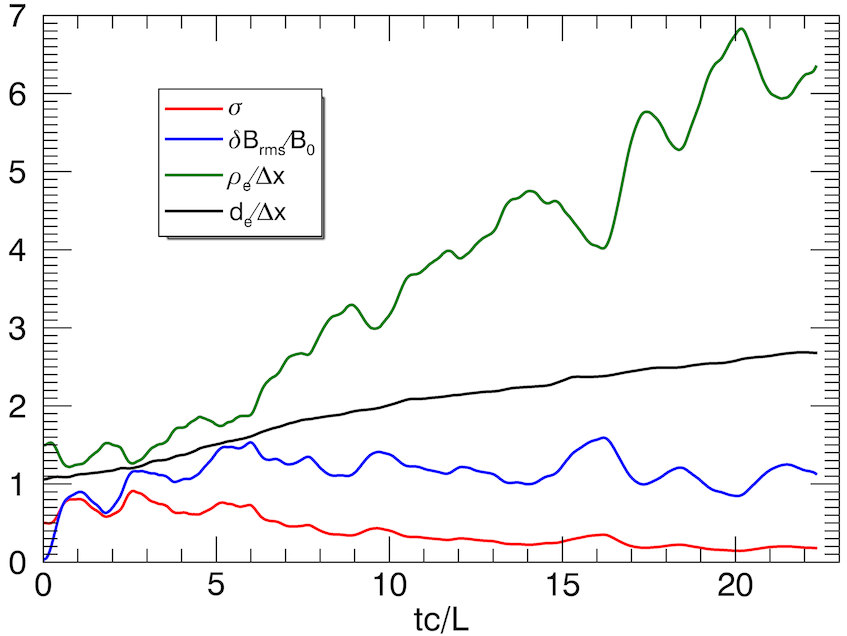}
      \centering
   \caption{\label{fig:evolution} Evolution of parameters: magnetization $\sigma$ (red), $\delta B_{\rm rms}/B_0$ (blue), skin depth relative to cell size $d_e/\Delta x$ (black), and Larmor radius relative to cell size $\rho_e/\Delta x$ (green).}
 \end{figure}
 
In this subsection, we begin the quantitative analysis of our simulations by discussing the time-evolution of physical parameters. This is critical to characterize because our simulations, lacking an energy sink, are inherently time-dependent. In Fig.~\ref{fig:evolution}, we show the evolution of the magnetization $\sigma(t)$, fluctuating-to-mean magnetic field ratio $\delta B_{\rm rms}(t)/B_0$, skin depth $d_e(t)$, and Larmor radius $\rho_e(t)$ for the fiducial case. In this example, heating causes $\rho_e$ to increase by a factor of $\sim 4$ and~$d_e$ to increase by a factor of $\sim 2.5$ over the duration of the simulation ($\sim 23 L/c$), while $\sigma = (d_e/\rho_e)^2$ decreases correspondingly. On the other hand, $\delta B_{\rm rms}(t)/B_0 \sim 1$ remains statistically constant.
 
The time dependence of physical parameters can be linked to plasma heating. The energy injection rate per unit volume is prescribed to be statistically constant, given by $\dot{{\cal E}}_{\rm inj} \sim \eta_{\rm inj} B_0^2 v_{A0} / 8 \pi L$, where $\eta_{\rm inj}$ is a dimensionless constant that describes the injection efficiency. Therefore, the mean particle Lorentz factor increases as
\begin{align} 
  \bar{\gamma} \sim \bar{\gamma}_0 \left( 1 + \frac{1}{2} \eta_{\rm inj} \sigma_0 \frac{v_{A0} t}{L} \right) \, , \label{eq:heating}
\end{align}
which causes~$\rho_e$ and~$d_e$ to increase in time and~$\sigma$ to decrease in time. The maximum duration of turbulence is set by the time it takes for the Larmor scale to grow to the driving scale ($\rho_e \sim L/2\pi$), thus eliminating the MHD inertial range. Estimating $\rho_e \sim \rho_{e0} + t \dot{{\cal E}}_{\rm inj} / n_0 e B_{\rm rms}$, we derive the duration $T c/L \sim (L / 2 \pi \rho_{e0} - 1) / \sigma_0 \to L/(2\pi\rho_{e0}\sigma_0)$ for large system sizes.

 \begin{figure*}
 \includegraphics[width=0.65\columnwidth]{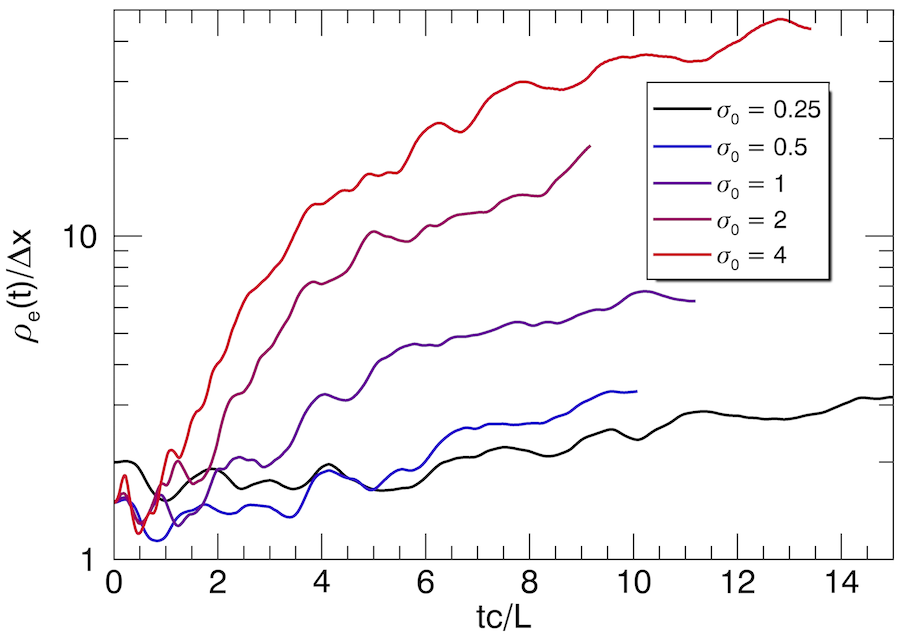}
\includegraphics[width=0.65\columnwidth]{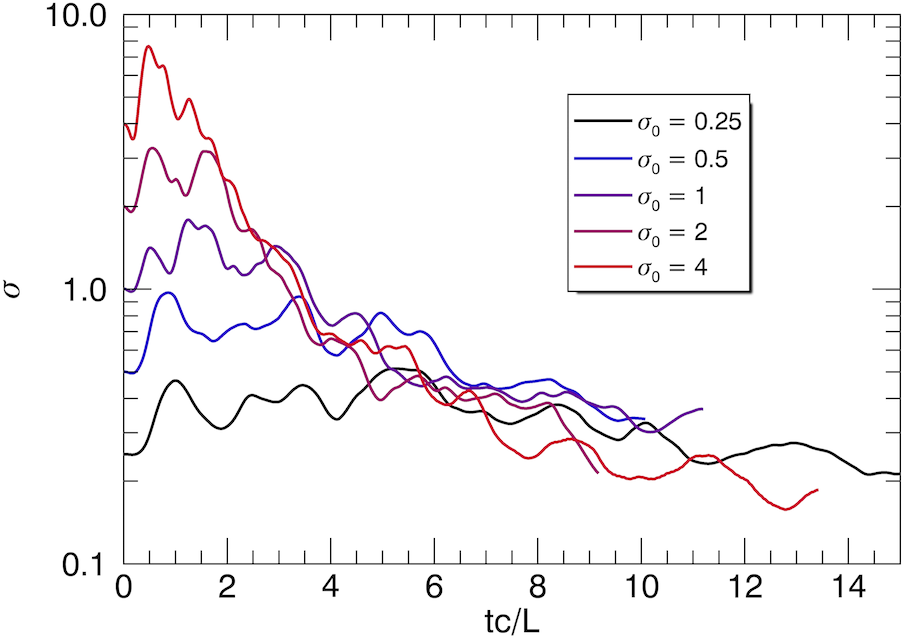}
 \includegraphics[width=0.65\columnwidth]{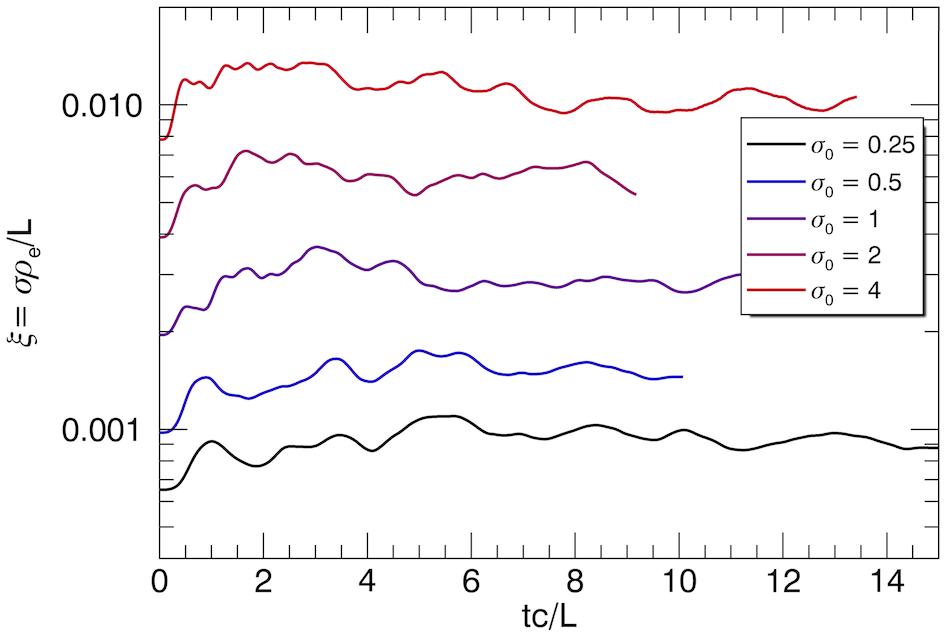}
      \centering
   \caption{\label{fig:parameter_sigma0} Characteristic Larmor scale $\rho_e(t)$ (relative to lattice cell size $\Delta x$), magnetization $\sigma(t)$, and the combination $\xi = \sigma \rho_e / L$ for the $768^3$ series of simulations with $\sigma_0 \in \{ 0.25, 0.5, 1, 2, 4\}$.}
 \end{figure*}
 
Due to stronger energy injection relative to the internal energy, simulations with high initial magnetization $\sigma_0$ exhibit rapid plasma heating, and consequently the Larmor radius $\rho_e$ increases more quickly than in the low $\sigma_0$ cases, as shown in the first panel of Fig.~\ref{fig:parameter_sigma0}. For example, whereas the kinetic scales increase by less than a factor of 2 for the $\sigma_0 = 0.25$ simulation, they increase by more than an order of magnitude for the $\sigma_0 = 4$ simulation. Likewise, the magnetization $\sigma$ rapidly decreases with time for high $\sigma_0$, which leads to late-time magnetizations that become comparable to cases with low $\sigma_0$. Indeed, all of the cases in Fig.~\ref{fig:parameter_sigma0} have $\sigma \lesssim 0.5$ at late times. Using the heating rate estimated in Eq.~\ref{eq:heating}, it can be shown that two simulations with different initial magnetizations, $\sigma_0$ and $\sigma'_0$ (where primes will indicate quantities for the second case), will acquire the same instantaneous magnetization, $\sigma(t) = \sigma'(t)$, at the time
 \begin{align}
 t_{\rm cross} = \frac{2 (\sigma_0 - \sigma'_0) c}{\eta_{\rm inj} \sigma_0 \sigma_0' (v_{A0} - v'_{A0})} \frac{L}{c} \label{eq:cross}
 \end{align}
For times $t > t_{\rm cross}$, the case with the {\it higher\rm} initial magnetization will have a {\it lower\rm} magnetization. The crossing occurs particularly early in time when $\sigma_0 \sim \sigma'_0 \gg 1$. For example, Eq.~\ref{eq:cross} predicts that for $\sigma_0 = 4$, $\sigma'_0 = 0.5$, and $\eta_{\rm inj} =1.7$ (as measured later in Fig.~\ref{fig:injection_1024cube}), the crossing time is $t_{\rm cross} \sim 6 L/c$, which agrees well with the observed crossing in Fig.~\ref{fig:parameter_sigma0}. In the limit of $\sigma_0 \gg \sigma_0' \gg 1$, Eq.~\ref{eq:cross} reduces to $t_{\rm cross} \sim 3 L / \eta_{\rm inj} c$, implying that all high $\sigma$ cases will approach $\sigma \sim 1$ within several light crossing times. The above considerations demonstrate that our present numerical set-up cannot be used to investigate fully-developed turbulence with sustained $\sigma \gtrsim 1$.

Whereas $L/\rho_e$ and $\sigma$ vary in time due to heating, the combination $\xi = \sigma \rho_e/L$ is independent of $\bar{\gamma}$ and is therefore statistically steady in time during developed turbulence. We explicitly demonstrate this for the $768^3$ series of simulations in the third panel of Fig.~\ref{fig:parameter_sigma0}. Hence, simulations with equal $\xi$ but different $\sigma$ can nominally be considered as different stages of evolution in a single long simulation. We also note that $\xi = E_{{\rm mag},n}/E_{\rm max}$, where $E_{\rm max} = L e B_{\rm rms} / 2 c$ is the energy associated with the most energetic particles (i.e., with Larmor radius equal to half the system size) and $E_{{\rm mag},n} = B_{\rm rms}^2/8\pi n_0$ is magnetic energy per particle. As noted in \cite{zhdankin_etal_2017}, this parameter may play an important role in particle acceleration.

\begin{figure*}
 \includegraphics[width=0.65\columnwidth]{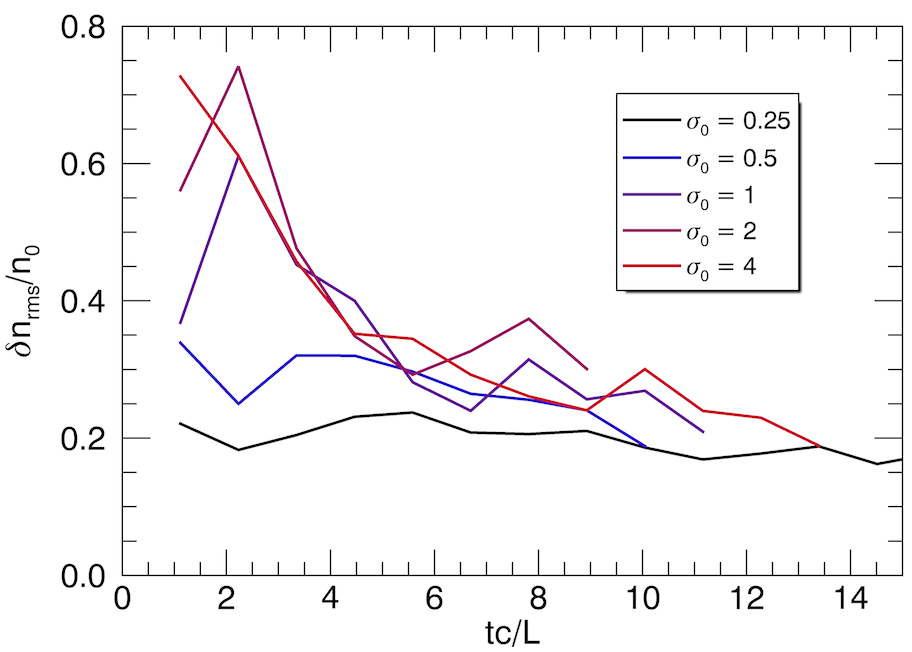}
 \includegraphics[width=0.65\columnwidth]{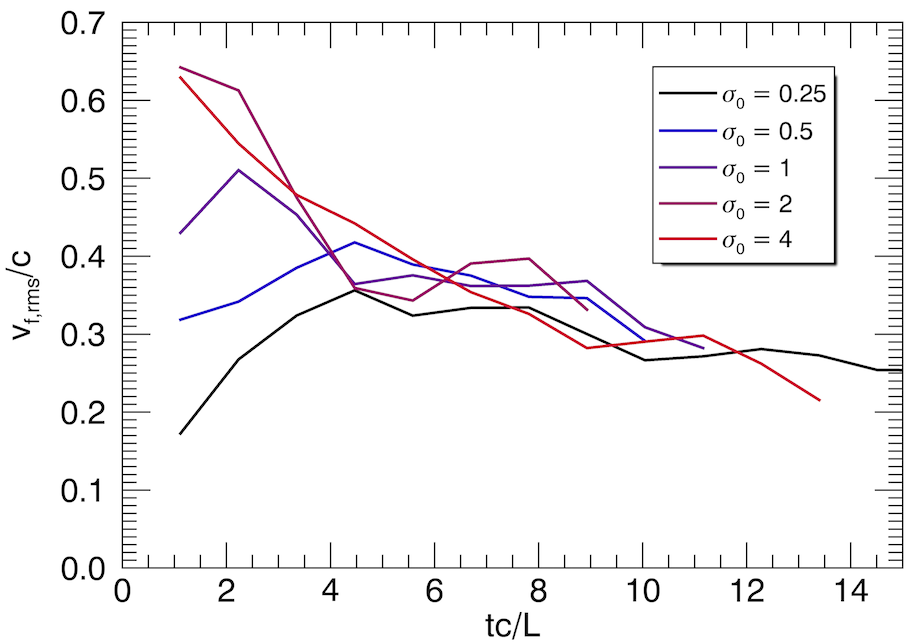}
 \includegraphics[width=0.65\columnwidth]{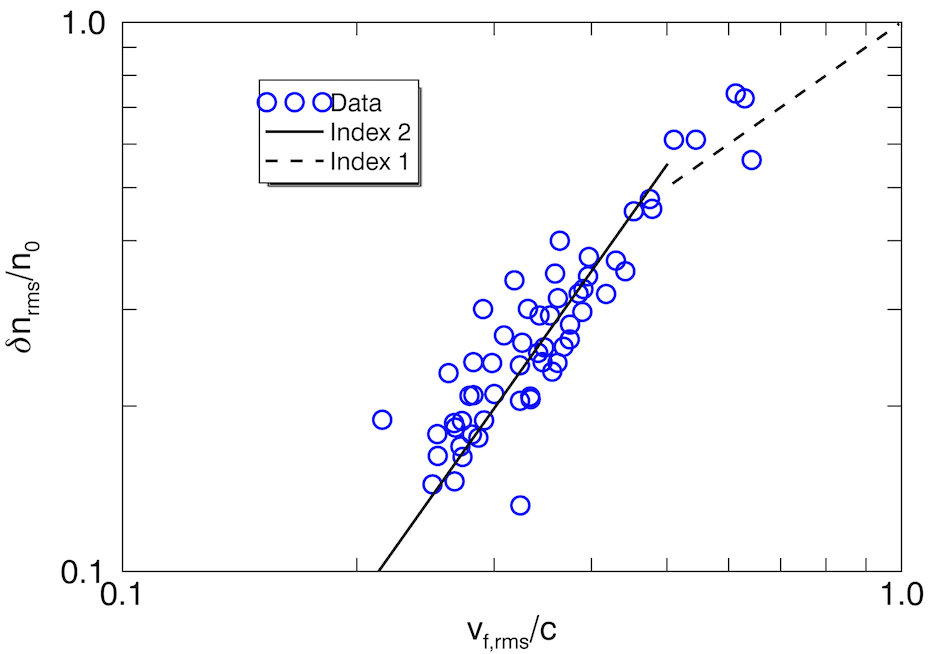}
      \centering
   \caption{\label{fig:dnrms} Left panel: Evolution of rms density fluctuations, $\delta n_{\rm rms}/n_0$, for the $768^3$ series of simulations with $\sigma_0 \in \{ 0.25, 0.5, 1, 2, 4\}$. Middle panel: similar for rms velocity fluctuations, $v_{f,{\rm rms}}/c$. Right panel: scatterplot of $\delta n_{\rm rms}/n_0$ versus $v_{f,{\rm rms}}/c$ in the same simulations, with quadratic (solid) and linear (dashed) compressible MHD scalings for comparison.}
 \end{figure*}

Finally, we comment on the evolution of the density fluctuations. In Fig.~\ref{fig:dnrms}, we show the evolution of the rms density fluctuations $\delta n_{\rm rms}$ (relative to the mean density $n_0$), measured from snapshots in the $768^3$ series of simulations. The density fluctuations evidently increase with $\sigma$, implying that the turbulence becomes increasingly compressive as it becomes more relativistic. For example, $\delta n_{\rm rms} \gtrsim 0.7 n_0$ at very early times for the cases with $\sigma_0 > 2$. The density fluctuations decrease rather quickly until $\delta n_{\rm rms} \lesssim 0.3 n_0$, after which there is a slower decrease in tandem with $\sigma$. Density fluctuations are not negligible even at $\sigma_0 = 0.25$, for which $\delta n_{\rm rms} \sim 0.2 n_0$. As also shown in Fig.~\ref{fig:dnrms}, the rms turbulent fluid velocity $v_{f,{\rm rms}}/c$ has a qualitatively similar evolution as for $\delta n_{\rm rms}$, with higher $\sigma_0$ cases having initially larger $v_{f,{\rm rms}}/c$, but $v_{f,{\rm rms}} \lesssim 0.3 c$ at late times for all the $768^3$ simulations. The correlation between $\delta n_{\rm rms}/n_0$ and $v_{f,{\rm rms}}/c$ is shown in the third panel of Fig.~\ref{fig:dnrms}, along with the quadratic scaling predicted (and measured) in compressible MHD turbulence with high plasma beta (low $\sigma$) \citep{cho_lazarian_2003}. We find good agreement with this predicted MHD scaling. Note that \cite{cho_lazarian_2003} predict a linear scaling for low plasma beta (high $\sigma$), which may explain the shallower scaling at high $v_{f,{\rm rms}}/c$ in our simulations; however, this scaling covers a very narrow range of our data.
 
  \subsection{Energetics} \label{sec:energetics}
  
  \begin{figure}
\includegraphics[width=0.9\columnwidth]{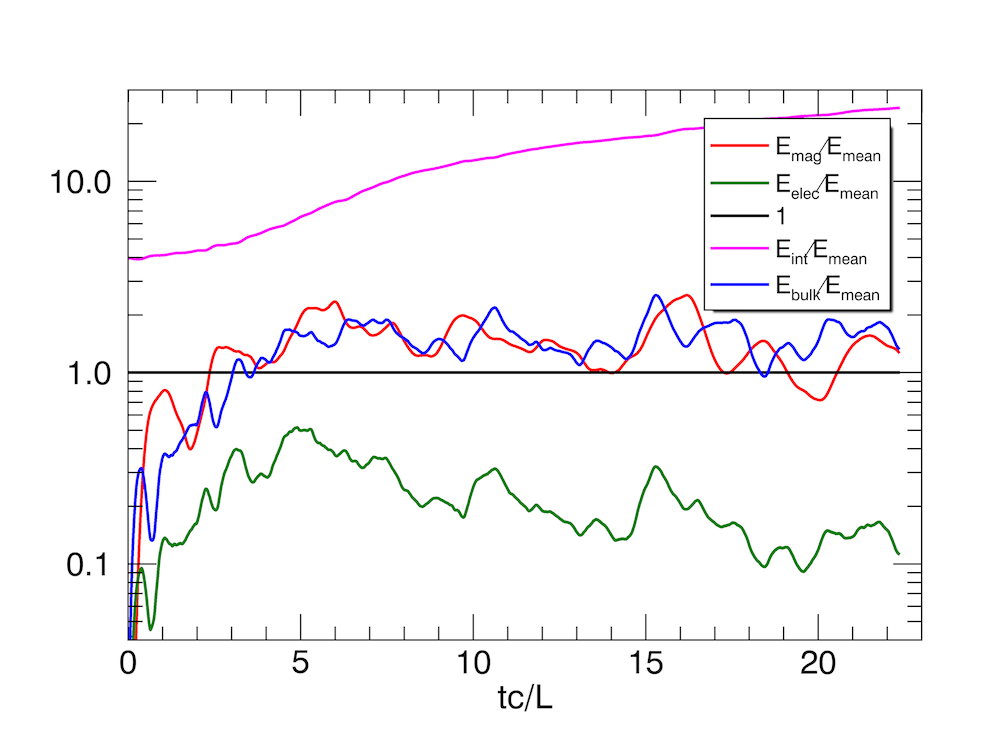}
      \centering
   \caption{\label{fig:energy} Evolution of turbulent magnetic energy $E_{\rm mag}$ (red), electric energy $E_{\rm elec}$ (green), internal energy $E_{\rm int}$ (magenta), and bulk fluid energy $E_{\rm bulk}$ (blue), all normalized to background magnetic energy $E_{\rm mean}$ (black).}
 \end{figure}
 
 We now discuss the overall energetics in the simulations. As described in Section~\ref{sec:defs}, we decompose total energy into background magnetic energy $E_{\rm mean}$, turbulent magnetic energy $E_{\rm mag}(t)$, electric energy $E_{\rm elec}(t)$, bulk fluid kinetic energy~$E_{\rm bulk}(t)$, and internal fluid energy~$E_{\rm int}(t)$. We show the evolution of these various contributions to total energy for the fiducial case in Fig.~\ref{fig:energy}. $E_{\rm mag}$ and $E_{\rm bulk}$ both quickly come into equipartition with $E_{\rm mean}$, as dictated by the driving, while $E_{\rm elec}$ is several times smaller than the other turbulence energies and slowly decreases in time. $E_{\rm int}$ dominates the other energies and slowly increases in time due to heating, giving the plasma significant inertia. The hierarchy of energies is a function of $\sigma$, but is qualitatively similar for all of our simulations in the developed stage (since $\sigma \lesssim 1$).
 
  \begin{figure}
  \includegraphics[width=0.9\columnwidth]{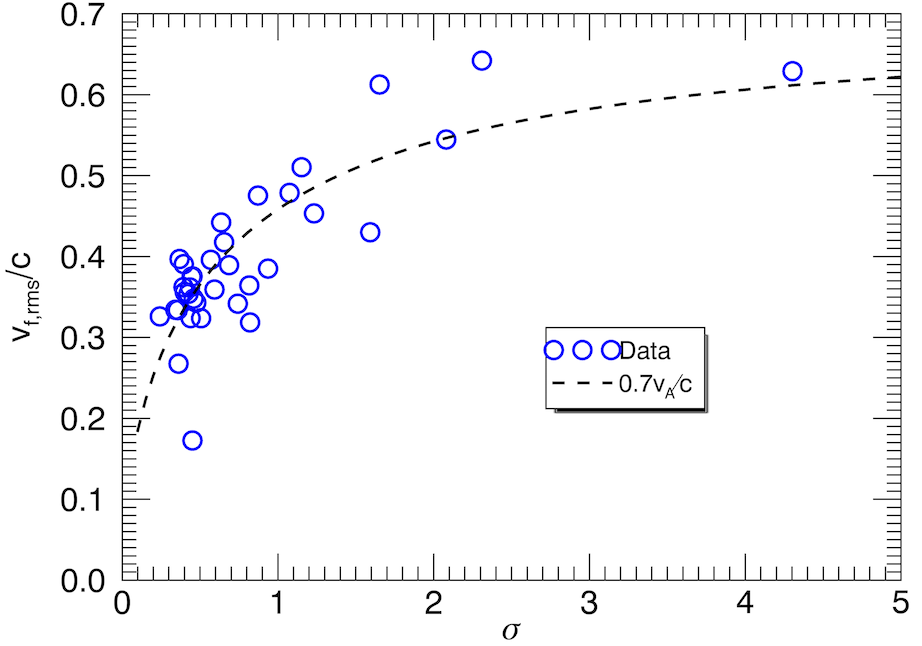}
\includegraphics[width=0.9\columnwidth]{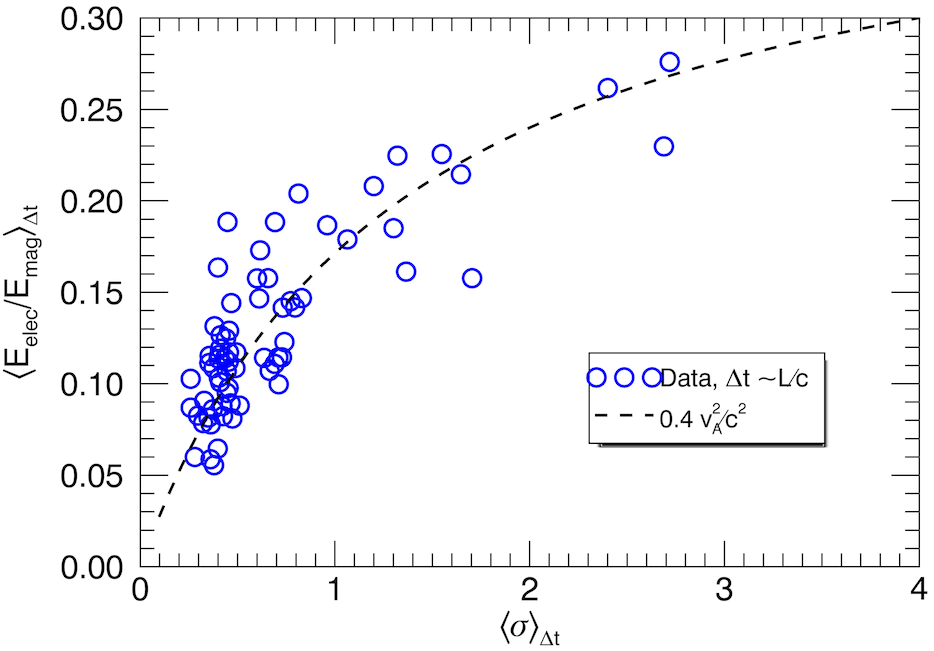}
      \centering
   \caption{\label{fig:ratio} Top panel: rms averaged flow velocity $v_{f,{\rm rms}}/c$ versus magnetization $\sigma$ for snapshots in the $768^3$ series of simulations (blue circles). The Alfv\'{e}nic scaling $0.7 v_A/c$ is also shown (black, dashed). Bottom panel: Ratio of the electric energy to magnetic energy, $\langle E_{\rm elec}/E_{\rm mag} \rangle_{\Delta t}$, averaged over intervals of duration $\Delta t \sim L/c$, versus the mean magnetization during the same intervals, $\langle \sigma \rangle_{\Delta t}$. Data points (blue circles) are taken from a sample of intervals in the $768^3$ series of simulations. The scaling $0.4 v_A^2/c^2$ is shown for comparison (black, dashed), for $v_A$ computed using $\langle \sigma \rangle_{\Delta t}$.}
 \end{figure}
 
The decrease of electric energy in time can be attributed to the turbulent motions becoming progressively slower as the effective fluid mass increases due to heating. In the ideal MHD approximation, $\boldsymbol{E} \sim (\boldsymbol{v}_f/c) \times \boldsymbol{B}$, so the ratio of the electric energy to magnetic energy can be estimated as
\begin{align}
\frac{E_{\rm elec}}{E_{\rm mag}} \sim \left(\frac{v_{\perp}}{c}\right)^2 \sim \left(\frac{v_A}{c} \right)^2 \sim \frac{\sigma}{\sigma + 4/3} \, , \label{eq:elecratio}
\end{align}
where $v_\perp$ is the flow velocity perpendicular to the magnetic field, which we estimate to be $v_A$ (up to a coefficient which describes the degree of alignment of the flow with the magnetic field). In Fig.~\ref{fig:ratio}, we verify that the flow fluctuations are Alfv\'{e}nic by showing that $v_{f,{\rm rms}} \sim v_A$ across a range of $\sigma$ in snapshots from the $768^3$ series of simulations. We also verify the estimate in Eq.~\ref{eq:elecratio} by measuring the ratio of electric to magnetic energy averaged across time intervals of duration $\Delta t$, which we denote $\langle E_{\rm elec} / E_{\rm mag} \rangle_{\Delta t}$, versus similar averages of the magnetization, $\langle \sigma \rangle_{\Delta t}$. Using the heating rate estimated in Eq.~\ref{eq:heating}, we can further estimate
\begin{align}
\frac{E_{\rm elec}}{E_{\rm mag}} \sim \frac{1}{1 + 4/3\sigma_0 + 2\eta_{\rm inj} v_{A0} t/3L}  \, .
\end{align}

\begin{figure}
  \includegraphics[width=0.9\columnwidth]{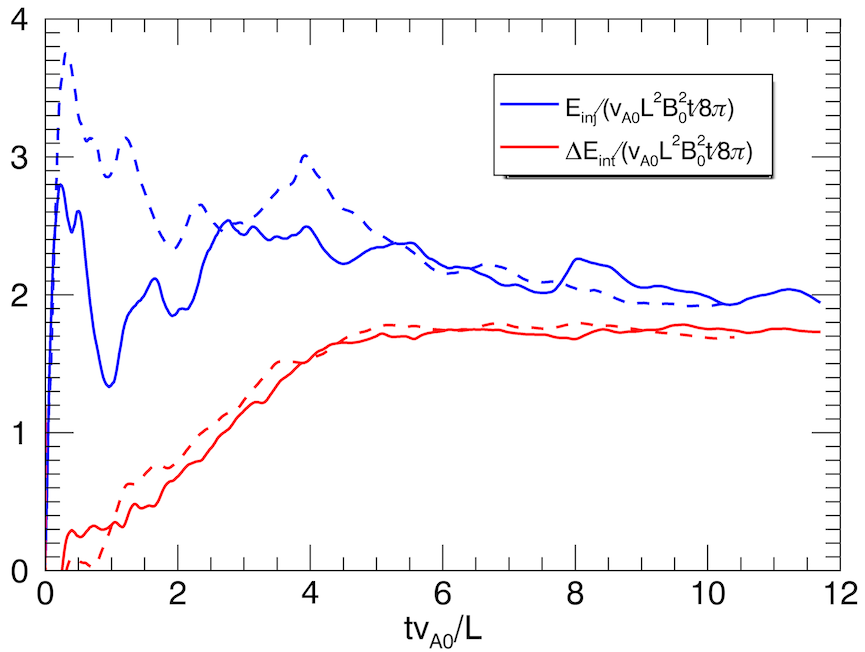}
      \centering
   \caption{\label{fig:injection_1024cube} Measurement of energy injection rate: evolution of injected energy~$E_{\rm inj}$ (blue) and increase in internal energy~$\Delta E_{\rm int} = E_{\rm int} - E_{\rm int, 0}$ (red), normalized to the nominal injected energy, $\dot{E}_{\rm inj} t \sim L^2 B_0^2 v_{A0} t/8\pi$, for $1024^3$ simulations with $\sigma = 0.5$ (solid) and $\sigma = 2$ (dashed). Both energies asymptotically approach similar values, implying~$\eta_{\rm inj} \sim 1.7$.}
 \end{figure}

Finally, we comment on the injected energy. The total injected energy $E_{\rm inj}$ increases linearly in time due to the constant energy injection rate, estimated to be~$\dot{E}_{\rm inj} \sim L^2 \eta_{\rm inj} B_0^2 v_{A0}/8\pi$. Therefore $E_{\rm inj}/(L^2 B_0^2 v_{A0} t/8\pi)$ approaches a constant value, which is confirmed in Fig.~\ref{fig:injection_1024cube} for our two largest simulations (Case A2 and Case A4); results are similar for the smaller simulations. In order to maintain a balanced energy budget, the heating rate of the plasma must equal the injection rate, which leads to a linear increase for $\Delta E_{\rm int} = E_{\rm int} - E_{\rm int, 0}$ in time. As also shown in Fig.~\ref{fig:injection_1024cube}, $\Delta E_{\rm int}/(L^2 B_0^2 v_{A0} t/8\pi) \approx 1.7$ during developed turbulence. This follows a more gradual accumulation of internal energy at early times, $t \lesssim 4 L/v_{A0}$ (corresponding to $t \lesssim 8 L/c$ in the fiducial case). The saturation of $\Delta E_{\rm int}/(B_0^2 v_{A0} t/8\pi)$ conveniently demarcates the beginning of fully-developed turbulence. As expected, $E_{\rm inj} \to \Delta E_{\rm int}$ as $t \to \infty$, with the residual energy difference associated with the energy contained in turbulent fluctuations. This implies an asymptotic injection efficiency of $\eta_{\rm inj} \sim 1.7$.
 
 \subsection{Probability distribution functions} \label{sec:pdfs}
 
We now proceed to discuss the spatial statistics of turbulence in our simulations. We first consider probability density functions (PDFs) for various quantities. We note that PDFs for large-scale quantities can be sensitive to driving, and hence statistics over a large number of snapshots are required to average over random fluctuations in the driving. Therefore, in this section, we present results from the $768^3$, $\sigma = 0.25$ simulation, rather than the fiducial simulation, for better statistics; the results are similar in both cases. The PDFs for small-scale quantities (such as charge density, current density) can be sensitive to particle noise, which sometimes dominates the physical signatures and makes the PDFs artificially appear as normal or log-normal. To reduce this noise and get results that are insensitive to $N_{\rm ppc}$, we coarse-grain the data onto a reduced lattice of $(N/2)^3$ cells. Each data point on the coarse-grained lattice represents an average across a cube of 8 cells in the original lattice.

 \begin{figure*}
 \includegraphics[width=0.65\columnwidth]{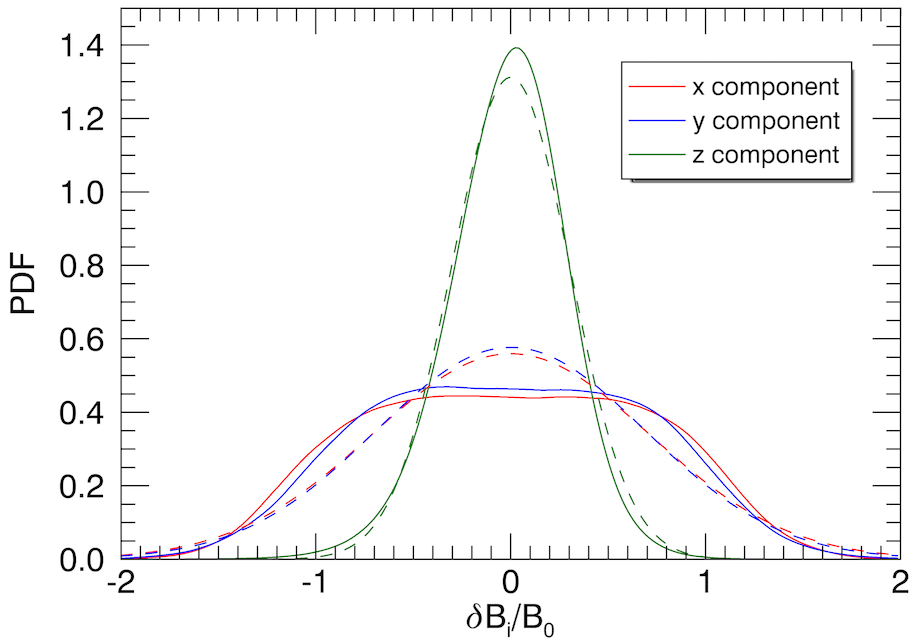}
  \includegraphics[width=0.65\columnwidth]{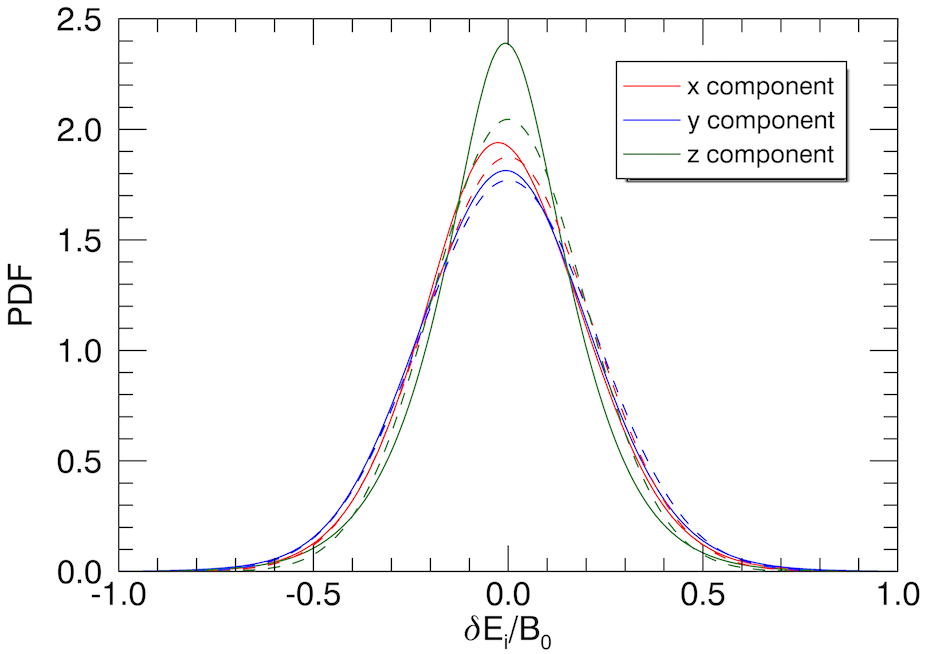}
\includegraphics[width=0.65\columnwidth]{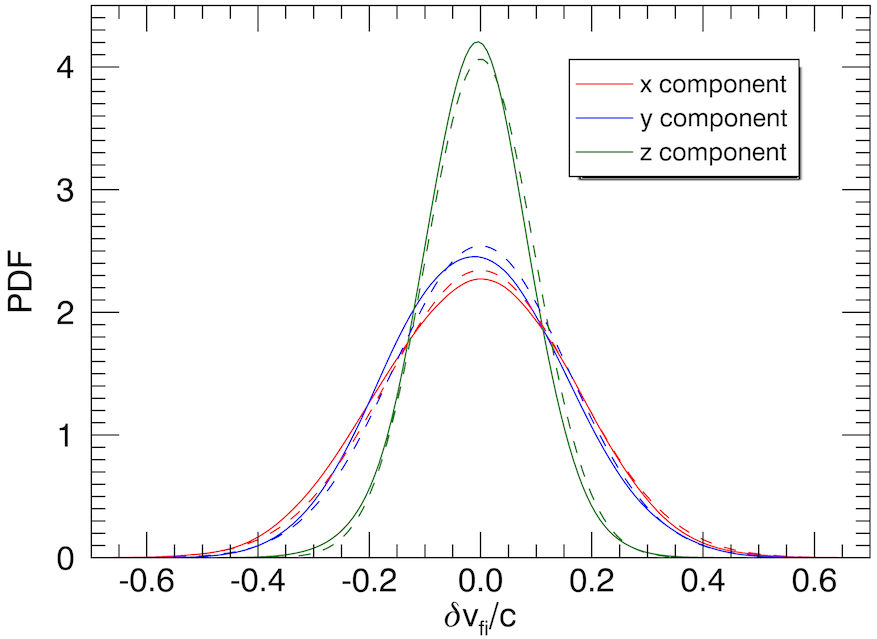}
      \centering
   \caption{\label{fig:pdfbe} Probability density function for components of magnetic field fluctuations $\delta \boldsymbol{B}$ (top left), electric field fluctuations $\delta\boldsymbol{E}$ (top right), fluid velocity fluctuations $\delta\boldsymbol{v}_f$ (bottom left). Solid lines indicate the measurements ($x$ component in red, $y$ component in blue, $z$ component in green) and dashed lines indicate the corresponding fits by a normal distribution.}
 \end{figure*}

We show the PDFs for the components of the magnetic field fluctuations $\delta \boldsymbol{B}$, electric field fluctuations $\delta\boldsymbol{E} = \boldsymbol{E}  - \langle \boldsymbol{E}  \rangle$, and fluid velocity fluctuations $\delta\boldsymbol{v}_f = \boldsymbol{v}_f - \langle \boldsymbol{v}_f  \rangle$ in Fig.~\ref{fig:pdfbe}. Here, angle brackets indicate the spatially-averaged field in the corresponding snapshot, which is statistically zero but generally nonzero in each snapshot. The magnetic field fluctuations are preferentially perpendicular to the mean field due to the anisotropy of the imposed driving, with~$\delta B_{\perp,{\rm rms}} \equiv (\delta B_{x,{\rm rms}}^2 + \delta B_{y,{\rm rms}}^2)^{1/2} \approx B_0$ and~$\delta B_{z,{\rm rms}} \approx 0.3 B_0$ for the given simulation. We find that $\delta B_z$ is consistent with a normal distribution, while~$\delta B_x$ and~$\delta B_y$ are platykurtic, i.e., have a broad, flat peak; this appears to be a result of the coherent driving mechanism. On the other hand, $\delta \boldsymbol{E}$ is more isotropic, with $\delta E_{x,{\rm rms}} \approx \delta E_{y,{\rm rms}} \approx \delta E_{z,{\rm rms}}$. The mean-field perpendicular components, $\delta E_x$ and $\delta E_y$, are well fit by a normal distribution while $\delta E_z$ deviates from the normal distribution, having a significantly narrower peak and broader tails, which may indicate intermittency in the parallel electric field. The PDF for flow fluctuations $\delta\boldsymbol{v}_f$ is very well fit by a normal distribution, with fluctuations preferentially perpendicular to $\boldsymbol{B}_0$; we measure $\delta v_{f\perp,{\rm rms}} \approx 0.23$ and $\delta v_{fz,{\rm rms}} \approx 0.10$.

 \begin{figure}
  \includegraphics[width=0.9\columnwidth]{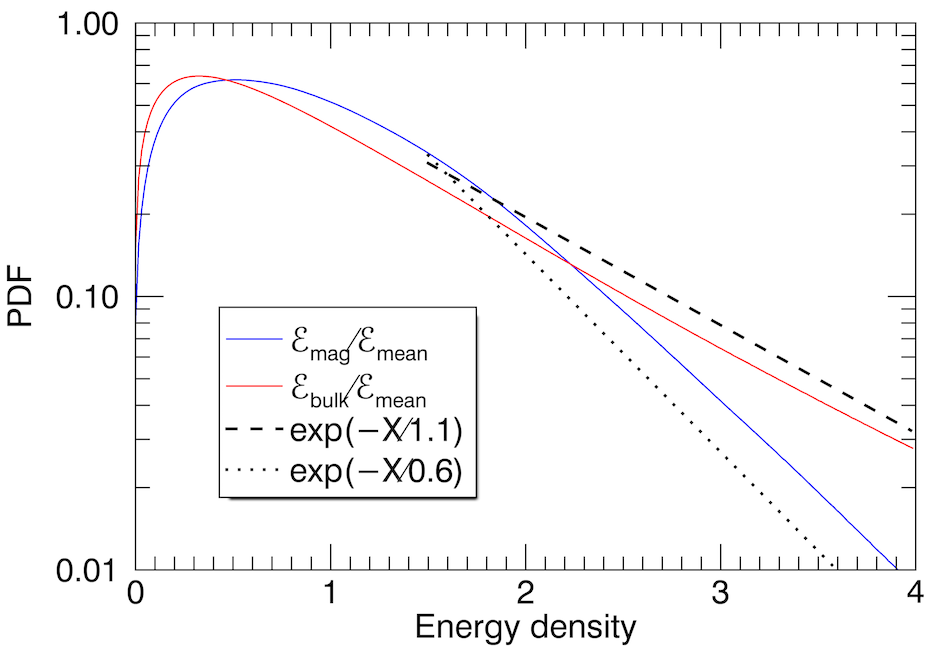}
   \includegraphics[width=0.9\columnwidth]{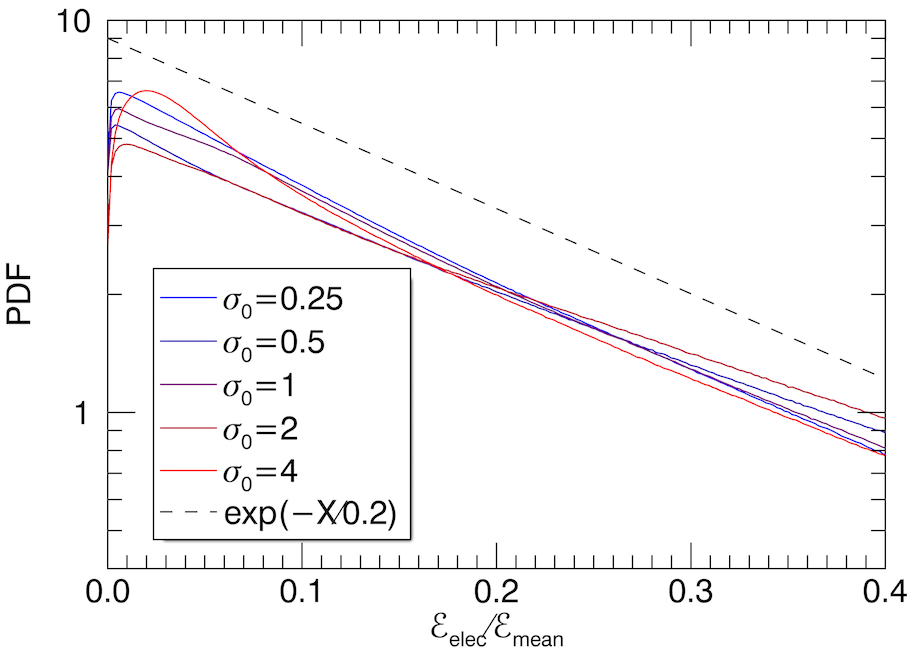}
      \centering
   \caption{\label{fig:pdf_energy} Top panel: PDF for turbulent magnetic energy density ${\mathcal E}_{\rm mag}$ and bulk fluid kinetic energy density ${\mathcal E}_{\rm bulk}$, normalized to the energy density in the mean field, ${\mathcal E}_{\rm mean}$. Bottom panel: PDF for electric energy density ${\mathcal E}_{\rm elec}$ normalized to ${\mathcal E}_{\rm mean}$, for $768^3$ series with $\sigma_0 \in \{0.25,0.5,1,2,4\}$. Exponential fits are shown in black lines.}
 \end{figure}

In Fig.~\ref{fig:pdf_energy}, we show the PDFs for turbulent magnetic energy density ${\mathcal E}_{\rm mag}$, bulk fluid energy density ${\mathcal E}_{\rm bulk}$, and electric energy density ${\mathcal E}_{\rm elec}$. We find that all three PDFs are qualitatively similar, with exponentially declining tails. The peak for ${\mathcal E}_{\rm elec}$ is at significantly smaller scales than the PDFs for ${\mathcal E}_{\rm mag}$ and ${\mathcal E}_{\rm bulk}$ (which are comparable to each other), consistent with the overall energetics heirarchy. The PDF for ${\mathcal E}_{\rm elec}$ appears to have some dependence on $\sigma$ (declining more rapidly for low $\sigma$) and is time-varying; however, the time-averaged PDF is qualitatively similar for all of the given simulations (going as $\sim \exp{(-5 {\mathcal E}_{\rm elec}/{\mathcal E}_{\rm mean})}$).
 
  \begin{figure}
  \includegraphics[width=0.9\columnwidth]{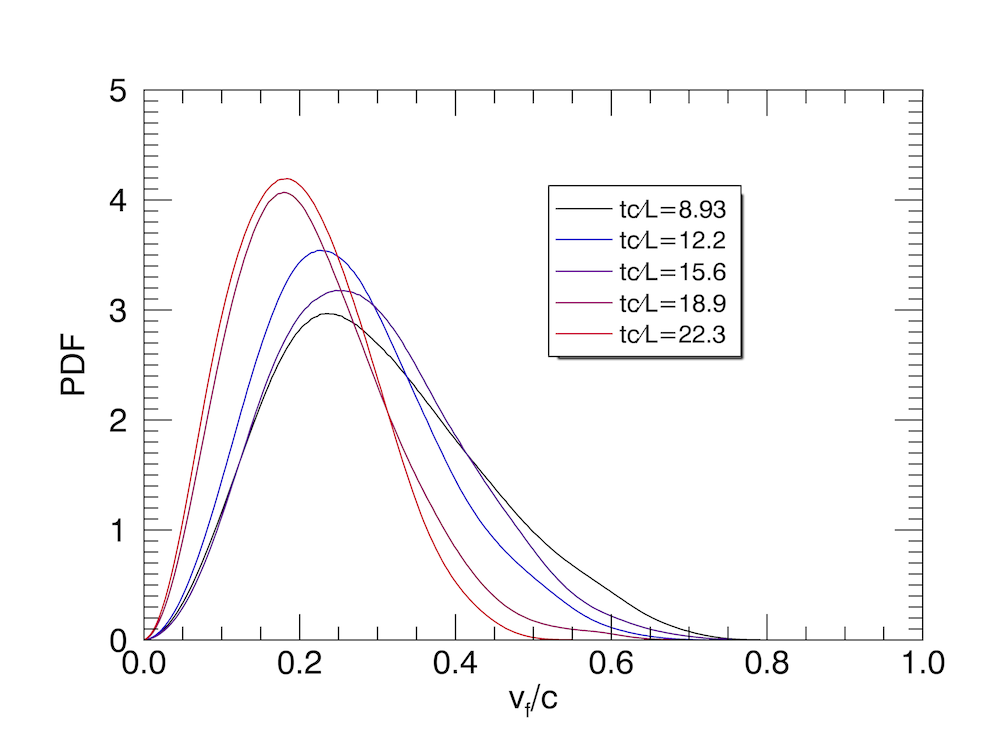}
\includegraphics[width=0.9\columnwidth]{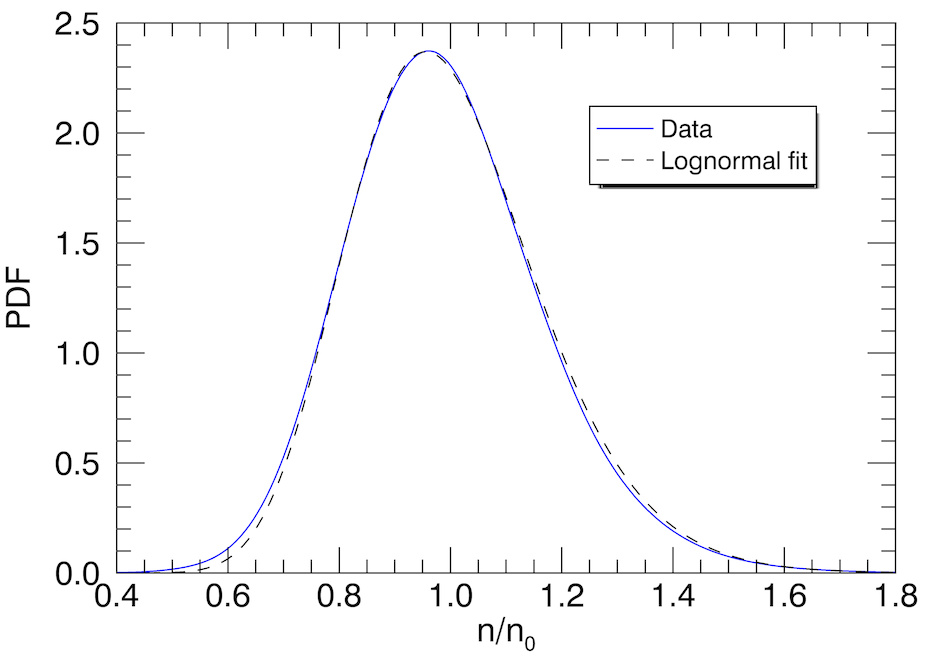}
      \centering
   \caption{\label{fig:pdfv} Top panel: PDF for the magnitude of flow velocity $v_f/c$ at varying times in the fiducial simulations. Bottom panel: PDF for density $n/n_0$ (blue) with a log-normal fit (black, dashed line).}
 \end{figure}
 
In Fig.~\ref{fig:pdfv}, we show the PDF for magnitude of flow velocity $v_f$ at several times in the fiducial simulation. As previously implied by Fig~\ref{fig:dnrms}, the peak of this PDF generally shifts toward lower $v_f$ with increasing time, with a more rapid evolution for high $\sigma_0$. Also in Fig.~\ref{fig:pdfv}, we show the PDF for lab-frame particle density $n$, which we find is fit very well by a log-normal distribution, as previously found in studies of compressible hydrodynamic and MHD turbulence \citep[e.g.,][]{nordlund_padoan_1999, kritsuk_etal_2007, lemaster_stone_2008, federrath_etal_2008, hopkins_2013}. A log-normal density distribution can be anticipated due to the continuity equation,
\begin{align}
(\partial_t + \boldsymbol{v}_f \cdot \nabla) \log{n} &= - \nabla \cdot \boldsymbol{v}_f \, .
\end{align}
Since $\nabla \cdot \boldsymbol{v}_f$ is small in our simulations, the density is randomly advected by the velocity field and $\log{n}$ takes a normal distribution (ignoring corrections due to intermittency and the backreaction of $n$ onto $\boldsymbol{v}_f$) \citep[e.g.,][]{biskamp_2003}.

  \begin{figure*}
   \includegraphics[width=0.65\columnwidth]{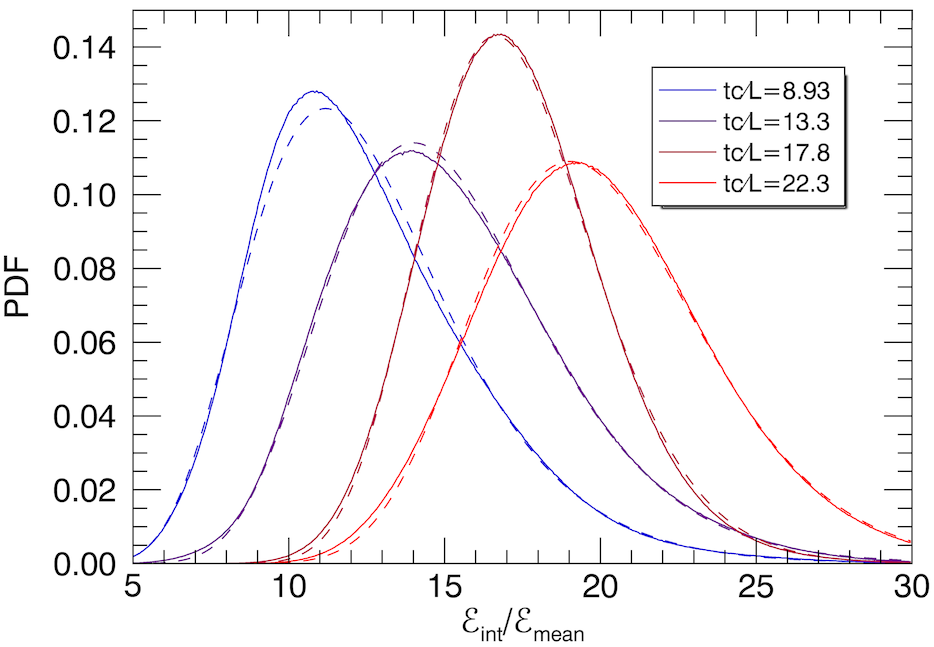}
  \includegraphics[width=0.65\columnwidth]{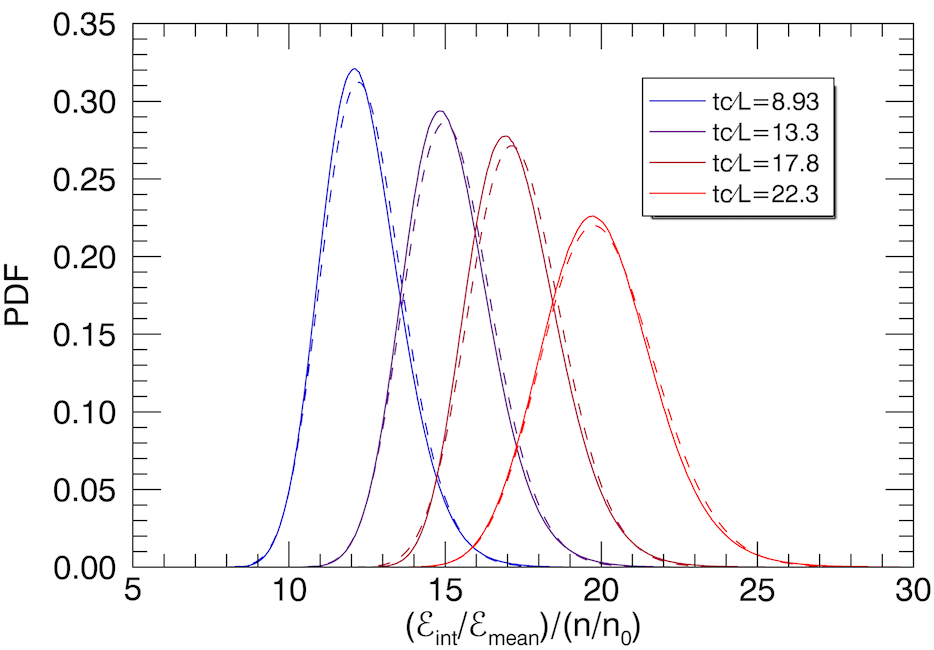}
  \includegraphics[width=0.65\columnwidth]{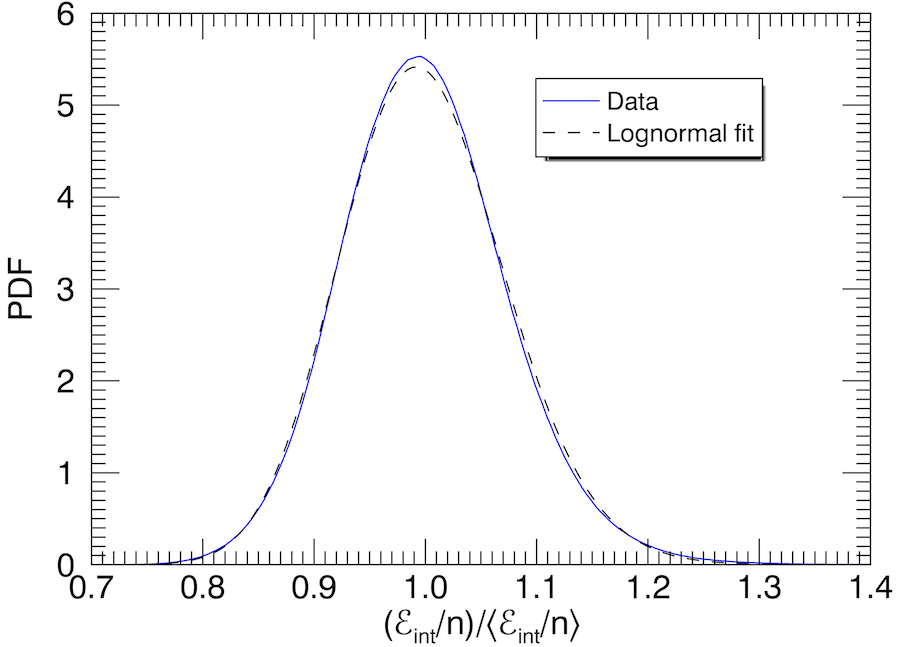}
      \centering
   \caption{\label{fig:pdf_internal} Top left panel: PDF for internal energy density ${\mathcal E}_{\rm int}$ (normalized to ${\mathcal E}_{\rm mean}$) at four different times. Top right panel: PDF for internal energy per particle, ${\mathcal E}_{\rm int}/n$ (normalized to ${\mathcal E}_{\rm mean}/n_0$) at same times. Bottom left panel: time-averaged PDF for $({\mathcal E}_{\rm int}/n)/\langle {\mathcal E}_{\rm int}/n \rangle$. Corresponding log-normal fits are shown in dashed lines.}
 \end{figure*}
 
 In Fig.~\ref{fig:pdf_internal}, we show the PDF for the internal energy density ${\mathcal E}_{\rm int}$ and the internal energy per particle ${\mathcal E}_{\rm int}/n$, which can be used as a proxy for temperature. We find that both quantities are very well fit by log-normal distributions. The PDF for ${\mathcal E}_{\rm int}$ is relatively broad, with the PDF half-width fluctuating significantly in time. On the other hand, ${\mathcal E}_{\rm int}/n$ is narrowly peaked and shows a clear trend in which the half-width increases in time. We find that when ${\mathcal E}_{\rm int}/n$ is normalized to its spatially-averaged value, $\langle {\mathcal E}_{\rm int}/n\rangle$, the PDF becomes time-independent; the time-averaged PDF of $({\mathcal E}_{\rm int}/n)/\langle {\mathcal E}_{\rm int}/n \rangle$ with log-normal fit is shown in Fig.~\ref{fig:pdf_internal}.
 
  \begin{figure}
  \includegraphics[width=\columnwidth]{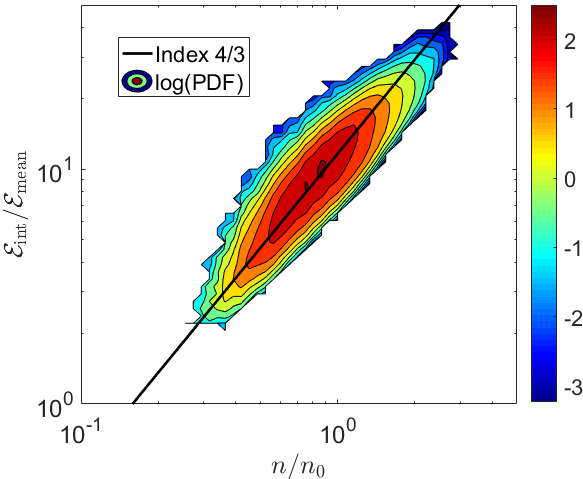}
      \centering
   \caption{\label{fig:correlation_eint_n} Demonstration of adiabatic equation of state: joint PDF of internal energy density ${\mathcal E}_{\rm int}$ (relative to ${\mathcal E}_{\rm mean})$ versus particle density $n$ (relative to $n_0$) measured in a snapshot of the fiducial simulation. Also shown is a power law with (adiabatic) index of $4/3$.}
 \end{figure}
 
 The fact that $n$ and ${\mathcal E}_{\rm int}$ are simultaneously fit by log-normal distributions can be explained by a power-law correlation between the two. Indeed, as shown in Fig.~\ref{fig:correlation_eint_n}, there is a tight empirical power-law correlation,
 \begin{align}
 \frac{{\mathcal E}_{\rm int}}{{\mathcal E}_{\rm mean}} &\sim \left(\frac{n}{n_0}\right)^{4/3} \, ,
 \end{align}
 which implies that the plasma can be described as an ideal gas with adiabatic index of $4/3$ on the timescale of turbulent fluctuations. This is the well-known equation of state for an ultra-relativistic gas \citep[e.g.,][]{weinberg_1972}, which emerges from first principles in our simulations. A more complete equation of state must account for pressure anisotropy, as in the Chew-Goldberger-Low equations \citep{cgl_1956} and its relativistic generalization \citep{gedalin_1991}; we leave an analysis of pressure anisotropy to future work. 
 
   \begin{figure}
  \includegraphics[width=0.9\columnwidth]{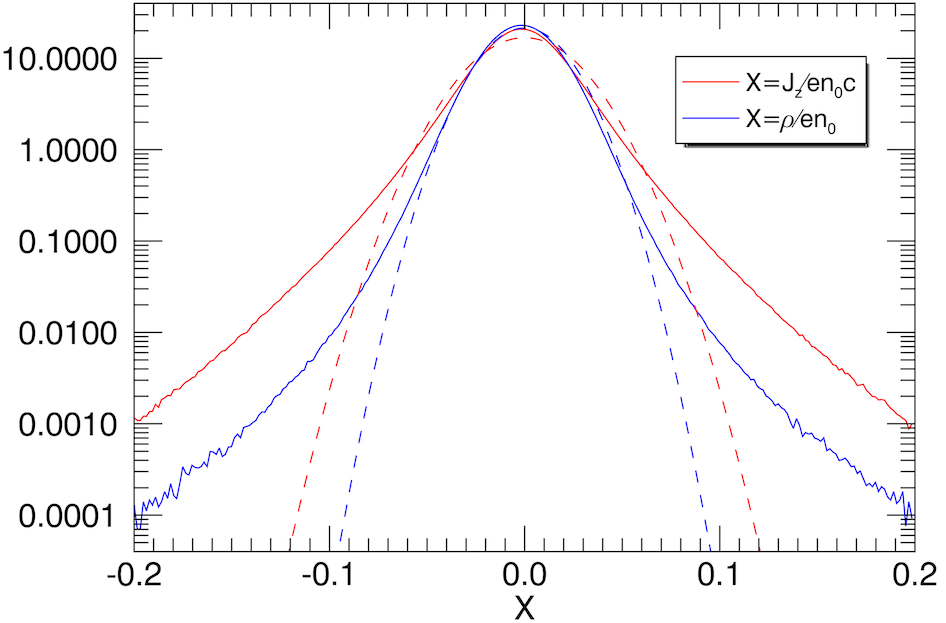}
   \includegraphics[width=0.9\columnwidth]{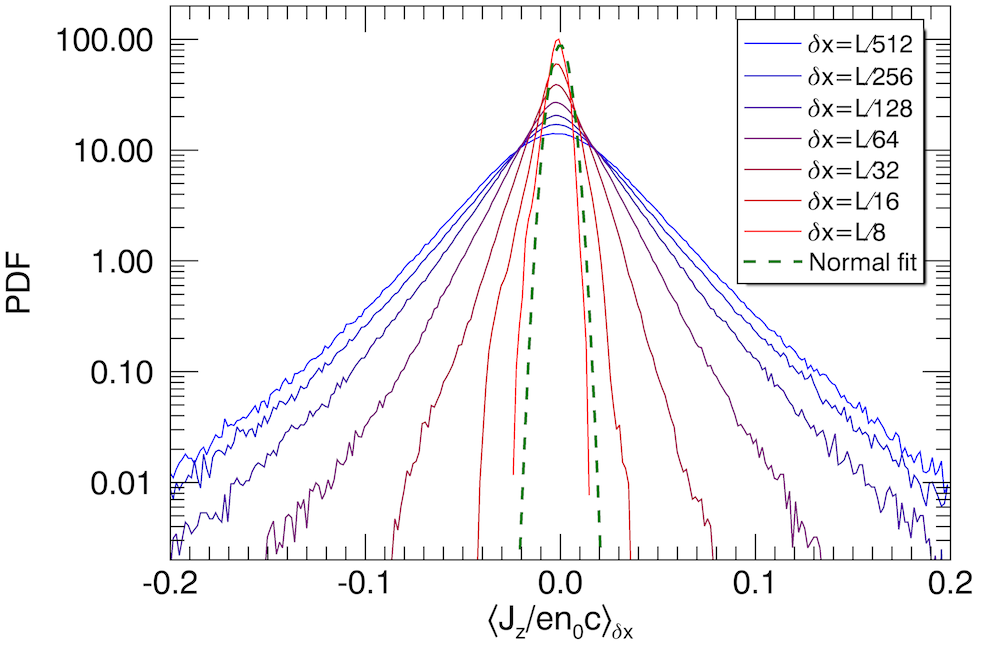}
      \centering
   \caption{\label{fig:pdf_rhoj} Top panel: PDF for the current density parallel to the mean magnetic field, $J_z/en_0c$ (red), and for the charge density $\rho/en_0$ (blue) in the fiducial simulation. Normal fits are shown in dashed lines. Bottom panel: PDF for coarse-grained current density $\langle J_z/en_0c \rangle_{\delta x}$ for $\delta x = L/512$ (blue) up to $\delta x = L/8$ (red).}
 \end{figure}
 
Finally, we discuss the PDFs for some representative small-scale quantities, in contrast to the large-scale quantities described above (the scale dependences will be discussed explicitly in the next subsection). In Fig.~\ref{fig:pdf_rhoj}, we show the PDF for the current density parallel to the mean field, $J_z$, and for the charge density $\rho$. Both PDFs show broad, non-Gaussian tails characteristic of intermittency.  To confirm that $J_z$ is classically intermittent, we measure the coarse-grained current density $\langle J_z \rangle_{\delta x}$, where $\langle \cdot \rangle_{\delta x}$ indicates an average performed over cubes of side length $\delta x$. The resulting PDF is shown in the second panel of Fig.~\ref{fig:pdf_rhoj}; as $\delta x$ decreases, the PDF broadens and deviates from a normal distribution. Hence, statistics for $J_z$ are scale-dependent, in qualitative agreement with current density statistics in MHD studies.

 \begin{figure}
  \includegraphics[width=0.9\columnwidth]{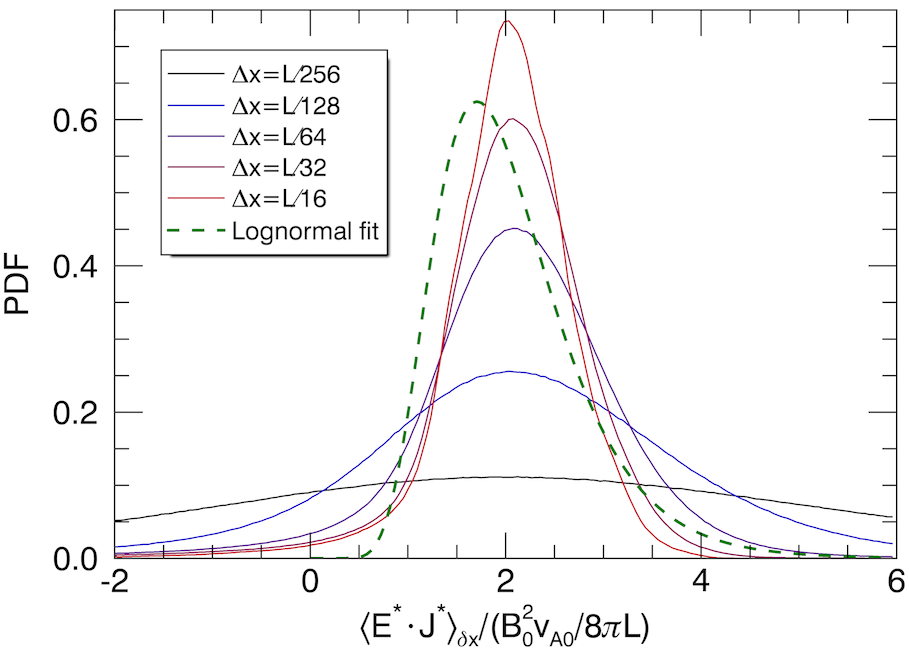}
      \centering
   \caption{\label{fig:pdf_diss} PDF for the coarse-grained dissipation rate proxy $\langle \boldsymbol{E}^* \cdot \boldsymbol{J}^* \rangle_{\delta x}$ in the fiducial simulation, for $\delta x \in \{L/256,L/128,L/64,L/32,L/16\}$. A log-normal fit is shown for reference (dashed line).}
 \end{figure}

An intermittent quantity of central importance in turbulence is the local heating rate; hence, we next comment on the PDF for the time derivative of internal energy. For simplicity, we consider the expression for the heating rate in the well-defined limit of non-relativistic motions ($V_{\rm eff}/c \ll 1$):
\begin{align}
\partial_t {\mathcal E}_{\rm int} &\to - \nabla \cdot \boldsymbol{\mathcal P}_f c +  \left( \boldsymbol{E} + \frac{\boldsymbol{v}_f}{c} \times \boldsymbol{B}  \right) \cdot \left( \boldsymbol{J} - \rho \boldsymbol{v}_f \right) \, . \label{eq:diss_nonrel}
\end{align}
Ignoring the flux terms, we therefore consider
\begin{align}
 \boldsymbol{E}^* \cdot \boldsymbol{J}^* &= \left( \boldsymbol{E} + \frac{\boldsymbol{v}_f}{c} \times \boldsymbol{B}  \right) \cdot \left( \boldsymbol{J} - \rho \boldsymbol{v}_f \right) \, ,
 \end{align}
 where asterisks denote the fluid-frame electric field and current density. Note that irreversible dissipation does not occur in the collisionless Vlasov-Maxwell system (Eqs.~\ref{eq:vlasov}) since entropy is conserved. However, $\boldsymbol{E}^* \cdot \boldsymbol{J}^*$ does represent local heating and cooling of the plasma due to electromagnetic fields. We show the PDF for the coarse-grained dissipation rate proxy, $\langle \boldsymbol{E}^* \cdot \boldsymbol{J}^* \rangle_{\delta x}$, in Fig.~\ref{fig:pdf_diss}. As with the current density, the PDF broadens as $\delta x$ decreases, indicating intermittency. In contrast to hydrodynamic and MHD turbulence, where the PDF of coarsed-grained energy dissipation rate is strictly positive and close to log-normal \citep[e.g.,][]{zhdankin_etal_2016a}, the PDF for $\langle \boldsymbol{E}^* \cdot \boldsymbol{J}^* \rangle_{\delta x}$ is not well fit by a log-normal in our simulations (except for at the largest scales). Also, the PDF significantly extends into negative values, implying local cooling of the plasma. We make no further remarks on the statistics of dissipation in this paper.

\subsection{Power spectra} \label{sec:spectra}

We next investigate the turbulence statistics as a function of scale by considering Fourier power spectra for the fluctuations. In the following,  $\tilde{\boldsymbol{y}}(\boldsymbol{k})$ denotes the Fourier transform of $\boldsymbol{y}(\boldsymbol{x})$, and the corresponding power spectrum is given by $E_{\boldsymbol{y}}(\boldsymbol{k}) = \langle|\tilde{\boldsymbol{y}}(\boldsymbol{k})|^2\rangle$. Here, angle brackets $\langle \cdot \rangle$ indicate an average over a specified data set; we mainly perform averages over a short period of time during the early stages of developed turbulence, when the intertial range is relatively long. In particular, for the fiducial case, we show results averaged over 4 snapshots spanning times $7.7 < t c / L < 10.0$, during which the Larmor scale increases from $\rho_e \approx 2.4 \Delta x$ to $\rho_e \approx 3.2 \Delta x$ (as previously shown in Fig.~\ref{fig:evolution}). Spectra at later times are similar but with an increasingly shorter inertial range. To account for global anisotropy with respect to $\boldsymbol{B}_0$, we focus on spectra with respect to the mean-field perpendicular wavenumber $k_\perp$, obtained by integrating the spectrum across $k_z$ and across angles in the $k_x$-$k_y$ plane, $E_{\boldsymbol{y}}(k_\perp) = \int dk_z d\phi k_\perp \langle|\tilde{\boldsymbol{y}}(k_\perp \cos{\phi}, k_\perp \sin{\phi}, k_z)|^2\rangle$.

We first consider the set of power spectra that characterize the turbulence energetics, which we will refer to as energy spectra. These include the magnetic energy spectrum~$E_{\rm mag}(\boldsymbol{k}) = \langle|\tilde{\boldsymbol{B}}(\boldsymbol{k})|^2/8\pi\rangle$, electric energy spectrum~$E_{\rm elec}(\boldsymbol{k}) = \langle|\tilde{\boldsymbol{E}}(\boldsymbol{k})|^2/8\pi\rangle$, bulk fluid kinetic energy spectrum~$E_{\rm bulk}(\boldsymbol{k}) = \langle|\tilde{\boldsymbol{\mathcal W}}(\boldsymbol{k})|^2\rangle$, and internal energy spectrum~$E_{\rm int}(\boldsymbol{k}) = \langle|\tilde{\mathcal E}_{\rm int}(\boldsymbol{k})|^2\rangle$. The motivation for considering energy spectra comes from the standard picture of the energy cascade in non-relativistic turbulence; namely, the inertial range is characterized by a constant flux of total energy in wavevector space.

  \begin{figure}
\includegraphics[width=\columnwidth]{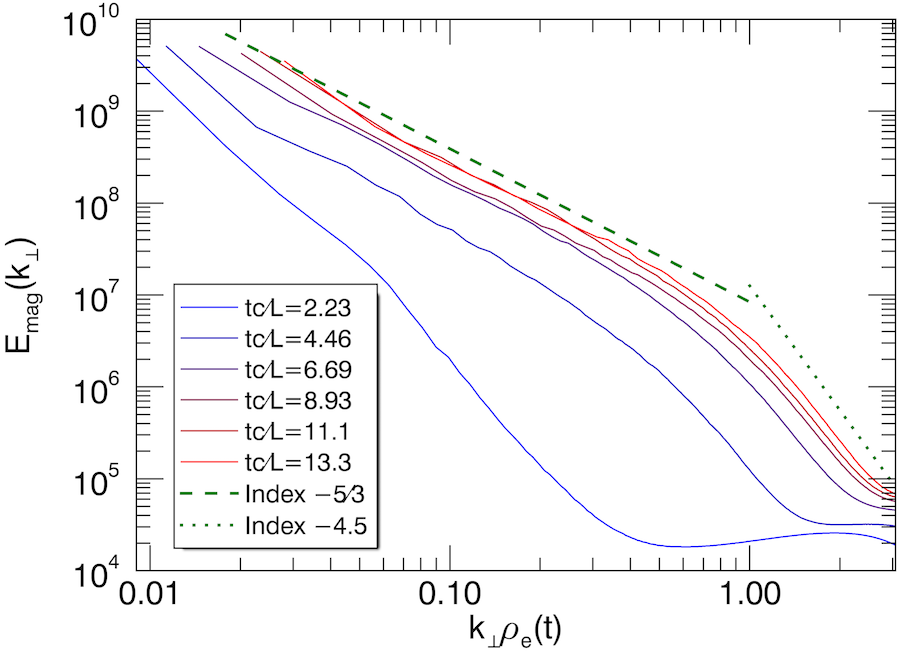}
\includegraphics[width=\columnwidth]{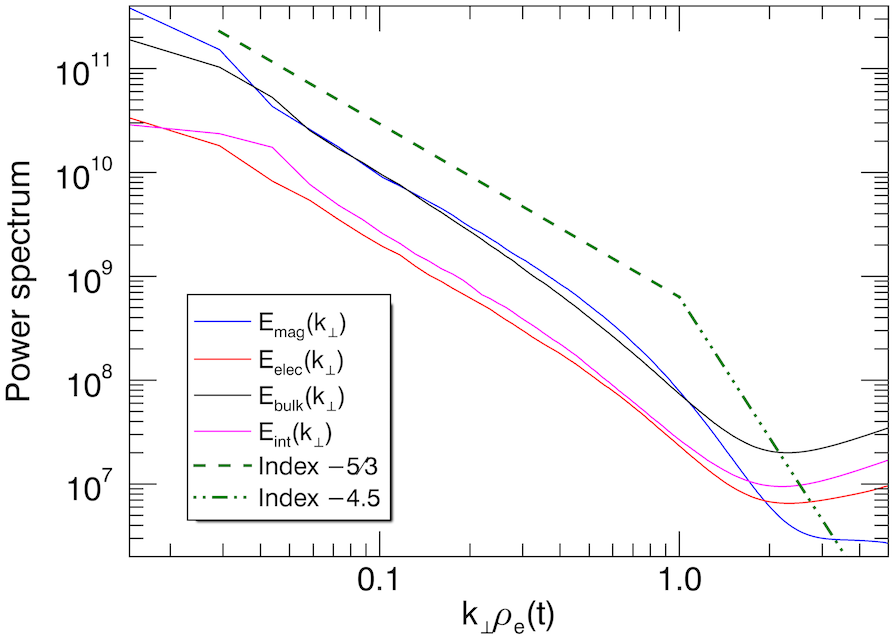}
   \centering
   \caption{\label{fig:spectrum_energy} Top panel: magnetic energy spectrum $E_{\rm mag}(k_\perp)$ for fiducial simulation at times $tc/L \in \{2.2, 4.5, 6.7, 8.9, 11.1, 13.3 \}$. Bottom panel: Energy spectra averaged over early times, including $E_{\rm mag}(k_\perp)$ (blue), electric energy spectrum $E_{\rm elec}(k_\perp)$ (red), bulk fluid energy spectrum $E_{\rm bulk}(k_\perp)$ (black), and internal fluid energy spectrum $E_{\rm int}(k_\perp)$ (magenta). Power-laws $k_\perp^{-5/3}$ (green, dashed) and $k_\perp^{-4.5}$ (green, dotted) are shown for reference.}
 \end{figure}

We first show the magnetic energy spectrum for the fiducial simulation at several different times in the top panel of Fig.~\ref{fig:spectrum_energy}. During the initial stages of the simulation, the spectrum broadens from low $k_\perp$ to high $k_\perp$, until it reaches $k_\perp \rho_e \sim 1$ (which occurs by the time $tc/L \sim 6.7$, or $t v_{A0}/L \sim 3.5$). The evolution of the spectrum then slows dramatically, indicating that the developed stage of turbulence has been reached. The developed magnetic energy spectrum is characterized by an approximate $E_{\rm mag}(k_\perp) \sim k_\perp^{-5/3}$ scaling in the inertial range ($k_\perp \rho_e < 1$) and a steeper scaling broadly consistent with $E_{\rm mag}(k_\perp) \sim k_\perp^{-4.5}$ in the kinetic range ($k_\perp \rho_e > 1$). The inertial range extends over nearly an order of magnitude in scale at the given times. The kinetic range scaling, on the other hand, is very limited, spanning roughly a factor of 3 (across $1 \lesssim k_\perp \rho_e \lesssim 3$); hence, the asymptotic scaling of the kinetic range is poorly constrained by the fiducial simulation. At even higher $k_\perp$, the magnetic energy spectrum flattens due to particle noise; this noise floor can be lowered by increasing the number of particles per cell (see Appendix~\ref{sec:ppc} for a discussion of the effect of $N_{\rm ppc}$ on power spectra).

In the bottom panel of Fig.~\ref{fig:spectrum_energy}, we compare the various types of energy spectra, averaged for times $7.7 < t c / L < 10.0$. We find that the bulk fluid energy spectrum $E_{\rm bulk}(k_\perp)$ is in excellent equipartition with $E_{\rm mag}(k_\perp)$ across the inertial range. However, the kinetic range is nearly absent in $E_{\rm bulk}(k_\perp)$ due to a high noise floor. In the vicinity of the spectral break at~$k_\perp \rho_e \sim 1$, there is an excess of magnetic energy over bulk fluid energy, which may be caused by energy exchange associated with kinetic instabilities \citep[e.g.,][]{kunz_etal_2014}. The electric energy spectrum $E_{\rm elec}(k_\perp)$ exhibits a similar inertial-range scaling as $E_{\rm mag}(k_\perp)$, but has lower amplitude by a factor of few (consistent with the overall energy partition measured in Sec.~\ref{sec:energetics}). This follows from the ideal MHD scaling, where $\boldsymbol{E} \sim - (\boldsymbol{v}_f/c)\times\boldsymbol{B}$ in Fourier space is a convolution dominated by coupling with small-wavenumber modes. Curiously, the internal energy spectrum $E_{\rm int}(k_\perp)$ matches $E_{\rm elec}(k_\perp)$ to a very good approximation.
 
   \begin{figure*}
\includegraphics[width=0.9\columnwidth]{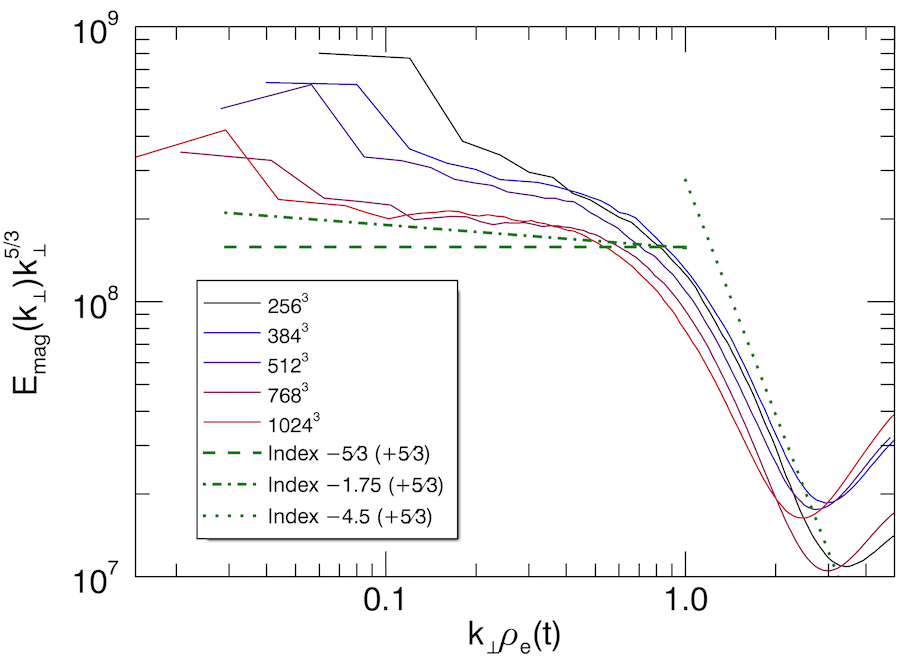}
\includegraphics[width=0.9\columnwidth]{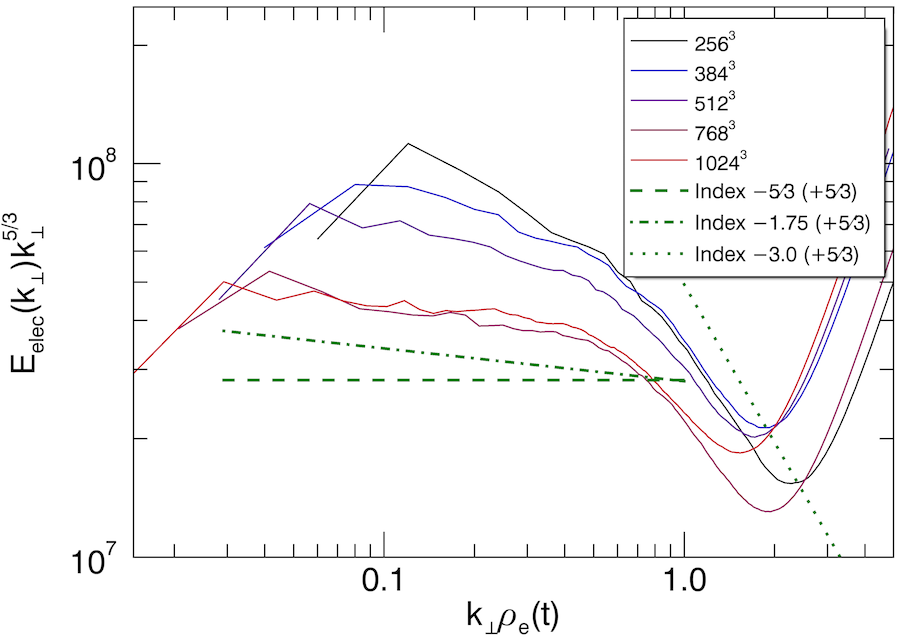}
\includegraphics[width=0.9\columnwidth]{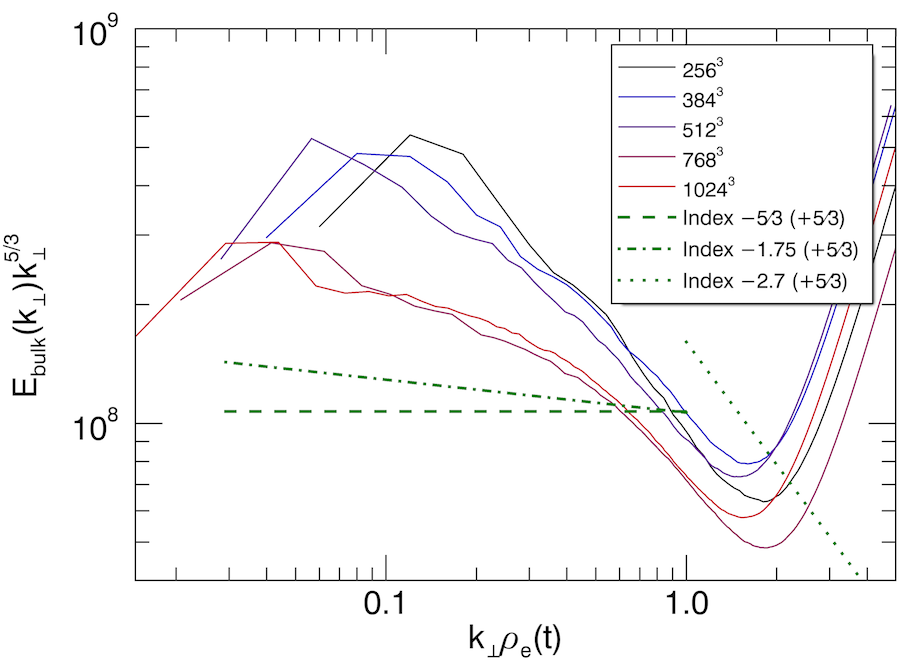}
\includegraphics[width=0.9\columnwidth]{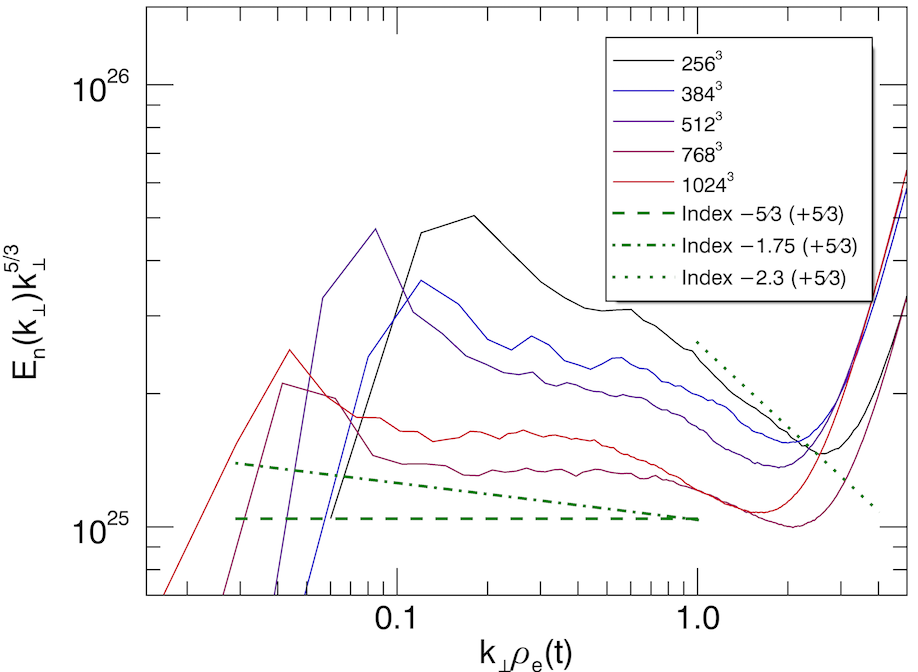}
   \centering
   \caption{\label{fig:spectra_compensated} Measurement of inertial range: spectra of magnetic energy $E_{\rm mag}(k_\perp)$, electric energy $E_{\rm elec}(k_\perp)$, bulk fluid kinetic energy $E_{\rm bulk}(k_\perp)$, and particle density $E_n(k_\perp)$, all compensated by $k_\perp^{5/3}$ and averaged over early stages of fully developed turbulence for $\sigma_0 = 0.5$ series of simulations with system sizes $L/2\pi\rho_{e0} \in \{ 27.2, 40.7, 54.3, 81.5, 108.6 \}$. Compensated power-law fits are shown for $k_\perp^{-5/3}$ (dashed line) and $k_\perp^{-1.75}$ (dashed-dotted line) in the inertial range ($k_\perp \rho_e < 1$). Approximate power-law fits to the kinetic range ($k_\perp \rho_e > 1$) are also shown (dotted lines).}
 \end{figure*}
 
To better characterize the inertial range scaling, we show several different power spectra compensated by $k_\perp^{5/3}$ in Fig.~\ref{fig:spectra_compensated}. In addition to the fiducial case, we show the complete~$\sigma_0 = 0.5$ series of simulations with varying size to demonstrate convergence. We find that~$E_{\rm mag}(k_\perp)$ and~$E_{\rm elec}(k_\perp)$ approach the classic~$k_\perp^{-5/3}$ scaling with increasing system size (with a fit close to $k_\perp^{-1.75}$ for the fiducial case); simulations with smaller system sizes have steeper apparent initial ranges. The asymptotic scaling cannot be verified without even larger simulations, although we note the scalings appear converged when comparing the $768^3$ and $1024^3$ cases. The bulk fluid kinetic energy spectrum $E_{\rm bulk}(k_\perp)$ is somewhat steeper than the other spectra and does not exhibit a clear inertial range. We also show the particle density spectrum~$E_n(k_\perp)$, which has a robust $k_\perp^{-5/3}$ scaling (dominated by particle noise below kinetic scales); this is consistent with predictions from passive advection of small density fluctuations in weakly compressible MHD turbulence \citep{montgomery_etal_1987, goldreich_sridhar_1995, schekochihin_etal_2009}.

\begin{figure*}
\includegraphics[width=0.65\columnwidth]{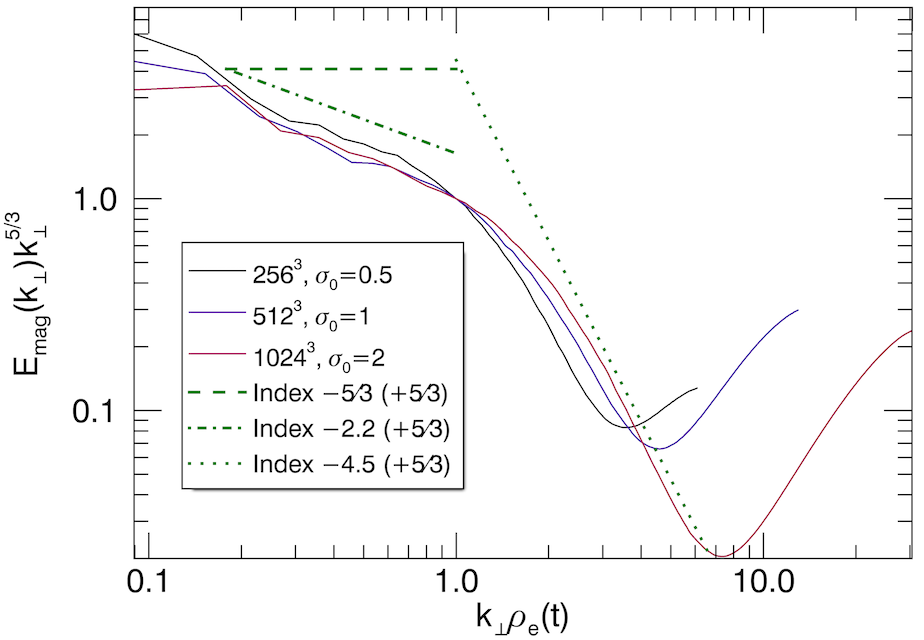}
\includegraphics[width=0.65\columnwidth]{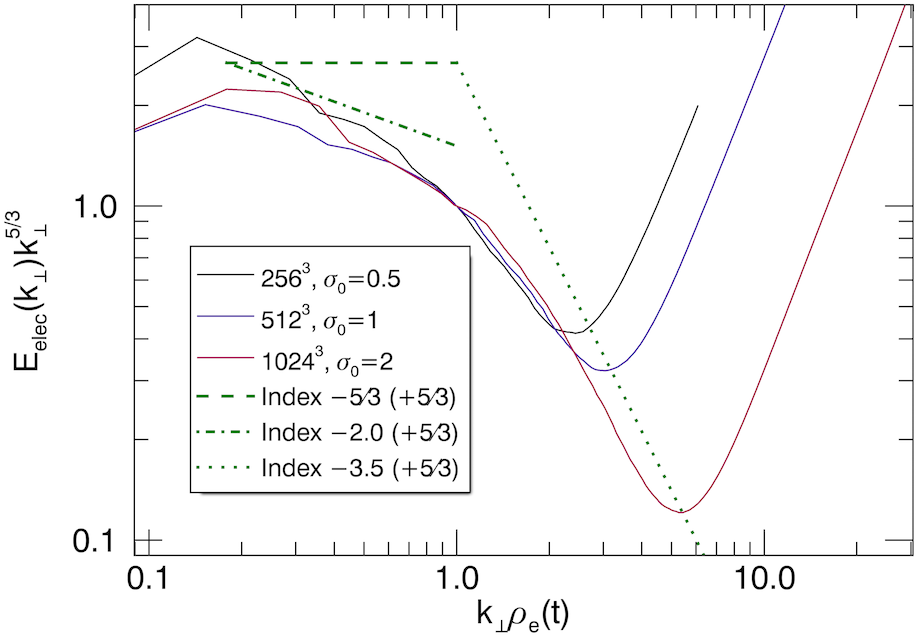}
\includegraphics[width=0.65\columnwidth]{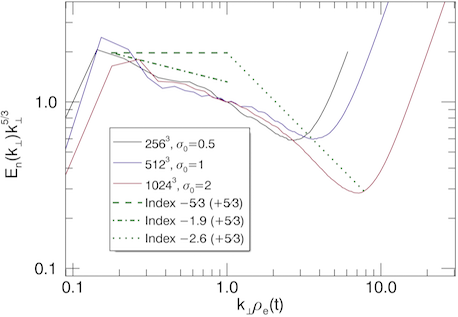}
   \centering
   \caption{\label{fig:spectrum_sigma2} Magnetic energy spectrum $E_{\rm mag}(k_\perp)$, electric energy spectrum $E_{\rm elec}(k_\perp)$, and density spectrum $E_n(k_\perp)$, all compensated by $k_\perp^{5/3}$, for $\sigma_0 = 2$, $1024^3$ simulation (red) along with smaller $\sigma_0 = 1$, $512^3$ (magenta) and $\sigma_0=0.5$, $256^3$ (black) simulations with equal values of $\xi_0 = \sigma_0 \rho_{e0}/L \approx 0.003$. Normalization is arbitrary. A power law with index (before compensation) of $-5/3$ is shown (dashed), along with empirical power-law fits in the inertial range (dashed-dotted) and in the kinetic range (dotted).}
 \end{figure*}
 
For comparison, we next consider power spectra for the $\sigma_0 = 2$, $1024^3$ simulation as a representative high-$\sigma_0$ case. Due to the rapid transient heating associated with high $\sigma_0$, the inertial range is significantly narrower than in the fiducial case. This is a general result: our simulations with high $\sigma_0$ tend to have steeper inertial-range spectra than the low $\sigma_0$ cases, consistent with a smaller effective system size ($L/2\pi\rho_e$). To illustrate this, in Fig.~\ref{fig:spectrum_sigma2}, we compare power spectra for magnetic, electric, and density fluctuations from the $\sigma_0 = 2$, $1024^3$ simulation to two smaller ($512^3$ and $256^3$) simulations with lower $\sigma_0$, such that all three cases have identical values of $\xi_0 = \sigma_0 \rho_{e0}/L \approx 0.003$. The inertial-range spectra in all three simulations are similar, with apparent index in the neighborhood of $-2$, demonstrating that simulations with equal values of $\xi_0$ tend to develop similar turbulence statistics. With a larger system size, we expect that the $\sigma_0 = 2$ simulation should converge to the spectrum at low $\sigma$.

The $\sigma_0 = 2$, $1024^3$ simulation has relatively well-resolved kinetic range, at the expense of the inertial range, which invites precision measurements of the kinetic-range spectra. As seen in Fig.~\ref{fig:spectrum_sigma2}, we find that the kinetic-range spectra are consistent with power laws having indices $-4.5$ for $E_{\rm mag}(k_\perp)$, $-3.5$ for $E_{\rm elec}(k_\perp)$, and $-2.6$ for $E_n(k_\perp)$. We cannot rule out steeper power laws or slow exponential decline due to the presence of particle noise. The existence of power-law spectra at sub-Larmor scales would imply a kinetic cascade, either mediated by kinetic modes or by phase-space structure. In this light, it is reasonable to compare our results to the theoretical predictions for an entropy cascade \citep{schekochihin_etal_2009}. In the absence of a relativistic theory, we compare to the predictions for non-relativistic, weakly-collisional, electron-proton plasma, which yields indices $-16/3$ for $E_{\rm mag}(k_\perp)$, $-4/3$ for $E_{\rm elec}(k_\perp)$, and $-10/3$ for $E_n(k_\perp)$. It is possible that, given a longer kinetic range, our magnetic and density spectra will further steepen to agree with these predictions, but the electric energy spectrum is much steeper than the prediction and does not show any signs of becoming shallower with increasing resolution. Hence, our results are not fully consistent with the non-relativistic entropy cascade predictions, motivating the future development of a relativistic theory.
 
\begin{figure}
\includegraphics[width=0.9\columnwidth]{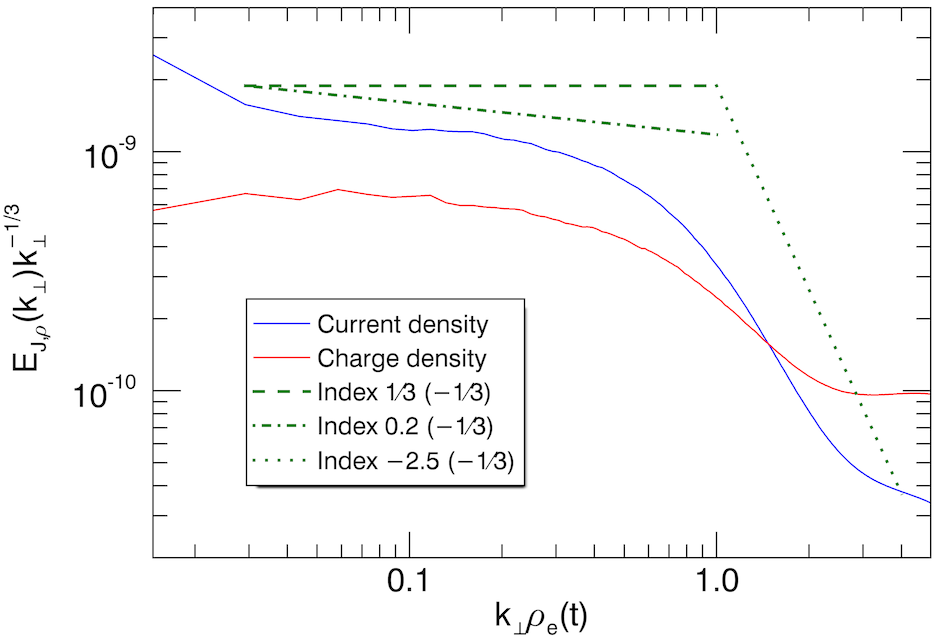}
\includegraphics[width=0.9\columnwidth]{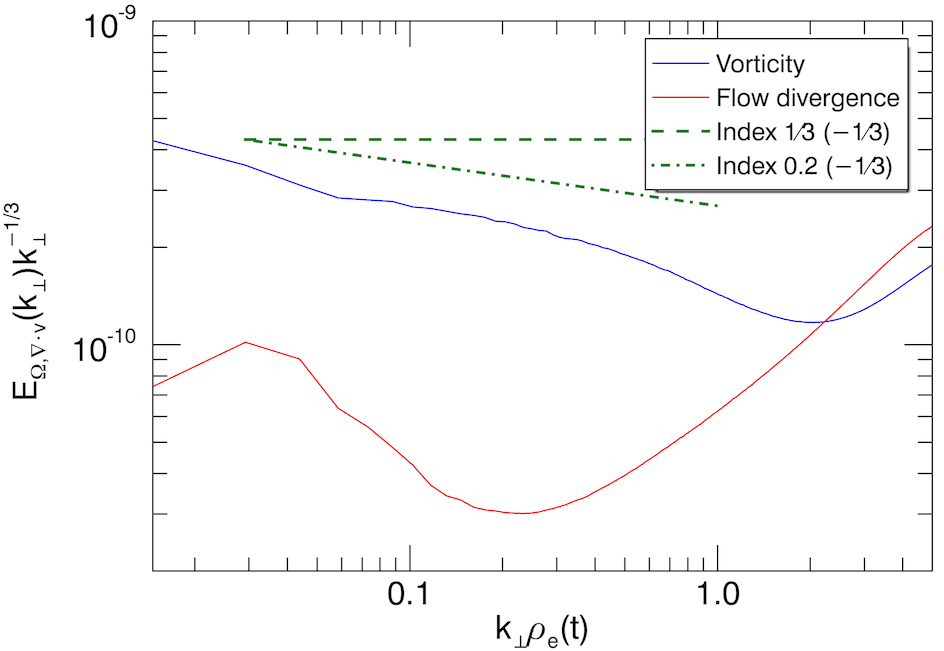}
   \centering
   \caption{\label{fig:spectrum_current} Top panel: Power spectrum for current density, $E_{\boldsymbol{J}}(k_\perp)$, and for charge density, $E_{\rho}(k_\perp)$, compensated by $k_\perp^{-1/3}$. Bottom panel: Power spectrum for vorticity, $E_{\boldsymbol{\Omega}}(k_\perp)$, and for divergence of flow, $E_{\nabla \cdot \boldsymbol{v}_f}(k_\perp)$, compensated by $k_\perp^{-1/3}$. Power-laws with indices $1/3$ (dashed), $0.2$ (dashed-dotted), and $-2.5$ (dotted) before compensation are shown for reference.}
 \end{figure}

The energy spectra discussed above are all peaked at small $k_\perp$, implying that most of the power is contained in large-scale fluctuations. In contrast, quantities made up of gradients of those fields can be peaked at small scales. Classical examples include the vorticity $\boldsymbol{\Omega} = \nabla \times \boldsymbol{v}_f$ and current density $\boldsymbol{J}$. If we neglect the displacement current, then $\tilde{\boldsymbol{J}} \sim (c/4\pi) i \boldsymbol{k} \times \tilde{\boldsymbol{B}}$, implying that the power spectrum of current density is $E_{\boldsymbol{J}}(k_\perp) \propto k_\perp^2 E_{\rm mag}(k_\perp)$, yielding a $k_\perp^{1/3}$ inertial range for MHD turbulence in the Goldreich-Sridhar picture. Similarly, since charge density is given by $\tilde{\rho} = i \boldsymbol{k} \cdot \tilde{\boldsymbol{E}}/4\pi$, the power spectrum for charge density is $E_{\rho}(k_\perp) \propto k_\perp^2 E_{\rm elec}(k_\perp)$, also yielding a $k_\perp^{1/3}$ scaling for standard inertial-range MHD turbulence. We confirm these scalings in the compensated spectra $E_{\boldsymbol{J}}(k_\perp)k_\perp^{-1/3}$ and $E_{\rho}(k_\perp)k_\perp^{-1/3}$ in Fig.~\ref{fig:spectrum_current}. The spectrum for $\boldsymbol{J}$ appears to be slightly less than $1/3$ (by about $\sim 0.1$, similar to offset of the magnetic energy spectrum from $-5/3$). The peak of both spectra is near $k_\perp \rho_e \sim 1$, beyond which they steepen. In particular, $E_{\boldsymbol{J}}(k_\perp) \sim k_\perp^{-2.5}$ below kinetic scales, in agreement with the $k_\perp^{-4.5}$ magnetic energy spectrum. For comparison, we also show in Fig.~\ref{fig:spectrum_current} the power spectra for vorticity $\boldsymbol{\Omega}$ and for divergence of flow $\nabla \cdot \boldsymbol{v}_f$, compensated by $k_\perp^{-1/3}$ for similar reasons. We find that $\nabla \cdot \boldsymbol{v}_f$, which represents the compressive component of turbulence, does not exhibit power-law spectrum, declining very rapidly at small $k_\perp$ and dominated by noise over a significant range of $k_\perp$. 

 \subsection{Structure functions} \label{sec:sf}
 
 In this subsection, we deal with structure functions, which complement power spectra as a tool for characterizing the turbulent fluctuations as a function of scale. The structure function of order $q$ for the field $g(\boldsymbol{x},t)$ is given by  
 \begin{align}
 S^{(q)}_{g}(\delta\boldsymbol{x},t) = \langle |g(\boldsymbol{x}+\delta\boldsymbol{x},t)-g(\boldsymbol{x},t)|^q \rangle_{\boldsymbol{x}} \, ,
 \end{align}
where the angle brackets indicate an average over positions $\boldsymbol{x}$. Here, $g$ can be any field, such as the magnetic field vector, particle density, and so on. It is often assumed that, in the inertial range, the structure functions are power-law functions of scale,
\begin{align}
 S^{(q)}_{g}(\delta\boldsymbol{x},t) \sim |\delta\boldsymbol{x}|^{\zeta^{(q)}_g} \, ,
\end{align}
where $\zeta^{(q)}_g$ are the scaling exponents for the corresponding structure function. The set of structure functions gives a very detailed statistical description of the dynamics. The second-order structure function is linked to the energy spectrum by a Fourier transform, and hence $\zeta^{(2)}_g = - \alpha_g - 1$, where $\alpha_g < 2$ is index of the corresponding Fourier power spectrum (e.g., $\alpha_{\boldsymbol{B}} = -5/3$ for the magnetic energy spectrum implies $\zeta^{(2)}_{\boldsymbol{B}} = 2/3$ for the second-order magnetic structure function). The third-order structure function can be linked to the energy cascade rate under certain assumptions; in particular, the mixed third-order structure function of Els\"{a}sser fields can be proven to equal the energy cascade rate in non-relativistic, incompressible MHD \citep{politano_pouquet_1998, politano_pouquet_1998b}. Higher-order structure functions ($q > 3$) give information about the intermittency of turbulence \citep[e.g.,][]{she_leveque_1994, politano_pouquet_1995, chandran_etal_2015}.

As discussed in Appendix~\ref{sec:ppc}, structure functions can be contaminated at small scales due to particle noise in PIC simulations. This noise is nullified by smoothing the data prior to analysis. In the following, we filter the data by setting Fourier modes with $k L/2 \pi > N/4$ (i.e., wavelengths $\lambda < 4 \Delta x$) to zero prior to the measurement of the structure functions. We also focus mainly on structure functions for the magnetic field, which is less sensitive to noise than the other quantities.

\begin{figure}
\includegraphics[width=0.9\columnwidth]{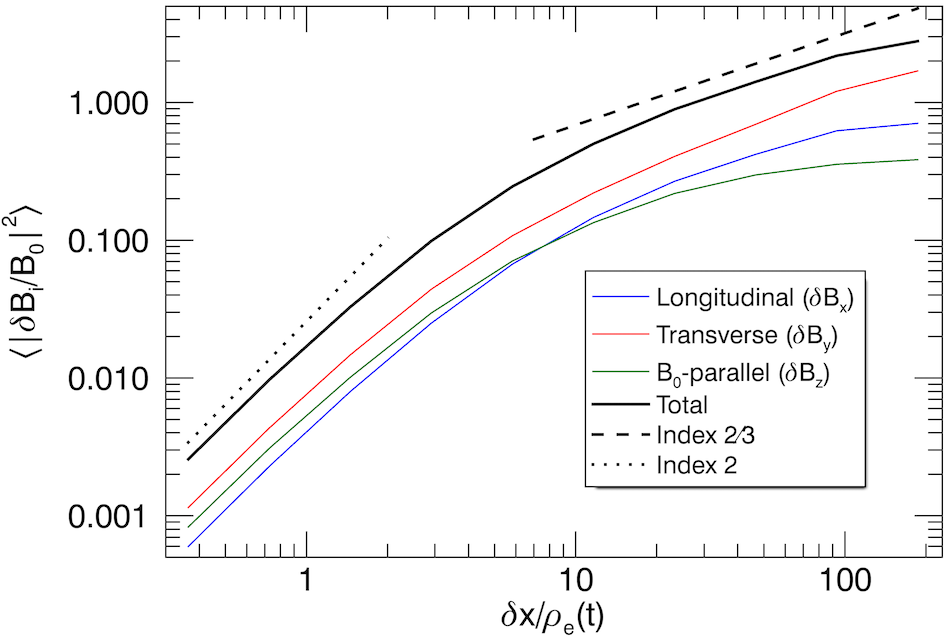}
\includegraphics[width=0.9\columnwidth]{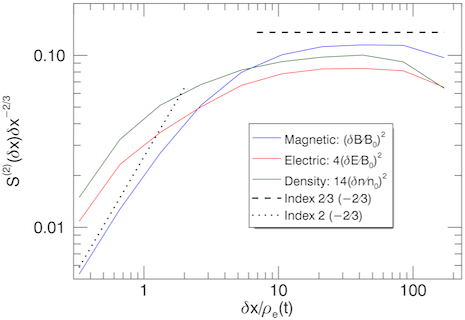}
   \centering
   \caption{\label{fig:sf1} Top panel: second-order structure functions for magnetic field components versus separation $\delta x$ (taken in the $x$ direction). These include the longitudinal component $S^{(2)}_{B_x}$ (blue), the transverse component $S^{(2)}_{B_y}$ (red), the $\boldsymbol{B}_0$-parallel component $S^{(2)}_{B_z}$ (green), and the total structure function $S^{(2)}_{\boldsymbol{B}}$ (black). For reference, $(\delta x)^{2/3}$ inertial-range scaling (dashed) and $(\delta x)^{2}$ sub-inertial scaling (dotted) are shown. Bottom panel: Second-order magnetic structure function $S^{(2)}_{\boldsymbol{B}}(\delta x)$ (blue), electric structure function $S^{(2)}_{\boldsymbol{E}}(\delta x)$ (red), and density structure function $S^{(2)}_{n}(\delta x)$ (green), all compensated by $\delta x^{-2/3}$. Normalization is arbitrary.}
   \end{figure}

We first consider the second-order magnetic structure function, $S^{(2)}_{\boldsymbol{B}}(\delta x)$, for separations arbitrarily taken in the $x$ direction (perpendicular to $\boldsymbol{B}_0$). In Fig.~\ref{fig:sf1}, we show the structure functions for the individual magnetic field components: the longitudinal component $S^{(2)}_{B_x}$, the transverse component $S^{(2)}_{B_y}$, and the $\boldsymbol{B}_0$-parallel component $S^{(2)}_{B_z}$. We also show the structure function for the entire vector, $S^{(2)}_{\boldsymbol{B}}$. We find that, to a good approximation, $S^{(2)}_{\boldsymbol{B}} \sim (\delta x)^{2/3}$ in the inertial range, consistent with the $-5/3$ index of the inertial-range magnetic energy spectrum. At sub-inertial scales, on the other hand, the structure function steepens to a scaling consistent with $S^{(2)}_{\boldsymbol{B}} \sim (\delta x)^{2}$. We interpret this scaling to be associated with the smooth variation of inertial-range fluctuations, which yields a $S^{(q)} \sim (\delta x)^{2q}$ scaling at small scales by Taylor expansion (as described in more detail in Appendix~\ref{sec:conjecture}). This scaling appears to dominate any signatures from the kinetic cascade; in this regard, the power spectrum is a more robust tool for characterizing the kinetic range. The break between the inertial and sub-inertial range appears to be rather broad, extending from $\delta x/\rho_e \sim 1$ to $\delta x/\rho_e \sim 10$.

We compare $S^{(2)}_{\boldsymbol{B}}(\delta x)$ to the second-order structure functions for electric field and density, $S^{(2)}_{\boldsymbol{E}}(\delta x)$ and $S^{(2)}_{n}(\delta x)$, respectively, in the second panel of Fig.~\ref{fig:sf1}. For clarity, we compensate the structure functions by $\delta x^{-2/3}$. All three structure functions are consistent in the inertial range: $S^{(2)}_{\boldsymbol{B}}(\delta x) \sim S^{(2)}_{\boldsymbol{E}}(\delta x) \sim S^{(2)}_{n}(\delta x) \sim (\delta x)^{2/3}$, consistent with the classical predictions and power spectra. Structure functions for other quantities and for other simulations are also consistent with the corresponding power spectra.
 
\begin{figure}
\includegraphics[width=0.9\columnwidth]{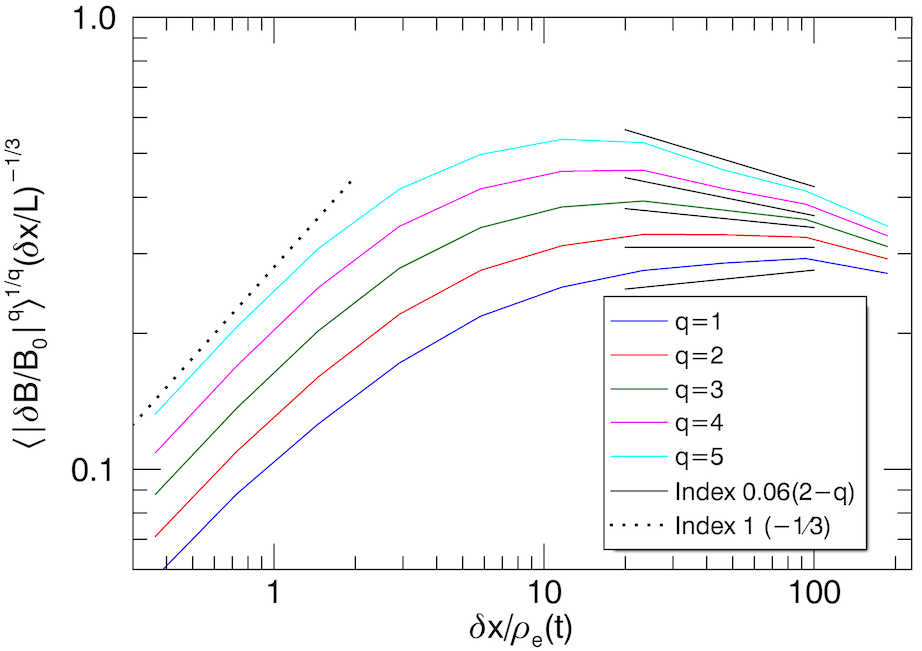}
   \centering
   \caption{\label{fig:multifractal} The multifractal scaling of structure functions: $(S^{(q)}_{\boldsymbol{B}})^{1/q} \delta x^{-1/3}$ versus $\delta x/\rho_e$ for $q \in \{ 1,2,3,4,5 \}$ (blue, red, green, magenta, and cyan, respectively). For reference, we also show corresponding empirical fits $\delta x^{0.06(q-2)}$ in the inertial range (solid black lines) and a $\delta x^{2/3}$ scaling corresponding to the sub-inertial range of smooth fluctuations (dotted black line).}
 \end{figure}

We next consider the magnetic structure functions of other orders. A self-similar cascade exhibits monofractal statistics, such that $S^{(q)}_{\boldsymbol{B}} \sim (\delta x)^{q \zeta^{(1)}_{\boldsymbol{B}}}$ for all orders $q$, i.e., there is only a single unspecified scaling exponent, which is associated with the fractal dimension of the turbulent field. In the Goldreich-Sridhar picture, the monofractal scaling yields exponents $\zeta^{(q)}_{\boldsymbol{B}} = q/3$. However, it has long been recognized that self-similarity is spontaneously broken due to intermittency. Hence, each structure function $S^{(q)}_{\boldsymbol{B}}$ is characterized by an independent scaling exponent, a property of multifractal statistics \citep[e.g.,][]{frisch_1995}. The complete spectrum of scaling exponents $\zeta^{(q)}$ can be used to characterize intermittency. Here, we present the scalings of $(S^{(q)}_{\boldsymbol{B}})^{1/q}$, compensated by $\delta x^{-1/3}$, for $1 \le q \le 5$ in Fig.~\ref{fig:multifractal}. It is evident that the higher-order structure functions depart from the $(\delta x)^{q/3}$ monofractal scaling in the inertial range, becoming relatively shallower with increasing $q$. In particular, we find that $(S^{(q)}_{\boldsymbol{B}})^{1/q} \sim (\delta x)^{1/3 + 0.06(2-q)}$ provides a good empirical fit\footnote{This empirical formula is reminiscent of the log-normal model \citep{kolmogorov_1962}, which provides a reasonable fit in MHD turbulence \citep[e.g.,][]{zhdankin_etal_2016a}.} to the given data, implying $\zeta^{(q)}_{\boldsymbol{B}} \sim q[1/3 + 0.06(2-q)]$. A proper measurement of the scaling exponents beyond this simple formula demands robust statistics and a broad inertial range, neither of which are available in our present simulations; for this reason, we defer a comparison to phenomenological theories to future work. In sub-inertial range, the structure functions become parallel, implying monofractal scaling (much like in the solar wind kinetic range; see \cite{kiyani_etal_2009}); however, this is likely an artifact of all structure functions capturing the smooth inertial-range fluctuations. In summary, the multifractal spectrum of structure functions confirms that our system is classically intermittent in the inertial range, qualitatively consistent with earlier numerical studies of MHD turbulence \citep{muller_biskamp_2000, biskamp_muller_2000, muller_etal_2003}. 
 
 \subsection{Fluctuation anisotropy} \label{sec:anisotropy}

Finally, we employ structure functions to characterize scale-dependent anisotropy. In particular, we aim to test critical balance \citep{goldreich_sridhar_1995}, which predicts that fluctuations are anisotropic with respect to the local background field, such that $k_\parallel \sim k_\perp^{2/3} L^{-1/3}$ for the bulk of magnetic fluctuations (as described in Sec.~\ref{sec:turb_back}). As noted in previous works on critical balance in MHD simulations \citep{cho_vishniac_2000, cho_etal_2002}, it is incorrect to test critical balance in the coordinate system relative to the global magnetic field $\boldsymbol{B}_0$, since the tilt of the local mean field from $\boldsymbol{B}_0$ can be sufficient to disrupt the measurement in the field-parallel direction. This prevents critical balance from being directly measurable in a 2D Fourier power spectrum or from structure functions in the global coordinate system. Instead, one must measure the structure functions in a coordinate system relative to the local magnetic field. Therefore, we consider the second-order magnetic structure function in local coordinates $(\delta x_\perp, \delta x_\parallel)$, given by
 \begin{align}
 S^{(2)}_{\boldsymbol{B}} (\delta x_\perp, \delta x_\parallel) =  \langle |\boldsymbol{B}(\boldsymbol{x}+\delta\boldsymbol{x})-\boldsymbol{B}(\boldsymbol{x})|^2 \rangle_{\boldsymbol{x}} \, , \nonumber \\
 \delta x_\parallel = \boldsymbol{\delta x} \cdot \hat{\boldsymbol{B}}_{\rm loc} \, , \nonumber \\
 \delta x_\perp = [(\delta x)^2 - (\delta x_\parallel)^2]^{1/2} \, ,
 \end{align}
 where the local mean field is defined by $\boldsymbol{B}_{\rm loc}(\boldsymbol{x},\delta \boldsymbol{x}) = [\boldsymbol{B}(\boldsymbol{x}+\delta\boldsymbol{x}) + \boldsymbol{B}(\boldsymbol{x})]/2$.

  \begin{figure}
  \includegraphics[width=0.9\columnwidth]{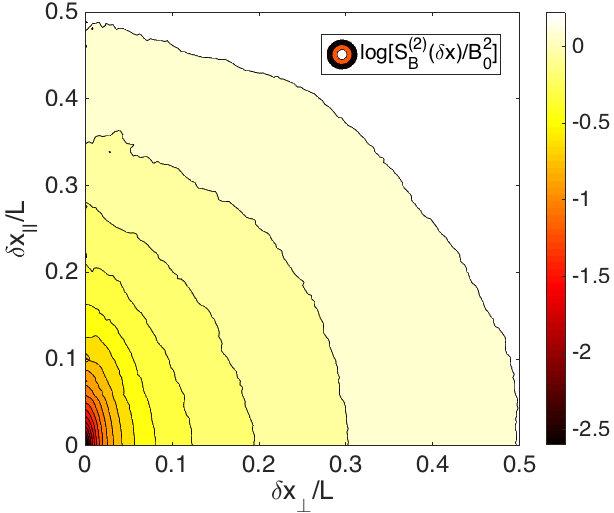}
   \centering
   \caption{\label{fig:ani_contour} Demonstration of scale-dependent anisotropy: contours of the second-order magnetic structure function $S^{(2)}_{\boldsymbol{B}}$ on the $(\delta x_{\perp}, \delta x_{\parallel})$ plane.}
 \end{figure}

We show the contours of the second-order magnetic structure function $S^{(2)}_{\boldsymbol{B}}$ on the $(\delta x_{\perp}, \delta x_{\parallel})$ plane in Fig.~\ref{fig:ani_contour}. The structure function is isotropic at large scales ($\delta x \sim L/2$), but becomes increasingly anisotropic at small scales. At small scales, the variations are weaker in the $\delta x_\parallel$ direction, indicating that fluctuations become elongated along the local mean magnetic field. This confirms a scale-dependent anisotropy.
 
   \begin{figure}
\includegraphics[width=0.9\columnwidth]{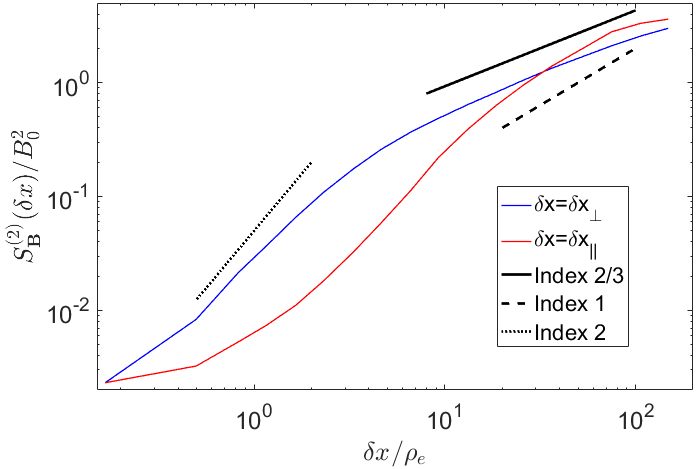}
\includegraphics[width=0.9\columnwidth]{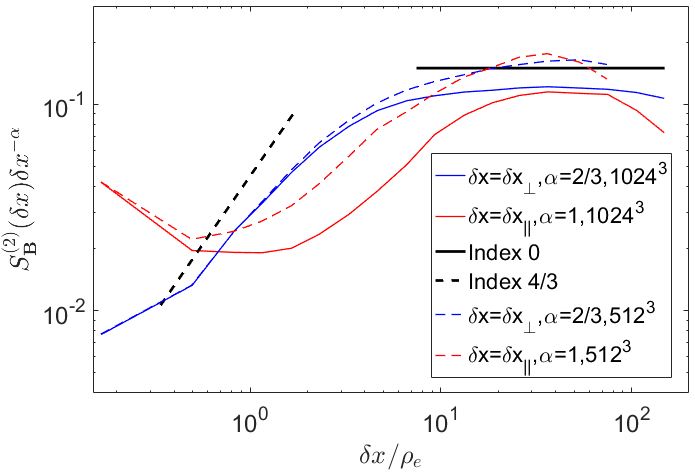}
\includegraphics[width=0.9\columnwidth]{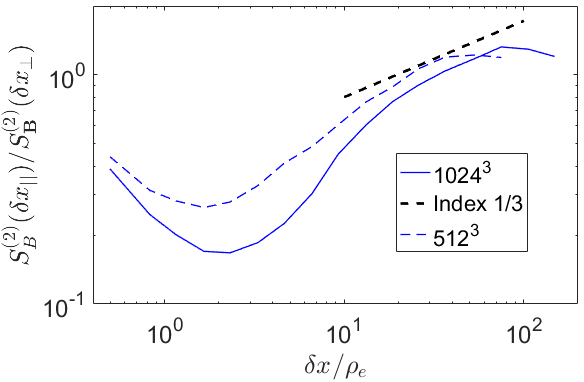}
   \centering
   \caption{\label{fig:ani_scaling} Verification of critical balance. First panel: Scaling of $S^{(2)}_{\boldsymbol{B}}$ in the perpendicular direction $\delta x_\perp$ (blue) and parallel direction $\delta x_\parallel$ (red) relative to the local mean field. For reference, we also show the predicted scalings: $\delta x_\perp^{2/3}$ (black solid line) and $\delta x_{\parallel}$ (dashed line) from critical balance, and $\delta x_\perp^2$ sub-inertial range scaling due to smooth fluctuations (dotted line). Second panel: compensated plots $S^{(2)}_{\boldsymbol{B}}(\delta x_\perp) \delta x_\perp^{-2/3}$ (blue) and $S^{(2)}_{\boldsymbol{B}}(\delta x_{\parallel}) \delta x_{\parallel}^{-1}$ (red) for fiducial $1024^3$ case (solid) and for smaller $512^3$ case with $\sigma_0 = 0.5$ (dashed). For reference, we also show a flat scaling corresponding to the critical balance prediction (black solid line) and a $\delta x_\perp^{4/3}$ scaling corresponding to a smooth sub-inertial range (dashed line). Third panel: ratio $S^{(2)}_{\boldsymbol{B}}(\delta x_{\parallel}) /S^{(2)}_{\boldsymbol{B}}(\delta x_\perp)$ for the fiducial case and the smaller $512^3$ case, along with the critical balance prediction of $\delta x^{1/3}$.}
 \end{figure}
 
In order to quantitatively test the critical balance conjecture, we measure $S^{(2)}_{\boldsymbol{B}}$ separately along the two axes of the $(\delta x_\perp,\delta x_{\parallel})$ plane, which we denote $S^{(2)}_{\boldsymbol{B}}(\delta x_\perp)$ and $S^{(2)}_{\boldsymbol{B}}(\delta x_\parallel)$. The result is shown in Fig.~\ref{fig:ani_scaling}, both without compensation and with compensation by the $\delta x_\perp^{2/3}$ and $\delta x_\parallel$ scalings predicted by critical balance. We find an excellent agreement with the predicted scalings in the inertial range - in particular, $S^{(2)}_{\boldsymbol{B}}(\delta x_\perp) \sim \delta x_\perp^{2/3}$ across an order of magnitude in scale, while $S^{(2)}_{\boldsymbol{B}}(\delta x_\parallel) \sim x_\parallel$ over roughly a factor of four in scale. The perpendicular break occurs near $\delta x_\perp \sim 6 \rho_e$ and the parallel break occurs near $\delta x_\parallel \sim 20 \rho_e$, also broadly consistent with critical balance. The sub-inertial range scaling appears to be consistent with the $\delta x_\perp^2$ smooth scaling associated inertial-range fluctuations, although possibly shallower due to the limited kinetic range. We also show the ratio $S^{(2)}_{\boldsymbol{B}}(\delta x_{\parallel}) /S^{(2)}_{\boldsymbol{B}}(\delta x_\perp)$ versus $\delta x$ (in the corresponding directions) in Fig.~\ref{fig:ani_scaling}, which agrees well with the critical balance prediction of $\delta x^{1/3}$ for the fidicual case.
  
  \section{Discussion}  \label{sec5}
  
Relativistic pair plasmas provide an opportune setting for the theoretical and numerical exploration of kinetic turbulence. The Vlasov-Maxwell equations take a simple form in this limit, with only two independent characteristic kinetic scales $\rho_e$ and $d_e$. This makes it possible to achieve sufficient scale separation in 3D PIC simulations to recover dynamics in the MHD inertial range, which is challenging for kinetic codes in the non-relativistic regime \citep[e.g.,][]{makwana_etal_2015, makwana_etal_2017} and even for relativistic fluid codes \citep{zrake_macfadyen_2011, zrake_macfadyen_2012, radice_rezzolla_2013}. Since MHD is a rigorous large-scale limit of the kinetic equations, any deviations of kinetic turbulence from the well-established MHD results would be of immense interest. PIC simulations are formulated from first principles and thus do not rely on the various MHD assumptions (isotropic pressure, prescribed equation of state, thermal equilibrium, collisional dissipation, etc.) which may be violated in certain parameter regimes or in localized regions.

In this work, we demonstrated that present-day computational resources are capable of bridging the gap between fluid and kinetic regimes in turbulent relativistic plasmas. The statistical properties of turbulence in our PIC simulations agree favorably with classical MHD turbulence phenomenology (in the non-relativistic Goldreich-Sridhar framework) and with previous MHD simulations in the literature. Going beyond this, our results also indicate that ultra-relativistic plasma temperatures and near-relativistic turbulence motions (i.e., $\sigma \sim 1$) do not substantially alter the nature of the turbulent cascade. Needless to say, larger simulations will be essential to fully explore turbulence in the large-system limit, including to identify the precise indices of the power spectra (e.g., to determine whether the magnetic energy spectrum wholly converges to the $k_\perp^{-5/3}$ scaling, or whether there are corrections due to intermittency, dynamic alignment, relativity, or kinetic effects) and to characterize higher-order statistics.
 
In addition to capturing the historically well-studied MHD inertial range, our PIC simulations unveil the transition of the cascade to the kinetic range at small scales. We performed pioneering measurements of power-law spectra at sub-Larmor scales, which may be explained by a kinetic cascade. The measured sub-Larmor spectra for magnetic and density fluctuations are slightly shallower than the non-relativistic predictions from an entropy cascade derived in \cite{schekochihin_etal_2009} (and tentatively measured in, e.g., \cite{schoeffler_etal_2014}), while the electric energy spectrum is substantially steeper (index $-3.5$ rather than $-4/3$). Hence, our results are not completely consistent with the non-relativistic entropy cascade predictions, pointing to the need for a relativistic extension of the theory. We do believe, however, that a qualitatively similar cascade process may occur in our system. Our measurements of the kinetic range are limited by (1) the kinetic scales being only minimally resolved ($\rho_e \sim d_e \gtrsim \Delta x$), due to our preference of maximizing the extent of the inertial range, and (2) contamination by particle noise, due to a limited number of particles per cell. In particular, power spectra and structure functions for quantities other than the magnetic field are strongly affected by particle noise at small scales. We defer a more concentrated investigation of the kinetic cascade to future work involving simulations with a larger number of particles and better resolved kinetic scales.
 
The simulations in our present study explored magnetizations in the neighborhood of unity, $\sigma \sim 1$, which is the most numerically tractable case. The regime of $\sigma \ll 1$ is characterized by (1) kinetic scale separation $\rho_e \gg d_e$ and (2) non-relativistic bulk motions due to $\delta v \sim v_A \ll c$ (requiring a larger number of timesteps to simulate for a given duration in terms of dynamical times). The regime of $\sigma \gg 1$ is characterized by (1) kinetic scale separation $d_e \gg \rho_e$, (2) strong compressibility ($\delta v \sim v_A \sim c_s \sim c$, where $c_s$ is the speed of sound, leading to load imbalance in the simulation), and (3) rapid plasma heating ($\sigma \to 1$ on timescales $t \sim L/c$). These issues make it numerically challenging to perform a broad $\sigma$ scan of developed turbulence, which could address important topics such as the location of the spectral break \citep{boldyrev_etal_2015}, the universality of the kinetic range, and the cascade of compressive fluctuations.

One implication of our results is that sustained relativistic turbulence, which requires $\sigma > 1$, is unrealizable in natural systems with constant energy injection rate and inefficient cooling mechanisms. This is understood as follows: to develop relativistic motions, the turbulent energy must exceed the effective plasma mass. However, the dissipation of this turbulent energy increases the mass density so that the two become comparable ($\sigma \lesssim 1$) within a turnover time. A study of fully-developed relativistic turbulence therefore requires either 1) an energy injection rate that steadily increases in time, to compensate for increasing plasma inertia or 2) a prescribed cooling mechanism, such as radiative cooling in an optically thin plasma. In either case, turbulence statistics at high $\sigma$ can be compared to force-free MHD, which resembles the non-relativistic case \citep{thompson_blaes_1998, cho_2005, cho_lazarian_2013}. We note these considerations do not preclude transient relativistic motions from developing in decaying turbulence without an energy sink, which may better represent many astrophysical systems (where turbulence occurs in outflowing plasma \citep[e.g.,][]{zrake_2016} or is impulsively triggered by instabilities \citep[e.g.,][]{nalewajko_etal_2016, yuan_etal_2016}).
  
  \section{Conclusions}  \label{sec6}
  
In this paper, we applied PIC simulations to investigate driven turbulence in magnetized, collisionless, relativistically-hot plasmas with magnetizations of order unity, $\sigma \sim 1$. Our PIC simulations successfully reproduce large-scale MHD turbulence, as indicated by (1) the inertial-range magnetic energy spectrum approaching a $k_\perp^{-5/3}$ scaling, with a similar bulk fluid energy spectrum; equivalently, the second-order magnetic structure function approaches a $(\delta x)^{2/3}$ scaling, (2) the magnetic structure functions exhibiting scale-dependent mean-field anisotropy consistent with critical balance ($\delta x_\parallel \sim \delta x_\perp^{2/3}$), and (3) the log-normal distribution of particle density and internal energy per particle, which are related by the $4/3$ adiabatic index predicted for an ultra-relativistic gas. We also identified signatures of intermittency, including the formation of current sheets and a multifractal spectrum of magnetic structure functions. This validates the standard MHD turbulence phenomenology for systems consisting of collisionless, relativistic plasmas with a population of nonthermal energetic particles \citep{zhdankin_etal_2017}.

In addition, we obtained new measurements of turbulent fluctuations in the kinetic regime (i.e., at scales below the characteristic Larmor scale). In particular, we measured approximate $k_\perp^{-4.5}$ magnetic energy spectrum, $k_\perp^{-3.5}$ electric energy spectrum, and $k_\perp^{-2.6}$ density spectrum in the kinetic regime. Structure functions are ill-suited for characterizing the kinetic range, since they capture the smooth variation of inertial-range fluctuations rather than the steep spectrum of the kinetic cascade; also, structure functions appear to be more strongly affected by particle noise than power spectra.

This work establishes that PIC simulation is a viable first-principles numerical approach to investigating turbulence in relativistic plasmas. Further numerical and theoretical work will be required to characterize fully the nature of turbulence in this regime and to explore the parameter space (including degrees of freedom associated with the driving mechanisms and system geometry). For the sake of brevity, we have only considered the statistics of turbulent fluctuations in this paper. In a follow-up paper, we will describe the particle statistics in the same set of simulations, addressing the properties of turbulent particle acceleration and implications for observations of high-energy astrophysical systems.
 
\section*{Acknowledgements}

The authors thank Stanislav Boldyrev, Jason TenBarge, Nuno Loureiro, Matt Kunz, and Alex Schekochihin for helpful discussions. The authors acknowledge support from NSF grant AST-1411879 and NASA ATP grants NNX16AB28G and NNX17AK57G. D.A.U. gratefully acknowledges the hospitality of the Institute for Advanced Study and the support from the Ambrose Monell Foundation. An award of computer time was provided by the Innovative and Novel Computational Impact on Theory and Experiment (INCITE) program. This research used resources of the Argonne Leadership Computing Facility, which is a DOE Office of Science User Facility supported under Contract DE-AC02-06CH11357.





\bibliographystyle{mnras}
\bibliography{refs_all}



\appendix

\section{Convergence with number of particles} \label{sec:ppc}

One of the free numerical parameters in our PIC simulations is the total number of particles per cell, $N_{\rm ppc}$. The numerical accuracy generally increases with $N_{\rm ppc}$, but the computational cost of the simulations is also proportional to $N_{\rm ppc}$. Hence, $N_{\rm ppc}$ must be optimized by convergence studies to attain a prescribed level of accuracy.

\begin{figure}
\includegraphics[width=0.9\columnwidth]{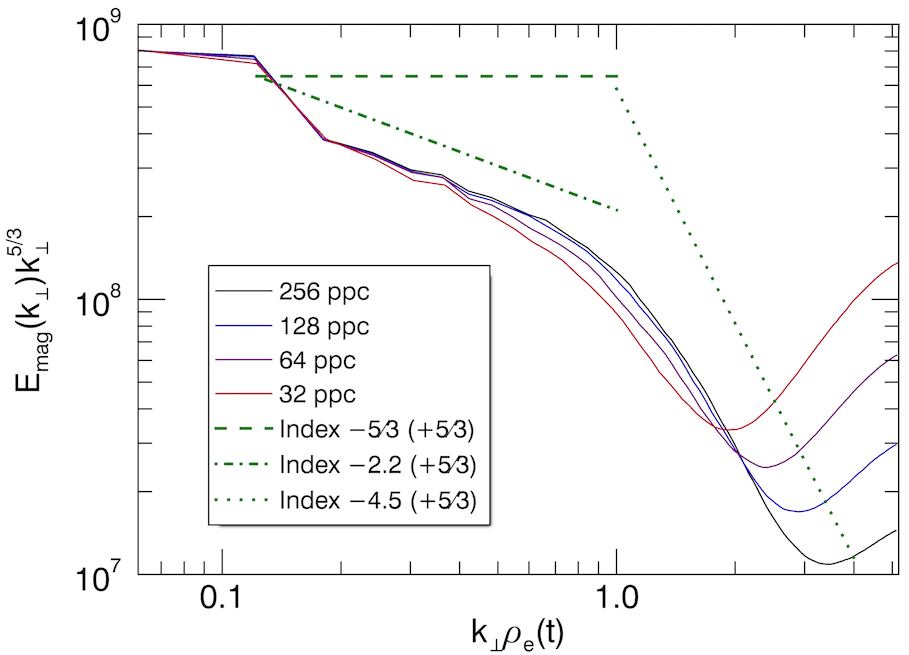}
\includegraphics[width=0.9\columnwidth]{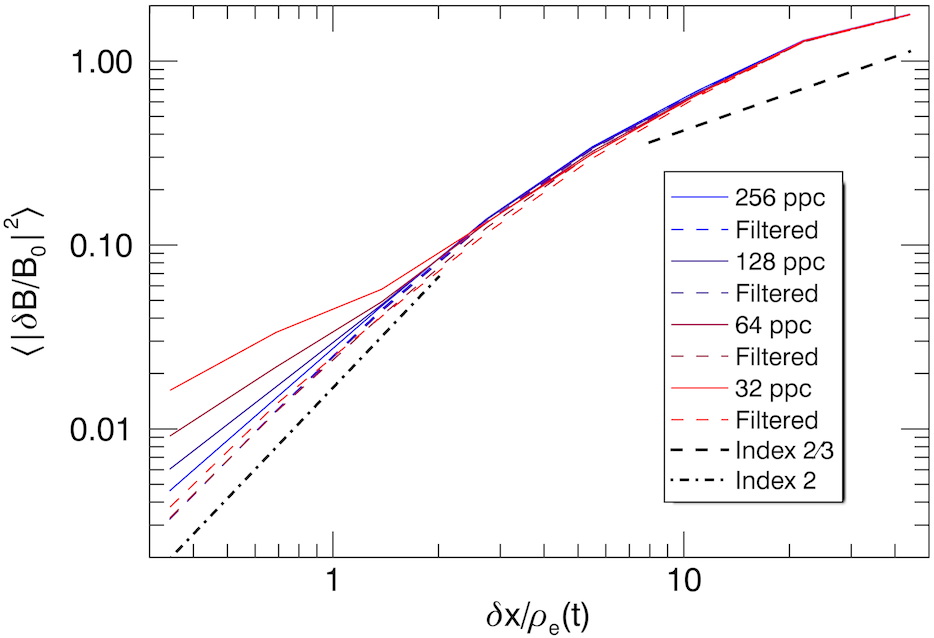}
   \centering
   \caption{\label{fig:spectrum_ppc} Left panel: Compensated magnetic energy spectrum, $E_{\rm mag}(k_\perp) k_\perp^{5/3}$, versus number of particles per cell in a series of $256^3$, $\sigma_0 = 0.5$ simulations. Power-law fits are shown for reference, as in Fig.~\ref{fig:spectrum_sigma2}. Right panel: Second-order magnetic structure function $S^{(2)}_{\boldsymbol{B}}(\delta x)/B_0^2$ for same series of simulations. Dashed lines indicate filtered cases.}
 \end{figure}

To demonstrate convergence of our results with respect to $N_{\rm ppc}$, we now describe results from a series of $256^3$, $\sigma_0 = 0.5$ simulations with $N_{\rm ppc} \in \{ 32, 64, 128, 256 \}$. In the first panel of Fig.~\ref{fig:spectrum_ppc}, we show the compensated magnetic energy spectrum for this series of simulations (averaged over $7.7 < t c / L < 10$). The most conspicuous difference between the simulations is the noise floor at large $k$, which has an amplitude $\sim 1/N_{\rm ppc}$. Hence, measurements of the kinetic range spectrum require relatively large $N_{\rm ppc}$ to make this noise floor negligible, especially for noisy quantities (e.g., particle density). The cases with $N_{\rm ppc} \ge 128$ show excellent agreement at scales above the noise floor, while the cases with $N_{\rm ppc} = 64$ and $N_{\rm ppc} = 32$ show deviations near the spectral break at $k_\perp \rho_e \sim 1$ and have steeper spectra than the higher $N_{\rm ppc}$ cases. This indicates that  $N_{\rm ppc} \sim 128$ is the optimal value to get robust power spectra for the given simulation resolution. In the second panel of Fig.~\ref{fig:spectrum_ppc}, we show the second-order magnetic structure function $S^{(2)}_{\boldsymbol{B}}$ for the same simulations. The particle noise once again affects the measurement at small scales, making the structure function scaling shallower than the converged scaling [$S^{(2)}_{\boldsymbol{B}} \sim (\delta x)^2$]. We find that filtering out Fourier modes with $k L/2 \pi > N/4$ (as applied in Sec.~\ref{sec:sf}) leads to a result that is insensitive to $N_{\rm ppc}$; coarse-graining gives a similar converged result. Spectra and structure functions for other quantities and for larger simulations have a similar dependence on $N_{\rm ppc}$.
 
 The error in energy conservation $R_{{\rm err},T}$ is another indicator of convergence with respect to $N_{\rm ppc}$. In Fig.~\ref{fig:spectrum_ppc}, we demonstrate accurate energy conservation by showing the total energy minus injected energy relative to its initial value, for the cases described above. Over the duration of $T \sim 10 L/c$, the cases with $N_{\rm ppc} \ge 64$ all have comparable errors, $R_{{\rm err},T} \sim 0.3 \%$. However, the case with $N_{\rm ppc} = 32$ has $R_{{\rm err},T} \sim 1 \%$, indicating that errors become dominated by the particle noise. At late times, energy slowly increases from the expected value, indicating a small amount of numerical heating. Empirically, for fixed $N_{\rm ppc}$ and duration (in terms of light-crossing times), we find that larger simulations have higher $R_{{\rm err},T}$ due to the accumulation of error over the larger number of timesteps (since the timestep is set by the Courant-Friedrichs-Lewy condition, $\Delta t \sim \Delta x/\sqrt{3} c$). Thus, many of our $384^3$ and $256^3$ simulations easily satisfy energy conservation to better than $1\%$. For the production simulations in Table.~\ref{table-sims}, several cases have $N_{\rm ppc} > 128$ to improve energy conservation and acquire better measurements of the kinetic range.
 
 \begin{figure}
\includegraphics[width=0.9\columnwidth]{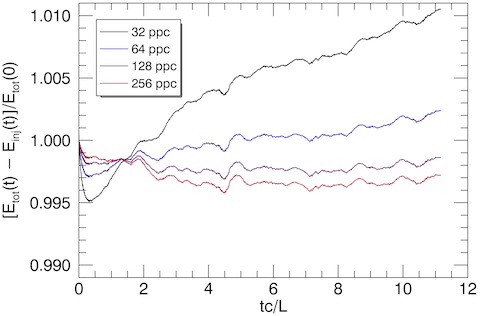}
   \centering
   \caption{\label{fig:spectrum_ppc} Total energy minus injected energy, $E_{\rm tot} - E_{\rm inj}$, relative to initial value for $256^3$, $\sigma_0 = 0.5$ simulations with varying number of particles per cell. Deviations from exact energy conservation are small, demonstrating numerical accuracy for $N_{\rm ppc} > 64$.}
 \end{figure}
 
 \section{Small-scale limit of structure functions} \label{sec:conjecture}
 
The second-order structure function and power spectrum are related as follows. First, expand the structure function to relate it to the autocorrelation function,
\begin{align}
S^{(2)}_{\boldsymbol{B}}(\delta\boldsymbol{x}) &\equiv \langle [\boldsymbol{B}(\boldsymbol{x} + \delta\boldsymbol{x}) - \boldsymbol{B}(\boldsymbol{x})]^2 \rangle_{\boldsymbol{x}} \nonumber \\
&= 2 \langle B^2 \rangle - 2 \langle \boldsymbol{B}(\boldsymbol{x} + \delta\boldsymbol{x}) \cdot \boldsymbol{B}(\boldsymbol{x}) \rangle_{\boldsymbol{x}} \, .
\end{align}
Fourier transform to obtain
\begin{align}
\int d^3\delta x S^{(2)}_{\boldsymbol{B}}(\delta\boldsymbol{x}) e^{-i\boldsymbol{k}\cdot\delta\boldsymbol{x}} = - 2 (2 \pi)^3 \langle B^2 \rangle \delta(\boldsymbol{k}) + 2 |\boldsymbol{B}(\boldsymbol{k})|^2 \, .
\end{align}
For a spectrum $E_{\boldsymbol{B}}(\boldsymbol{k}) \equiv |\boldsymbol{B}(\boldsymbol{k})|^2 \propto k^{-\alpha}$ with $\alpha < 2$, we obtain $S^{(2)}_{\boldsymbol{B}} \sim |\delta x|^{\alpha - 1}$. However, the result for $\alpha > 2$ is nontrivial. In particular, for $\alpha > 3$, structure functions will be dominated by the smooth variation of large-scale modes. For scales small relative to the fluctuations, we can perform a Taylor expansion of the magnetic field
\begin{align}
\boldsymbol{B}(\boldsymbol{x} + \delta\boldsymbol{x}) \sim \boldsymbol{B}(\boldsymbol{x}) + \delta\boldsymbol{x} \cdot \nabla \boldsymbol{B} (\boldsymbol{x}) \, .
\end{align}
The second-order structure function becomes
 \begin{align}
S^{(2)}_{\boldsymbol{B}}(\delta\boldsymbol{x}) = \langle [\boldsymbol{B}(\boldsymbol{x} + \delta\boldsymbol{x}) - \boldsymbol{B}(\boldsymbol{x})]^2 \rangle_{\boldsymbol{x}} \sim \delta x_j \delta x_i \left\langle \frac{\partial B_k}{\partial x_j} \frac{\partial B_k}{\partial x_i} \right\rangle_{\boldsymbol{x}} \, . \label{eq:smooth}
 \end{align}
For separations perpendicular to the mean magnetic field, $\delta \boldsymbol{x} \cdot \boldsymbol{B}_0 = 0$, only the term proportional to $\delta_{ij}$ survives in Eq.~\ref{eq:smooth}, leading to $S^{(2)}_{\boldsymbol{B}} \propto (\delta x)^2$. This quadratic scaling dominates any sufficiently steep kinetic spectrum. We note that the smoothness of turbulent fluctuations was previously exploited to predict the scaling of structure functions near the dissipation range in hydrodynamic turbulence, where the spectrum drops off exponentially \citep[e.g.,][]{stolovitzky_etal_1993, sirovich_etal_1994}.


\bsp	
\label{lastpage}
\end{document}